%% file: TEHM.tex
\documentclass[%
aps,
prd,
amssymb,
amsmath,
amsfonts, 
eqsecnum,
showpacs,
nofootinbib,
floatfix,
twocolumn,
twoside,
a4paper,
superscriptaddress,
longbibliography
]{revtex4-2}

\input{UsePackages}
\usepackage{graphicx}% Include figure files
\usepackage{dcolumn}% Align table columns on decimal point
\usepackage{bm}% bold math
\usepackage{tabularx}
\usepackage{multirow}
\usepackage{capt-of}
\usepackage{color}
\usepackage{url}
\usepackage[utf8]{inputenc}
\usepackage{placeins}
\usepackage{ulem}

\usepackage{graphicx}
\usepackage{booktabs}
\usepackage{adjustbox}
\usepackage{amsmath}

\usepackage{arydshln}
\usepackage{natbib}

\usepackage[nolist,nohyperlinks]{acronym}
\usepackage{xspace}
\usepackage[colorlinks]{hyperref}

\usepackage[caption=false]{subfig}

\usepackage{float}

\DeclareMathAlphabet{\mathbfsf}{\encodingdefault}{\sfdefault}{bx}{sl}

\newcommand{\be}{\begin{equation}}
\newcommand{\ee}{\end{equation}}
\newcommand{\bea}{\begin{eqnarray}}
\newcommand{\eea}{\end{eqnarray}}

\newcommand{\x}{\mathbf{\hat{x}}}
\newcommand{\y}{\mathbf{\hat{y}}}

\newcommand{\phT}{\textsc{IMRPhenomT}\xspace}
\newcommand{\phTE}{\textsc{IMRPhenomTEHM}\xspace}
\newcommand{\phTe}{\textsc{IMRPhenomTE}\xspace}
\newcommand{\phTHM}{\textsc{IMRPhenomTHM}\xspace}

\newcommand{\NRSur}{\textsc{NRSur7dq4}\xspace}

\newcommand{\seobe}{\textsc{SEOBNRv5EHM}\xspace}
\newcommand{\teobdali}{\textsc{TEOBResumDali}\xspace}

\newcommand{\chieff}{\chi_\mathrm{eff}}

\newcommand{\chirpMass}{\mathcal{M}}

\definecolor{dodgerblue}{HTML}{1E90FF}
\definecolor{viennared}{HTML}{DA0A14}
\definecolor{ctorange}{HTML}{FF6C0C}
\definecolor{granadagreen}{HTML}{078931}
\definecolor{wales}{HTML}{ff0038}
\definecolor{valenciacfred}{HTML}{ee3524}
\definecolor{barcelonafcgold}{HTML}{edbb00}
\definecolor{jam}{HTML}{A50B5E}
\definecolor{austriawien}{HTML}{441678}

\AtBeginDocument{%
  \hypersetup{
    citecolor=dodgerblue,
    linkcolor=dodgerblue,   
    urlcolor=dodgerblue}}

\newcommand{\UIB}{Departament de F\'isica, Universitat de les Illes Balears, IAC3 -- IEEC, Crta. Valldemossa km 7.5, E-07122 Palma, Spain}

\newcommand{\AEI}{Max Plank Institut für Gravitationsphysik (Albert Einstein Institut), Am M\"uhlenberg 1, Potsdam, Germany}

\newcommand{\ICE}
{Institut de Ci\`encies de l'Espai (ICE, CSIC), Campus UAB, Carrer de Can Magrans s/n, 08193 Cerdanyola del Vall\`es, Spain}

\newcommand{\Nikhef}
{Dutch National Institute for Subatomic Physics (Nikhef), Science Park 105, 1098 XG, Amsterdam, The Netherlands}

\newcommand{\ICS}{Astrophysik-Institut, Universit\"{a}t Z\"{u}rich, Winterthurerstrasse 190, 8057 Z\"{u}rich, Switzerland}

\allowdisplaybreaks[1]

\begin{document}

% ~~~~~~~~~~ Title & Abstract ~~~~~~~~~~ %
\title[ML]
{Time-domain phenomenological multipolar waveforms for
aligned-spin
binary black holes \\
in elliptical orbits}

\author{Maria de Lluc Planas}
\affiliation{\UIB}

\author{Antoni Ramos-Buades}
\affiliation{\UIB}

\author{Cecilio García-Quirós}
\affiliation{\ICS}
%\affiliation{\APC}

\author{Héctor Estellés}
\affiliation{\AEI}

\author{Sascha Husa}
\affiliation{\ICE}
\affiliation{\UIB}

\author{Maria Haney}
\affiliation{\Nikhef}

\date{\today}

\begin{abstract}
We introduce \phTE, a new phenomenological time-domain model for eccentric aligned-spin binary black holes.
Building upon the accurate quasi-circular \phTHM model, \phTE integrates the eccentric post-Newtonian (PN) dynamics, and introduces eccentric corrections into the waveform multipoles up to 3PN including spin effects. The model incorporates the dominant $(2,\pm2)$ spherical harmonic mode, as well as the subdominant modes $(2,\pm1), (3,\pm3), (4,\pm4)$ and $(5,\pm5)$, with the assumption that the binary has circularized by the time of merger. This approach ensures a smooth transition to the non-eccentric limit, and it is shown to provide an accurate quasi-circular limit against the \phTHM model.
When comparing against 28 public eccentric numerical relativity simulations from the Simulating eXtreme Spacetimes catalog, \phTE achieves lower than $2\%$ unfaithfulness, confirming its accurate description without calibration to numerical relativity eccentric datasets.
\phTE provides a reliable description of the evolution of eccentric black hole binaries with aligned spins and eccentricities lower than $e=0.4$ at a frequency of $10$ Hz, making it suitable for upcoming gravitational wave observing runs. 
We validate the model’s accuracy through parameter estimation studies, recovering injected parameters within 90\% credible intervals for three NR eccentric simulations and reanalyzing GW150914 and GW190521, obtaining results consistent with the literature.\\
\end{abstract}


\maketitle

%%%%%%%%%%%%%%%%%%%%%%%%%%%%%%%%%%%%%%%%%%%%%%%%%%%%%%%%%%%%%%%%%
%               INTRODUCTION
%%%%%%%%%%%%%%%%%%%%%%%%%%%%%%%%%%%%%%%%%%%%%%%%%%%%%%%%%%%%%%%%%
\section{Introduction}\label{sec:Introduction}

After several observation runs, the ground-based gravitational-wave detectors LIGO, Virgo and KAGRA (LVK) \cite{Aasi_2015, Acernese_2015, kagra_2021} have identified numerous binary black hole (BBH), binary neutron star (BNS), and black hole-neutron star (BH-NS) events \cite{gwtc1, gwtc2, gwtc21, gwtc3}.
With increasing detector sensitivity, future observations by LVK and upcoming detectors such as the Einstein Telescope (ET) \cite{Maggiore:2019uih}, Cosmic Explorer (CE) \cite{Reitze:2019iox}, and space-based detectors like LISA \cite{LISA:2022yao, PhysRevD.101.024024} are expected to reveal more source properties of these systems.
This will provide valuable insights into strong gravitational fields as well as support studies in gravitational theory, galaxy formation, cosmology, and stringent tests of general relativity~\cite{LISA:2022yao}.

Compact binary mergers are the only source of gravitational waves (GWs) observed by ground-based GW detectors. Their dynamics and associated GWs have been studied extensively, and detecting and analyzing these waves relies on constructing accurate waveform templates. %\cite{gwtc1, gwtc2, gwtc21, gwtc3}.  
Most inspiraling binaries detected by these ground-based detectors are likely formed through isolated binary evolution \cite{Stevenson:2017tfq} and are expected to circularize by the time they enter the detector frequency band \cite{PhysRev.136.B1224}. 
Indeed, most of the GW events detected by the LVK collaboration, as well as external groups \cite{Nitz_nov2021, Nitz_2023, Venumadhav:2019lyq, Olsen:2022pin,Wadekar:2023gea,Mehta:2023zlk}, are consistent with quasi-circular binaries.
However, some binaries may retain significant orbital eccentricity if they form via dynamical interactions in dense stellar environments like globular clusters or galactic nuclei \cite{Zevin:2018kzq, Stone:2016ryd}, or through the Kozai-Lidov mechanism in triple systems \cite{Wen:2002km, VanLandingham:2016ccd}.
Measuring eccentricity in GW signals can provide crucial insights into the origins and properties of binary systems. Several attempts have been made to search for signatures of eccentricity in these systems \cite{Abbott:2019hdd, Romero-Shaw:2019itr, Nitz:2019spj, Romero-Shaw:2020thy, Gayathri:2020coq, Favata:2021vhw, OShea:2021faf, Romero-Shaw:2021lcs, Ramos-Buades:2023yhy, Gupte:2024jfe}, and as the detector sensitivity increases, the detection of binaries with non-negligible orbital eccentricity is expected to increase. Thus, developing accurate waveform models is essential for understanding the astrophysical origin of the observed GW events.

State-of-the-art gravitational inspiral-merger-ringdown (IMR) waveform models accurately describe BBHs in quasi-circular (QC) orbits, including also the effects of black holes spins. These models are categorized into three main families: the Effective-One-Body (EOB)~\cite{Buonanno:1998gg, Buonanno:2000ef} formalism , including the SEOBNR models~\cite{Bohe:2016gbl, Cotesta:2018fcv,Ossokine:2020kjp, Pompili_2023, Ramos-Buades:2023ehm, Gamboa:2024hli, Liu:2019jpg, Cao:2017ndf} and TEOBResumS models~\cite{Akcay:2020qrj, Nagar:2018plt, Nagar:2018zoe, Nagar:2020pcj, Nagar_2024, nagar2025, Gamba:2024cvy}; the NRSurrogate models~\cite{Blackman:2017dfb, Varma:2019csw, Varma:2019vhw, Islam:2022laz, Rink:2024swg}, which interpolate between NR datasets; and the IMRPhenom approach~\cite{Husa:2015iqa,Khan:2015jqa,London:2017bcn,Pratten:2020ceb,Garcia-Quiros:2020qpx,Estelles:2021gvs,Husa:2015iqa,Khan:2015jqa,London:2017bcn,Pratten:2020ceb,Garcia-Quiros:2020qpx,Estelles:2020twz, Estelles:2020osj, Estelles:2021gvs}, known for its computational efficiency due to the use of closed-form expressions. For the QC aligned-spins subspace, the previously mentioned models have been calibrated to NR simulations, showing strong agreement in the region of parameter space where NR data is available (see e.g. Ref.~\cite{Colleoni:2020tgc}). 
However, in the precessing-spin case the more complex waveform morphology and the larger parameter space pose challenges for each modeling approach. For example, the \NRSur~\cite{Varma:2019vhw} model is limited in its coverage of mass ratio, spin magnitudes and waveform length. Instead, the state-of-the-art precessing models in the SEOBNR~\cite{Ramos-Buades:2023ehm} and IMRPhenom~\cite{Pratten:2021pro, Estelles:2021gvs} families rely on approximations to model precession without NR calibration, which offers broader applicability but at the cost of reduced accuracy. Nonetheless, NR calibrated precessing phenomenological models have recently been developed ~\cite{Eleanor_2021, Thompson:2023ase}.

Similar challenges exist for eccentric binaries. Publicly available NR eccentric simulations are limited, complicating both the development and validation of eccentric waveform models. However, recent progress has been made in this area. 
Eccentric spinning binaries are often characterized using the post-Newtonian (PN) formalism~\cite{Arun:2009mc, Ebersold:2019kdc, Henry:2023tka}, with the quasi-Keplerian (QK) parametrization commonly employed to describe their orbital motion~\cite{AIHPA_1985__43_1_107_0,Damour:1988mr, SCHAFER1993196}. Gravitational waveforms for inspiraling eccentric binaries have been developed both for the non-precessing case~\cite{Tiwari:2020hsu, Liu:2019jpg, Tanay:2019knc, Moore:2018kvz, Huerta:2014eca, Paul_2023, Sridhar_2024} and the most general systems~\cite{Klein:2018ybm, Klein:2021jtd, Klein:2013qda, Arredondo:2024nsl, morras2025}.

Assuming circularization before merger, hybrid IMR waveforms have been constructed by combining PN-inspiral waveforms with merger and ringdown signals derived from NR or the EOB formalism, which has become the common method for incorporating eccentricity into quasi-circular IMR waveform models for low eccentricity~\cite{Ramos-Buades:2019uvh, Hinder:2017sxy, Huerta:2017kez}. 
Within the EOB formalism, significant advancements have been made in this direction~\cite{Placidi:2021rkh, Khalil:2021txt, Albanesi:2021rby, Nagar:2021xnh, Nagar:2021gss, Liu:2021pkr, Chiaramello:2020ehz, Liu:2019jpg, Cao:2017ndf, Hinderer:2017jcs, Ramos-Buades:2021adz, Carullo:2023kvj}.
Additionally, a pioneering attempt has been made to produce generic IMR waveforms within the TEOB framework~\cite{Gamba:2024cvy, Liu_2023}.
There has also been development within the surrogate family; however, due to the limited number of NR eccentric simulations, the only available surrogate model is restricted to comparable-mass, non-spinning binaries~\cite{Islam:2021mha}.
Outside of these primary model families, other approaches to eccentric waveform modeling have been explored~\cite{Huerta:2016rwp, Huerta:2017kez, Chattaraj:2022tay, Manna:2024ycx, Paul:2024ujx, Carullo:2024smg}. Additionally, Ref.~\cite{Setyawati:2021gom} introduced a method to add eccentricity modulations to existing BBH QC models.
Similarly, the \texttt{gwNRHME} framework enables the conversion of multi-modal quasi-circular waveforms into multi-modal eccentric waveforms given the quadrupolar non-spinning eccentric contribution~\cite{Islam:2024bza, Islam:2024rhm,Islam:2024zqo}.

In this paper, we develop the first time-domain eccentric phenomenological waveform model, \phTE, which includes the $\{(l,|m|)=(2,2),(2,1),(3,3),(4,4),(5,5)\}$ spherical harmonics multipoles for aligned spins with two eccentric parameters, which is crucial for avoiding biases in parameter estimation~\cite{Clarke:2022fma,Ramos-Buades:2023yhy}. 
The model is based on the quasi-circular \phTHM model~\cite{Estelles:2020osj, Estelles:2020twz} and incorporates eccentric 3PN corrections in the non-spinning terms~\cite{Arun:2009mc, Ebersold:2019kdc}, and up to 3PN spin-orbit and spin-spin interactions~\cite{Henry:2023tka}. However, the default model limits spin contributions to 2PN for consistency with the underlying \phTHM QC framework.
We find that \phTE maintains the accuracy of the underlying QC model within its expected precision, and shows good agreement with publicly available eccentric NR simulations from the Simulation eXtreme Spacetimes (SXS) collaboration~\cite{Boyle:2019kee}.

This paper is structured as follows: In Sec.~\ref{sec:construction}, we describe the building blocks of the \phTE model. We first review the underlying aligned spin model \phTHM in Sec.~\ref{subsec:THM}. Then we summarise the main eccentric PN expressions used in this project and outline the main modifications done to include eccentric corrections to both the frequency and the amplitude of the modes. We assess the accuracy and performance of the model in Sec.~\ref{sec:validation} by assessing its accuracy in the quasi-circular limit, and comparing it against 28 public eccentric NR waveforms from the SXS waveform catalogue and two other eccentric IMR models, \seobe and \teobdali. 
To assess the robustness of the model, in Sec.~\ref{subsec:robustness} we compute mismatches against \seobe, the most accurate eccentric model currently available. In Sec.~\ref{subsec:benchmarks}, we evaluate the computational efficiency of \phTE by benchmarking its performance against the other eccentric waveform models used in this work. 
In Sec.~\ref{sec:PE} we demonstrate the model's applicability in parameter estimation (PE) studies by analyzing eccentric NR injections from the SXS catalog and reanalyzing two real GW events, GW150914 and GW190521.
In Sec.~\ref{sec:conclusions}, we summarise our main conclusions and discuss future work.

Throughout this paper, component masses are denoted by $m_i$. We define the mass ratio $q = m_1/m_2 \geq 1$, and the symmetric mass ratio
$\eta = {m_1 m_2}/{(m_1+m_2)^2}$. The total mass is denoted by $M = m_1 + m_2$ and serves as a scale parameter.
The $z$-component of the dimensionless spin magnitudes are denoted~$\chi_i=S_i^z/m_i^2$.

\section{Model construction}\label{sec:construction}

In Sec.~\ref{subsec:summary_construction} we provide a brief outline of the new \phTE model. 
We provide an overview of the underlying quasi-circular model in Sec.~\ref{subsec:THM}, while in Sec. ~\ref{subsec:PN} we provide a detailed description of the procedure to include eccentric PN corrections in the model. 

\subsection{Summary of the \phTE model} \label{subsec:summary_construction}
The GW strain can be described by a complex scalar $h$ of spin-weight $-2$, corresponding to two independent polarizations, $(h_+)$ and $(h_{\times} )$. As is common practice, we decompose the strain using spin-weighted spherical harmonics (SWSH):
\begin{equation}
h(d_L, t, \iota, \varphi; \Sigma) = h_+ - i h_{\times} = \frac{1}{d_L}\sum_{l,m} h_{lm}(t; \Sigma) \, {}_{-2}Y_{lm}(\iota, \varphi).
\label{eq:SWSK}
\end{equation}
Here, the modes $h_{lm}(t; \Sigma)$ reflect the time dependence and intrinsic physical properties of the source, $d_L$ is the luminosity distance, $\Sigma$ denotes the intrinsic parameters of the binary, and the SWSH basis functions ${}_{-2}Y_{lm}(\iota, \varphi)$ capture the angular dependence on the celestial sphere in the source frame.

This work focuses on eccentric spinning BBH systems, where the spins are aligned or anti-aligned with the orbital angular momentum direction, which remains preserved. 
The intrinsic degrees of freedom are then $\Sigma = \{m_1, m_2, \chi_1, \chi_2, e_{\mathrm{ref}}, l_{\mathrm{ref}}\}$, where $m_1$ and $m_2$ are the component masses and $\chi_{\{1,2\}}= (\mathbf{S}_{\{1,2\}} / m_{\{1,2\}}^2) \cdot \hat{\mathbf{L}}$ are the dimensionless spin components in the direction of the orbital angular momentum $\hat{\mathbf{L}}$. 
In our model, the elliptical orbit is defined in terms of the orbit-averaged orbital eccentricity $e_{\mathrm{ref}}$ and mean anomaly $l_{\mathrm{ref}}$ at a particular orbit-averaged frequency of the $(2,2)$ mode $f_{\mathrm{ref}}$ (which can be related to a reference time $t_{\mathrm{ref}}$.

Each mode $h_{lm}$ is a complex function that can be decomposed into an amplitude $H_{lm}$ and a phase $\phi_{lm}$ as
\begin{equation}
h_{lm} = H_{lm} e^{-i\phi_{lm}}\approx H_{lm}e^{-im\phi},
\end{equation}
with $\phi$ being the orbital phase.
%$\phi_{lm}(t_{\mathrm{ref}})=0$ by convention, and .
%$\dot{\phi}_{lm}(t) \equiv \omega_{lm}(t) > 0$ for $m > 0$ and $\omega_{lm}(t) < 0$ for $m < 0$. \toni{Why $\omega_{lm}$ is introduced here? The sign of $\omega_{lm}$ is a convention choice maybe we could state here the convention choice for $\phi_{lm}$ and introduce $\omega_{lm}$ later, when required.}
Both the \phTE and \phTHM models construct the $m > 0$ modes phenomenologically, while the $m < 0$ modes are computed using the symmetry relation
\begin{equation}
h_{lm} = (-1)^l h^*_{l, -m},
\label{eq:symmmodes}
\end{equation}
which holds due to the preservation of the orbital plane for non-precessing systems,
where $*$ denotes the complex conjugate. 

The \phTE model is constructed by adding the eccentric post-Newtonian corrections detailed in Sec.~\ref{subsec:PN} to both the amplitude and frequency (and hence phase) of the \phTHM modes, described in Sec.~\ref{subsec:THM}. 
The corrections, which only impact the inspiral phase of the non-eccentric model, are calculated using higher-order PN expansions of the quadrupole formula. 
These modes are usually provided in the literature in terms of the mean anomaly $l$ (and hence $\phi(l)$), which leads to a dependency on the arbitrary gauge constant $b$, which features below in Eq.~(\ref{eq:x0p}). A commonly used approach to eliminate this dependence is to redefine the phase following Refs.~\cite{LBlanchet_1993, PhysRevD.48.4757, Blanchet:1995ez, Arun:2004hn,  Kidder:2007rt} for circular orbits and generalized for eccentric orbits in~\cite{Boetzel:2019nfw}. 
This approach can be understood as a redefinition of the mean anomaly $l$ such that
\begin{equation}
    \xi = l -\frac{3G\mathcal{M}}{c^3}n\log\left(\frac{x}{x'_0}\right),
    \label{eq:xi}
\end{equation}
where $n$ is the radial angular frequency, and $x$ is an orbital parameter defined as
\begin{equation}
    x = \left(\frac{GM\omega}{c^3}\right)^{2/3},
    \label{eq:x}
\end{equation}
$\mathcal{M}= M(1-\eta x/2)+\mathcal{O}(x^2)$ is the ADM mass corresponding to initial data with orbital parameter $x$, and $x'_0$ is given by
\begin{equation}
    x'_0=\left[ \frac{e^{11/12-\gamma_E}}{4b} \right]^{2/3},
    \label{eq:x0p}
\end{equation}
where $\gamma_E$ denotes Euler's constant. 
This results in a new orbital phase that depends on the redefined mean anomaly $\xi$ as
\begin{align}
    \psi &= \phi -W(l)+W(\xi) -\frac{3G\chirpMass}{c^3}\omega\ln\left(\frac{x}{x_0'}\right) \\
    &= \lambda_l -\frac{3G\chirpMass}{c^3}\omega\ln\left(\frac{x}{x_0'}\right) + W(\xi) \\
    &= \lambda_{\xi} + W(\xi),
    \label{eq:psi}
\end{align}
where $\lambda$ is the orbit averaged orbital phase and $W$ is the oscillatory phase contribution. Note that for the non-eccentric case, $W$ vanishes and $\lambda$ is directly the orbital phase.
In terms of $\xi$, the gauge dependency disappears, and from now one we always refer to the redefined modes $h_{lm}^{\xi}$ (and amplitudes $H_{lm}^{\xi}$ and phase $\psi(\xi)$), removing $\xi$ to simplify notation:
\begin{equation}
    h_{lm}^{\mathrm{PN}}=\frac{8GM\eta}{c^2R}x\sqrt{\frac{\pi}{5}}H_{lm}e^{-im\psi} = \kappa\ x\ H_{lm}e^{-im\psi},
    \label{eq:hlmPN}
\end{equation}
where $\kappa=\frac{8GM\eta}{c^2R}\sqrt{\frac{\pi}{5}}$.
Since we are only interested in adding the eccentric PN contribution to the \phTHM inspiral, this leads to
\begin{equation}
    h_{lm}^{ \mathrm{insp \ ecc}}=h_{lm}^{\mathrm{PN}} - h_{lm}^{{{\rm PN, }}e=0} = \kappa \left[xH_{lm}-x^{e=0}H_{lm}^{e=0}\right]e^{-im\psi}.
    \label{eq:eccPN}
\end{equation}
The complete \phTE inspiral is then obtained by substituting the phase $\phi$ in  Eq.~\eqref{eq:eccPN} according to Eq.~\eqref{eq:psi}, adding the \phTHM inspiral as the non-eccentric contribution:
\begin{equation}
    \begin{aligned}
    h_{lm}^{\mathrm{insp\ TEHM}}&=h_{lm}^{ \mathrm{insp\ THM}} + h_{lm}^{\mathrm{insp\ ecc}}= \\
    &\kappa \left(x\tilde{H}_{lm}^{\mathrm{THM}} + xH_{lm}-x^{e=0}H_{lm}^{e=0}\right)e^{-im\lambda_{\xi}}e^{-imW}.
    \end{aligned}
    \label{eq:TEHM_insp}
\end{equation}
Both $W$ and $H_{lm}^{\mathrm{ecc}}$ vanish after the inspiral regime as they are proportional to the orbital eccentricity, and the \phTE model assumes that the orbit circularizes. As a consequence, \phTE falls back to \phTHM for the merger/ringdown phase. This allows us to write the complete \phTE modes as
\begin{equation}
    \begin{aligned}
    h_{lm}^{\mathrm{TEHM}}= \left(\tilde{h}_{lm}^{\mathrm{THM}} + \frac{8GM\eta}{c^2R}\sqrt{\frac{\pi}{5}}\left[xH_{lm}-x^{e=0}H_{lm}^{e=0}\right]e^{-im\lambda_{\xi}}\right)e^{-imW},
    \end{aligned}
    \label{eq:TEHM_modes}
\end{equation}
where the $\{\tilde{h}/\tilde{H}\}^{\mathrm{THM}}$ denote the \phTHM modes with an orbital frequency modified by the eccentric corrections.

\subsection{Quasi-circular baseline: \phTHM}
\label{subsec:THM}

The \phTHM model \cite{Estelles:2020twz} is the extension of the time-domain IMR phenomenological model \phT \cite{Estelles:2020osj} for GWs from quasi-circular BBHs which includes the subdominant multipoles $(l,m)= \{(2,\pm 1), (3,\pm 3), (4,\pm 4), (5,\pm 5)\}$. 
It is calibrated to NR simulations, specifically SEOB-NR hybrids, up to mass ratio $18$ and to numerical solutions of the Teukolsky equation for the extreme mass ratio limit, which leads to an accurate and fast time-domain model, which has been used as a standard tool for GW astronomy \cite{Colleoni:2020tgc, Mateu-Lucena:2021siq, Estelles:2021jnz}.
 
The waveform model provides analytical expressions for amplitude and phase as a function of time, which are divided in three distinct regions:
\begin{enumerate}
    \item \textit{Inspiral region}: based on PN expansions augmented with higher order terms calibrated to NR information.
    \item \textit{Plunge-Merger region}: PN expansions break down and the model is directly calibrated to NR in this regime using a phenomenological ansatz.
    \item \textit{Ringdown region}: phenomenological ansatz informed by ringdown and damping frequencies from perturbation theory.
\end{enumerate}
The boundary between the inspiral and merger regions is chosen depending on the mode and it varies for the amplitude and phase/frequency. The boundary between the merger and ringdown regions for both the frequency and amplitude models is set at the peak amplitude for each mode. The peak amplitude of the $(2,2)$ mode, $H_{22}(t)$, is placed at $t = 0$ by convention.
%For a complete description of the model we point to Refs.~\cite{Estelles:2020osj, Estelles:2020twz}, 
In the following, we provide an overview on its inspiral construction. We focus on this region because the eccentric model only incorporates eccentric corrections up to the merger of the QC waveform.

\subsubsection{Inspiral construction}
%%%%%%%%%%%%%%%%%%%%%%%%%%%%%%%%%%%%%
When the binary objects are far apart, the weak field and low velocity conditions allow the use of the PN framework to accurately describe the binary dynamics \cite{Blanchet:2013haa}. This involves numerical integration of the PN equations of motion, or alternatively using the adiabatic approximation, one arrives at the balance equations (see e.g. Ref.~\cite{Buonanno:2009zt}):
\begin{align}
\frac{d\phi}{dt} - \frac{v^3}{M} &= 0,\\
\frac{dv}{dt} + \frac{F(v)}{ME'(v)} &= 0,
\end{align}
where $\omega=d\phi/dt$ is the orbital frequency, $v = \omega^{1/3}$,
 $F(v)$ is the GW luminosity, and $E(v)$ is the system's binding energy. \phTHM relies on TaylorT3~\cite{Blanchet:2001ax,Buonanno:2009zt}, which provides an explicit expression for the orbital phase as a function of time:
\begin{align}
\phi_{n/2}^{\text{T3}}(t) &= \phi_{\text{ref}} + \phi_N(t) \sum_{k=0}^{n/2} \hat{\phi}_k \theta^k,\\
\omega_{n/2}^{\text{T3}}(t) &= \omega_N(t) \sum_{k=0}^{n/2} \hat{\omega}_k \theta^k,
\label{eq:TaylorT3}
\end{align}
where $\theta(t) = \left[\frac{\eta(t_0 - t)}{5M}\right]^{-1/8}$, $\omega_N = \theta^{3}/8$, and $n/2$ corresponds to the PN order.
Although TaylorT3 provides closed-form expressions, it is less accurate than other approximants
at higher frequencies, see e.g.~\cite{Buonanno:2009zt}, and becomes singular at $t = t_0$, which depends on the intrinsic parameters of the source and the PN order employed. 
To improve the description of the inspiral phase, the 3.5 PN ($n=7$ in Eq.~\eqref{eq:TaylorT3}) TaylorT3 ansatz is expanded by adding 5 extra pseudo-PN terms, which are calibrated to NR simulations.%~\cite{Estelles:2020twz}. 
Furthermore we set the merger time at $t_0 = 0$. This way one ensures an accurate description of the frequency and the phase until at least the minimum energy circular orbit time (MECO)~\cite{Cabero:2016ayq}. 
The model then assumes
\begin{equation}
    \omega_{22}^{\text{insp}}=2\omega = 2\left(\omega_{3.5}^{\text{T3}}(t)+\omega_N(t)\sum_{k=8}^{13}\hat{c}_k\theta^k\right).
    \label{eq:omega22}
\end{equation}
The frequency of the higher modes (HMs), $\omega_{lm}=\dot{\phi}_{lm}$, is obtained by rescaling the frequency and phase of the $(2,2)$-mode such that $\omega_{lm}=\frac{m}{2}\omega_{22}$ and $\phi_{lm}=\frac{m}{2}\phi_{22}$.

The amplitude of the emitted radiation during the inspiral phase can also be calculated using higher-order PN expansions of the quadrupole formula. 
For non-spinning BBHs, the model uses 3PN corrections from \cite{Blanchet:2008je} and the 3.5PN from~\cite{Faye:2012xt}. For aligned-spin systems, 1.5PN ~\cite{Arun:2008kb} and 2PN~\cite{Buonanno:2012rv} expansions are employed. 
These amplitude corrections also include three unknown higher-order PN terms, which are calibrated to NR to improve accuracy up to the MECO time:
\begin{equation}
    H^{\text{insp}}_{lm}(t) = 2\eta\sqrt{\frac{16\pi}{5}} x \left(\sum_{k=0}^{7} \hat{h}^k_{lm} x^{k/2} + \sum_{k=8}^{10} \hat{d}^k_{lm} x^{k/2} \right).
\end{equation}
Here, $\hat{d}_k$ are the unknown coefficients determined by matching the amplitude at NR-calibrated values.
The modes are then given by
\begin{equation}
h_{lm}^{\text{insp}}(t)=H_{lm}^{\mathrm{insp}}(t)\exp{\left[-i\left(\frac{m}{2}\phi_{22}^{\text{insp}}(t)-\Delta \phi_{lm}\right)\right]},
\end{equation}
where $\Delta \phi_{lm}$ is a fixed rotation performed such that the dominant contribution is given in the real part of each mode (see Eq.~(13) in Ref.~\cite{Estelles:2020twz}).

\subsection{Eccentric post-Newtonian corrections}
\label{subsec:PN}

We incorporate PN eccentric corrections into the \phTHM model using the quasi-Keplerian parametrization \cite{AIHPA_1985__43_1_107_0,Damour:1988mr, SCHAFER1993196} and the covariant spin-supplementary condition  \cite{AIHPA_1969__11_2_221_0, 1979igsg.conf..156D}.
The \phTE model includes the full 3PN orbit-averaged dynamics, accounting for both non-spinning and spinning corrections~\cite{Henry:2023tka}, implemented in both modified harmonic (MH) PN (referred to as PN) and EOB coordinates. Additionally, we modify the waveform modes of \phTHM by incorporating non-spinning and nonprecessing-spin eccentric corrections up to 3PN order~\cite{Ebersold:2019kdc, Henry:2023tka}, formulated in MH coordinates.

We evolve the orbital elements using the secular evolution equations driven by radiation reaction derived by connecting the balance equations with the orbit-averaged energy and angular momentum fluxes \cite{Henry:2023tka}. Specifically, the orbital elements consist of the temporal eccentricity $e_t\equiv e $, the orbital parameter $x$ (Eq. ~\eqref{eq:x}), the mean anomaly $l$, and the orbit-averaged orbital phase $\lambda = \int \overline{\omega} \, \mathrm{d}t$, where the upper line indicates orbit-averaging.  We incorporate the 3PN non-spinning contributions from Refs.~\cite{Arun:2009mc, Ebersold:2019kdc} in modified-harmonic coordinates, as well as the 3PN spin contributions from~\cite{Henry:2023tka}. 
As detailed in Sec.~\ref{subsec:THM}, the \phTHM model uses the TaylorT3 approximation to derive a closed-form expression for the orbital phase and frequency, which, by construction, corresponds to half of the $(2,2)$-mode phase and frequency (see Eq.~\eqref{eq:omega22}).
Ideally, we would construct the new \phTE model by extending the functional form of the TaylorT3 approximant to include eccentric terms of sufficient accuracy. However, existing literature only provides a PN expansion applicable to low-eccentricity systems~\cite{Moore:2016qxz, Sridhar_2024}, limiting the applicability of the model. Additionally, the eccentric TaylorT3 approximant has been shown to exhibit a non-monotonic frequency evolution, decreasing before the innermost stable circular orbit (ISCO), leading to a recommendation against its practical use~\cite{Moore:2016qxz}. 

The choice of coordinates in the secular evolution equations can substantially affect the evolution of the binary. To evaluate this impact, apart from working with PN expressions in modified-harmonic coordinates, we also incorporate the use of EOB coordinates. Comparisons between the two gauges are presented throughout the paper, with a more detailed discussion provided in the next section, where we detail how to solve the orbit-averaged secular evolution equations and obtain the modified eccentric orbital phase and frequency corrections to the \phTHM baseline. Finally, Sec.~\ref{subsubsec:amp} briefly introduces the PN expressions used for the modes' amplitudes, concluding the construction of the \phTE modes.

\subsubsection{Orbital secular evolution}
\label{subsubsec:orbfreq}

To overcome the limitations of the TaylorT3 approximant discussed earlier, we proceed with a different approach to incorporate eccentric information to the \phTE evolution:
\begin{enumerate}
    \item Compute the quasi-circular $(2,2)$ mode frequency $\omega_{22}$ from \phTHM, and the time derivative $\dot{x}^{\mathrm{THM}}=\frac{1}{3}\left(\omega_{22}/2\right)^{-1/3}\dot{\omega}_{22}$. 
    \item Interpolate $\dot{x}^{\mathrm{THM}}$ as a function of the orbital frequency parameter $x$, $\dot{x}^{\mathrm{THM}}(x)$, so that the time dependency is removed. This is essential since the correspondence between frequency and time is different in the quasi-circular and eccentric cases. 
    \item Evolve the secular evolution equations $\{\dot{x},\dot{e},\dot{l},\dot{\lambda}\}$ using $\dot{x}^{\mathrm{THM}}(x)$ as the quasi-circular contribution of the secular evolution equation of frequency parameter $\dot{x}$. This implies removing all the non-eccentric terms from the PN expansion for $\dot{x}$, which are included in $\dot{x}^{\mathrm{THM}}$, and only keeping the eccentric terms of the secular equation, i.e., $\delta x_{e\neq 0}=x^{\rm PN}-x^{\rm PN}_{e\neq 0}$.
\end{enumerate}
The system of ordinary differential equations (ODEs) can be schematically expressed in terms of the PN orders included as
\begin{align}\label{eq:EDOs}
    \dot{x} &= \dot{x}^{\mathrm{THM}} + \delta \dot{x}_{e\neq 0} ,\\
   \delta \dot{x}_{e\neq 0} &= \delta  \dot{x}^{\mathrm{Newt}}_{e\neq0} + \delta \dot{x}^{\mathrm{1PN}}_{e\neq0} + \delta \dot{x}^{\mathrm{1.5PN}}_{e\neq0} + \delta \dot{x}^{\mathrm{2PN}}_{e\neq0} + \delta \dot{x}^{\mathrm{2.5PN, tail}}_{e\neq0} \nonumber\\
    &+ \delta \dot{x}^{\mathrm{2.5PN, SS}}_{e\neq0} + \delta \dot{x}^{\mathrm{3PN}}_{e\neq0} + \delta\dot{x}^{\mathrm{3PN, SS}}_{e\neq0} + \delta\dot{x}^{\mathrm{3PN, tail}}_{e\neq0},\\
    \dot{e} &=  \dot{e}^{\mathrm{Newt}} + \dot{e}^{\mathrm{1PN}} + \dot{e}^{\mathrm{1.5PN}} + \dot{e}^{\mathrm{2PN}} + \dot{e}^{\mathrm{2.5PN}}_{\mathrm{SS}} + \dot{e}^{\mathrm{3PN}}_{\mathrm{SS}} \nonumber\\
    &+ \dot{e}^{\mathrm{3PN}}_{\mathrm{inst}} + \dot{e}^{\mathrm{3PN}}_{\mathrm{tail}}, \\
    \dot{l} &=  \dot{l}^{\mathrm{Newt}} + \dot{l}^{\mathrm{1PN}} + \dot{l}^{\mathrm{1.5PN}} + \dot{l}^{\mathrm{2PN}}_{\mathrm{inst}} + 
    \dot{l}^{\mathrm{2.5PN}}_{\mathrm{tail}} + \dot{l}^{\mathrm{3PN}}_{\mathrm{SS}} + \dot{l}^{\mathrm{3PN}}_{\mathrm{inst}}, \\
    \dot{\lambda} &= x^{3/2}.
\end{align}

The secular equations include 3PN non-spinning and spinning contributions, and we consider two different coordinates. Specifically, we use the expansions for $\dot{x}$, $\dot{e}$ and $\dot{l}$ in modified-harmonic (MH) PN coordinates from Refs.~\cite{Arun:2009mc, Ebersold:2019kdc, Henry:2023tka}, as well as $\dot{e}$ and $\dot{x}$ in EOB coordinates, which were derived recently in Ref.~\cite{Gamboa:2024imd} for the development of the state-of-the-art eccentric EOB model, \seobe~\cite{Gamba:2024cvy}. 
Since $x$ is an almost gauge-independent quantity, we transform the eccentricity evolved in EOB coordinates, $e^{\mathrm{EOB}}$, into its counterpart in MH PN coordinates, $e^{\mathrm{PN}}$, using the relation provided in Ref.~\cite{Gamboa:2024imd}. Once $e^{\mathrm{PN}}$ is obtained, we also recompute the mean anomaly accordingly.

One of the most interesting results of this work is the observation that different gauges can lead to significantly different system evolutions. This is because the eccentricity is a gauge-dependent quantity, meaning that certain gauges may be more suitable for aligning with NR data. While the GW signal itself is gauge-independent, differences in how eccentricity is defined and evolved (together with the coupled frequency) can affect the accuracy of the binary's trajectory and its agreement with NR waveforms.
Gauge-invariant approaches to defining eccentricity have been proposed based on waveform modulations~\cite{Shaikh:2023ypz, Islam:2025oiv} and catastrophe theory~\cite{Boschini:2024scu}.
Throughout the paper, we use the waveform-based eccentricity definition of Ref.~\cite{Shaikh:2023ypz}, $e^{\rm GW}$, as implemented in the \texttt{gw\_eccentricity} package~\cite{Ramos-Buades:2022lgf}.
We also explore the two implemented eccentricity gauges and discuss their respective implications. 
Based on its better agreement with NR eccentric simulations and its closer correspondence to  $e^{\rm GW}$, we adopt the EOB secular evolution equations as the default choice for modeling the eccentric dynamics.
This is the assumption for all results unless otherwise specified.

The initial conditions for the evolution of the dynamics are the mean anomaly $l_{\mathrm{ref}}$ and the eccentricity $e_{\mathrm{ref}}$ given at a concrete reference orbit-averaged frequency of the $(2,2)$-mode, $f_{\mathrm{ref}}$, which can be translated into $x_{\mathrm{ref}}$. Following the LALSuite~\cite{lalsuite} conventions specified in Ref.~\cite{schmidt2017}, we set $\lambda_{\mathrm{ref}}=0$.

Solving this set of coupled differential equations represents one of the main computational bottlenecks of the model, and the key to improving efficiency lies in the interpolation of $\dot{x}^{\mathrm{THM}}(x)$.
An effective approach is to perform the interpolation on a $\theta$-grid, as defined in Eqs~\eqref{eq:TaylorT3}. This parameter offers the advantage of creating a sparse grid at low frequencies, where the frequency curve evolves slowly, while providing higher resolution at high frequencies - particularly near the merger phase.
To solve the system of equations, we use the ``DOP853'' method, an 8th-order Runge-Kutta solver, implemented within the \texttt{solve\_ivp} function in the SciPy Python package~\cite{Virtanen_2020}. We set the absolute and relative tolerances to $10^{-12}$ for this work, but can be changed as suited depending on the particular application.
These parameters were selected to balance computational cost with accuracy, specifically ensuring the accurate recovery of the frequency from the quasi-circular model. More details on this are given in Sec.~\ref{subsec:nonecclimit}. 

Since the \phTE model assumes circularization by the time of merger, the stopping conditions of the ODEs solver are set when the eccentricity vanishes or when the frequency reaches the \phTHM peak frequency. We then taper the eccentricity at $t=-20M$, ensuring that all corrections reduce to zero at merger, thus recovering the \phTHM merger/ringdown phase.

The model also supports specifying a minimum frequency $f_{\mathrm{min}}$ lower than the reference frequency $f_{\mathrm{ref}}$. In such cases, we start from the reference values and solve the system of equations backwards until $x(t)$ reaches $x_{\mathrm{min}}$. This feature provides significant flexibility, particularly for parameter estimation studies, as demonstrated in Sec.~\ref{subsec:GWevents}. 
A key benefit is ensuring sufficiently long waveforms, preventing failures in the integration of eccentric dynamics. Through testing, we established that the initial separation, computed at the lowest PN order (which corresponds to the QC separation at initial time, $r_{\min}\approx 1/x_{\min}$) must always be greater than $7M$. This threshold guarantees enough integration points, even for the highest eccentricities within the model's validity range, to avoid failures.
Currently, the model does not support $f_{\mathrm{ref}}<f_{\min}$ as there is no way to ensure $r_{\mathrm{ref}}>7M$ without internally modifying the reference frequency.
Another advantage of this feature is enabling the addition of extra cycles to the time-domain waveform. This prevents tapering effects in Fourier transforms from removing relevant information within the frequency range of interest. By default, we extend the underlying QC waveform by three cycles when computing the Fourier transform.

Once the secular equations are solved, the orbit-averaged frequency and phase of the $(2,2)$ \phTE mode are obtained directly as $\bar{\omega}_{22}^{\mathrm{TEHM}} = 2x^{3/2}$ and $\bar{\phi}_{22}^{\mathrm{TEHM}}=2\lambda$ up to the peak frequency of (2,2)-mode of the \phTHM model.
It is important to remark two main features involved in the calculation of the orbit-average $(2,2)$-mode phase. First, the use of the quasi-circular \phTHM (2,2)-mode frequency to evolve the PN/EOB secular evolution equations, which implies the use of NR-calibrated information in the $e=0$ limit for the evolution of the $x$ parameter. 
Second, a consequence of the use of the frequency of the \phTHM model is related to the fact that the $(2,2)$-mode frequency is a wave quantity introduced in the source dynamics. This implies an inconsistency in the usual redefinition of the orbital phase to take into account tail effects and remove gauge ambiguities (see Refs.~\cite{Boetzel:2019nfw,Ebersold:2019kdc,Henry:2023tka}). In order to recover the quasi-circular phase, we do not perform the phase redefinition as described in Eq.~\eqref{eq:xi}. Instead, we assume that, due to the inclusion of NR information, the resulting $l$ (and hence $\phi$) can be understood as $\xi$ and $\psi$. We further investigate the implications of this phase redefinition in Sec. ~\ref{sec:validation} by comparing against NR waveforms.
To maintain consistency in the binary evolution when applying the phase redefinition, the input mean anomaly for the model will be set to $\xi_{\mathrm{ref}}$. Since we solve the evolution of $l(t)$ in Eqs.~\eqref{eq:EDOs}, we first obtain $l_{\mathrm{ref}}$ which is derived from $\xi_{\mathrm{ref}}$. We then obtain $\xi(l)$ and $\lambda_{\xi}$, as given in Eqs.~\eqref{eq:xi} and \eqref{eq:psi}.

The construction of the higher modes follows a process similar to that of \phTHM: the inspiral region is constructed by assuming that $\omega_{lm} = \frac{m}{2}\omega_{22}$, and hence $\phi_{lm} = \frac{m}{2}\phi_{22}$, while the intermediate and merger/ringdown regions are built in the same manner as \phTHM, see  Sec. II.B.3 in Ref~\cite{Estelles:2020twz} for details.
The orbit-averaged frequency and phase also enter in the non-eccentric contribution of the model, which is denoted as $\{\tilde{h}/\tilde{H}\}^{\mathrm{THM}}$ in Eq.~\eqref{eq:TEHM_modes}.
This approach allows us to recover the \phTHM frequency in the limit $e\rightarrow 0$, since all the PN terms in Eq.~\eqref{eq:EDOs} vanish and we are only left with $\dot{x}^{\mathrm{THM}}(x)$.

Finally, as shown in Eq.~\eqref{eq:TEHM_modes}, we add the eccentric oscillatory contribution $W(\xi)$ to the modes' phases. This variable includes all 3PN order instantaneous non-spinning, spin-orbit and spin-spin contributions for eccentric binaries eccentricity expanded up to $\mathcal{O}(e^{12})$ \cite{Henry2025}.

\subsubsection{Mode amplitudes}
\label{subsubsec:amp}
Once the evolution of the orbital quantities is completed, we can compute the amplitudes of the \phTE modes following Eq.~\eqref{eq:TEHM_modes}. 
The eccentric contributions are decoupled from the \phTHM modes, which have been modified by introducing the new orbit-averaged eccentric orbital frequency and phase, impacting both the amplitudes and phases of all modes. 

Hence, we first compute the modified \phTHM modes, denoted as $\tilde{h}_{lm}^{\mathrm{THM}}$ in Eq.~\eqref{eq:TEHM_modes}, which depend on $x(t)$, computed as explained in Sec.~\ref{subsubsec:orbfreq}. Then, we evaluate the 3PN expressions for the modes' amplitudes $H_{lm}^{\mathrm{ecc}}(x(t),e(t),\xi(t))$ up to merger. As explained in Sec.~\ref{subsec:summary_construction} these eccentric amplitudes vanish by the time of the merger, so the peak and ringdown phases of \phTE revert to \phTHM. 

We incorporate the next-to-leading order instantaneous non-spinning, spin-orbit and spin-spin contributions to the waveform modes and hereditary (tail and memory) contributions to the modes for eccentric orbits, using the 3PN non-spinning terms from Ref.~\cite{Ebersold:2019kdc} and the 3PN spinning contributions from Ref.~\cite{Henry:2023tka}.
As mentioned in Sec.~\ref{subsec:THM}, the underlying quasi-circular \phTHM model includes only up to 2PN order for the spinning contributions. Hence, we set the same PN order for the \phTE model for consistency. In Sec.~\ref{subsec:SXS} we investigate the effect of incorporating these terms. 
The \phTE model uses eccentricity-expanded expressions for both instantaneous and tail contributions up to $\mathcal{O}(e^6)$ for all the available modes, $\{(2,|2|),(2,|1|),(3,|3|),(4,|4|),(5,|5|) \}$, since a non-expanded in eccentricity expression for the tail terms is not available.
For the $(2,2)$ modes, we derived expansions up to $\mathcal{O}(e^{12})$~\cite{Henry2025}. 
Since it does not provide a significant improvement in accuracy and only increases computational cost, this option is not set as the default in the model. 
We leave for future work the use of expressions not expanded in eccentricity for the waveform multipoles leveraging for instance recently developed re-summation techniques for the tail contributions~\cite{Gamboa:2024imd}.

\section{Model validation}\label{sec:validation}

In this section, we evaluate the validity of the model across the BBH eccentric nonprecessing-spin parameter space.
For these systems, the GW signal depends on six intrinsic parameters: the mass ratio $q$, the total mass $M$, the spin components in the direction of the orbital angular momentum $\chi_1$ and $\chi_2$, and two parameters describing the ellipse, which we choose to be the orbital eccentricity $e$ and the mean anomaly $l$ at a reference time. The intrinsic parameters are denoted as $\Pi$.

To describe the source in relation to a ground-based GW detector one needs seven extrinsic parameters: the angular position of the line of sight in the source frame $(\iota, \varphi)$, the polarization angle $\Psi$, the luminosity distance of the source $d_L$, the sky location of the source in the detector frame $(\alpha, \delta)$ and the coalescence time $t_c$.
The signal can then be expressed using the antenna-pattern functions $F_{+/\times}$~\cite{Schutz_2011}, which account for the detector's response to the incoming waveform - described in terms of the polarizations in Eq.~\eqref{eq:SWSK} - as
\begin{align}
    h(t)=&F_{+}(\alpha, \delta, \Psi)h_{+}(\iota, \varphi, d_{L}, \Pi, t_c;t )+\\
    &F_{\times}(\alpha, \delta, \Psi)h_{\times}(\iota, \varphi, d_{L}, \Pi, t_c;t ).
    \label{eq:antenna}
\end{align}
Given two waveforms, one can define the noise-weighted inner product or overlap as~\cite{PhysRevD.44.3819, Finn:1992xs}
\begin{equation}
(h_1(t)|h_2(t))=4\Re\int_{f_{\mathrm{min}}}^{f_{\mathrm{max}}}\mathrm{d}f \frac{\tilde{h}_1(f)\tilde{h}_2^*(f)}{S_n(f)},
\end{equation}
where the tilde denotes Fourier transform, the star complex conjugate, and $S_n(f)$ is the one-sided power-spectral density (PSD) of the detector's noise. 
We use the zero-detuned high power Advanced LIGO PSD, representing the expected sensitivity of the fifth observing run (O5)~\cite{PSDo5}, with frequency limits set to $f_{\mathrm{min}}=10$Hz and $f_{\mathrm{max}}=2048$Hz. For NR waveforms where the orbit-averaged starting frequency of the $(2,2)$ mode, $\bar{f}_{\mathrm{start}}$, exceeds $f_{\mathrm{min}}$, we set $f_{\mathrm{min}}=1.35\bar{f}_{\mathrm{start}}$. 
This choice is consistent with Refs.~\cite{Ramos-Buades:2021adz,Gamboa:2024hli} and ensures the inclusion of as much of the inspiral phase as possible, where the effects of eccentricity are most pronounced.
Given the oscillatory nature of the GW frequency in eccentric signals, we consistently refer to the orbit-averaged frequency $\bar{f}$ if not stated otherwise. To simplify notation, we omit the bar and use $f$ throughout the text, while always referring to the orbit-averaged value.

The \textit{faithfulness} or \textit{match} is defined as the maximization of the normalized overlap over selected parameters.
For general aligned-spin QC systems including higher modes, the standard choice of parameters to optimize the template with respect to the signal includes the coalescence time $t_c$, azimuthal angle $\varphi$, and an effective polarization angle $\kappa$ which depends on the sky location of the source. When considering only the dominant harmonic, the angular dependence of $\mathcal{Y}^{-2}_{2,\pm 2}$ introduces a degeneracy between the inclination and polarization angle~\cite{Capano:2013raa}, and in this case the optimization is performed only over $t_c$ and $\kappa$.
Eccentric waveforms introduce additional complications due to the gauge-dependent nature of eccentricity in general relativity. In particular, we employ two different gauge choices for the eccentricity in our model, which, as we show later, lead to significantly different definitions of this parameter. Consequently, the signal and the template may be characterized by distinct eccentricity parameters, since both $e$ and $l$ are gauge dependent.
The comparisons of eccentric waveforms with different definitions of eccentricity involve computing a mapping to connect them. For instance, Ref.~\cite{Shaikh:2023ypz} implements a definition of eccentricity based on the waveform which reduces to the Newtonian definition of eccentricity~\cite{Ramos-Buades:2022lgf}, while Ref.~\cite{Bonino:2024xrv} introduces an algorithm based on the same definition of eccentricity to map different eccentric waveforms.
In our comparisons against NR we adopt an optimization method to identify the best-matching waveform for a given eccentric NR waveform. This requires optimizing over the mean anomaly and eccentricity at the initial frequency to align both waveforms at merger, following Ref.~\cite{Gamboa:2024hli}, to mitigate contamination from spurious frequencies.
Hence, the \textit{faithfulness} is defined as 
\begin{equation}
    \mathcal{F}_{\Sigma} = \underset{\Sigma}{\mathrm{max}} \frac{(h_1(t)|h_2(t))}{\sqrt{|h_1(t)||h_2(t)|}},
    \label{eq:faith}
\end{equation}
where $\Sigma$ stands for the set $\Sigma=\{t_c, \varphi, \kappa, e, l, f\}$ in the eccentric case and it reduces to $\Sigma_{\mathrm{QC}}=\{t_c, \varphi,\kappa\}$ for the quasi-circular case. In the case of optimizing only over the $(2,2)$ mode, $\Sigma_{22}=\Sigma-\{\kappa\}$. 
The optimizations are performed numerically using the \texttt{dual\_annealing} function of the \texttt{scipy} \cite{scipy} python package, except for the polarization $\kappa$, which is done analytically as described in \cite{Harry:2017weg, Capano:2013raa}. 
We finally define the mismatch as the deviation of the match from unity $\mathcal{M} = 1 - \mathcal{F}_{\Sigma}$.

In the following, we will use the mismatch as our metric to assess the accuracy of the \phTE model across parameter space. Specifically, we first assess in Sec.~\ref{subsec:nonecclimit} the recovery of the QC limit by comparing it to the underlying  \phTHM model. In Sec.~\ref{subsec:SXS}, we compare the model against the 28 publicly available eccentric simulations from the SXS catalog, performing mismatch calculations both for the dominant harmonic and including all available modes. We also compare our results with those obtained with other state-of-the-art eccentric EOB models, \seobe~\cite{Gamboa:2024hli} and \teobdali~\cite{Gamba:2024cvy}, using our same optimization procedure.
In Sec.~\ref{subsec:modellingstrategies}, we summarize the different modeling choices introduced in Sec.~\ref{sec:construction} and evaluate their impact on accuracy by comparing them to the same set of NR simulations.
In Sec.~\ref{subsec:robustness} we test the model's robustness across parameter space by comparing it to the state-of-the-art \seobe model.
Finally, in Sec.~\ref{subsec:benchmarks} we provide benchmark comparisons across parameter space with direct comparisons with the \seobe and \teobdali models. 

\subsection{Quasi-circular limit}\label{subsec:nonecclimit}
We validate the eccentric aligned-spin model \phTE in the quasi-circular limit by comparing it to the underlying non-eccentric \phTHM model.
The \phTE model is built upon the \phTHM framework within the new \texttt{phenomxpy} python package~\cite{phenomxpy}. Consequently, the underlying aligned-spin model differs in some aspects from the published LALSuite implementation~\cite{Estelles:2020twz}, introducing minor bug fixes. 
One of the differences involves incorporating analytical derivatives wherever possible to mitigate instabilities associated with numerical differentiation. This enhancement improves the model's stability, particularly at positive effective spin, where minor numerical inaccuracies can lead to larger differences in the values of the NR-calibrated collocation points. However, this cannot be applied to the eccentric model due to the numerical integration of the frequency evolution. As a result, the \phTHM and \phTE models inherently exhibit slight discrepancies, constrained to numerical errors affecting the calculation of the collocation points. 
Furthermore, the numerical error associated with the integration of the frequency compared to the analytical expressions used in the quasi-circular \phTHM model is more significant at lower masses, where the longer waveform duration allows for greater phase accumulation over the binary evolution.
\begin{figure}
    \centering
    \includegraphics[width=1\linewidth]{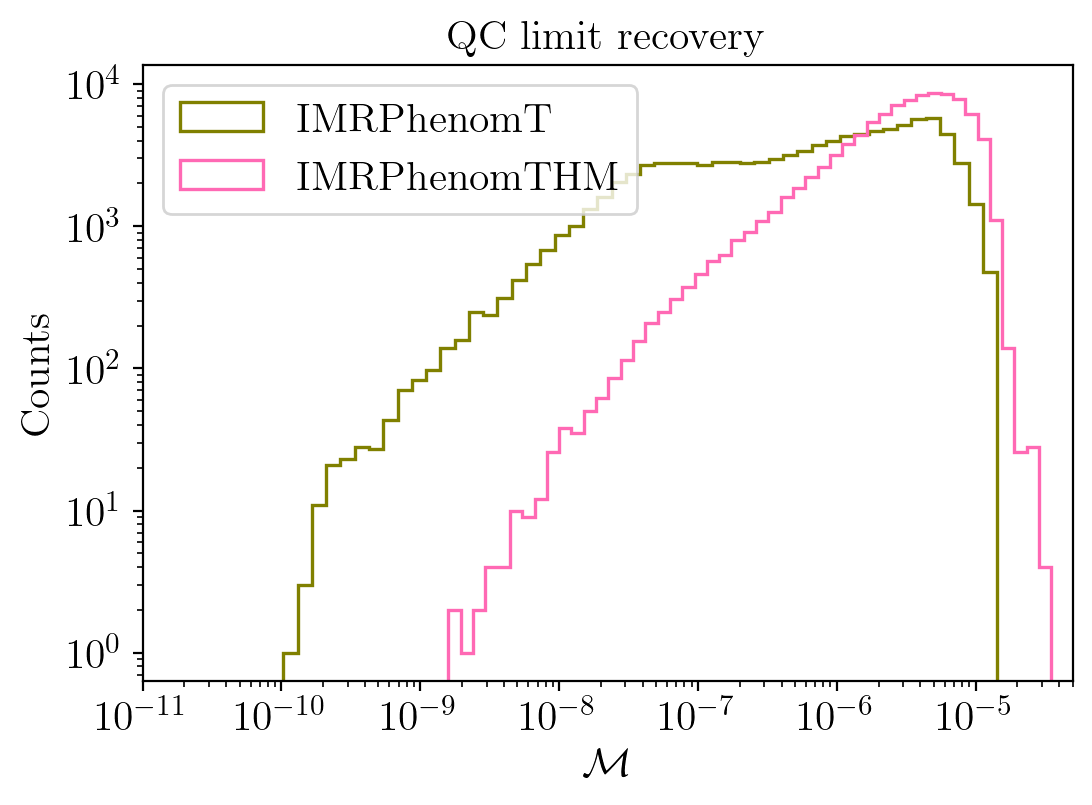}
    \caption{
    Mismatch distribution between the \phTHM  and \phTE models computed over $10^5$ non-eccentric waveforms for the dominant (2,|2|)-modes (green curve) and including higher order modes (pink curve). The configurations are randomly distributed in the \phTHM model's validity region: mass ratio $q\in [1,20]$, total mass $M\in [10,300]\ M_{\odot},$ z-component dimensionless spin vectors $ \chi_i\in[-0.995,0.995],$ azimuthal phase $ \varphi\in[0,2\pi]$ and inclination angle $\iota\in[0,\pi]$.
    }
    \label{fig:TEvsT}
\end{figure}
In Fig.~\ref{fig:TEvsT} we present mismatch calculations for \phTE against \phTE for the dominant (2,|2|)-modes and including higher order modes across $10^6$ cases distributed within the \phTHM models' validity region: 
 mass ratio $q\in [1,20]$, total mass $M\in [10,300]\ M_{\odot},$ z-component dimensionless spin vectors $ \chi_i\in[-0.995,0.995],$ azimuthal phase $ \varphi\in[0,2\pi]$ and inclination angle $\iota\in[0,\pi]$. 
Both the mismatches for the dominant harmonic and all higher modes remain below $10^{-5}$, with some outliers reaching up to $3\cdot10^{-5}$ due to the specific construction of the models. Specifically, higher mismatches occur in the positive effective spin region due to the construction of the collocation points in the eccentric model compared to the QC baseline. 
This is mainly because positive aligned spins slow down the binary evolution, allowing higher frequencies near merger. This results in a steeper frequency slope, which can cause instabilities in the numerical derivatives needed for the collocation points. Additionally, as in the low-mass case, longer evolutions accumulate more numerical integration errors in the orbital frequency.
Finally, these effects have a greater impact on mismatches including all modes, as expected. Since they influence the baseline construction, the accumulation of errors increases with the number of modes considered.

These results combined with the accuracy against quasi-circular NR simulations of \phTHM (see Fig.~7 in Ref.~\cite{Estelles:2020twz}) demonstrate that the eccentric \phTE model faithfully recovers the quasi-circular limit with minor discrepancies against the quasi-circular \phTHM model due to some small differences in the construction of both models.

\subsection{Comparison against numerical relativity waveforms}\label{subsec:SXS}
\begin{figure}
    \centering
    \includegraphics[width=\linewidth]{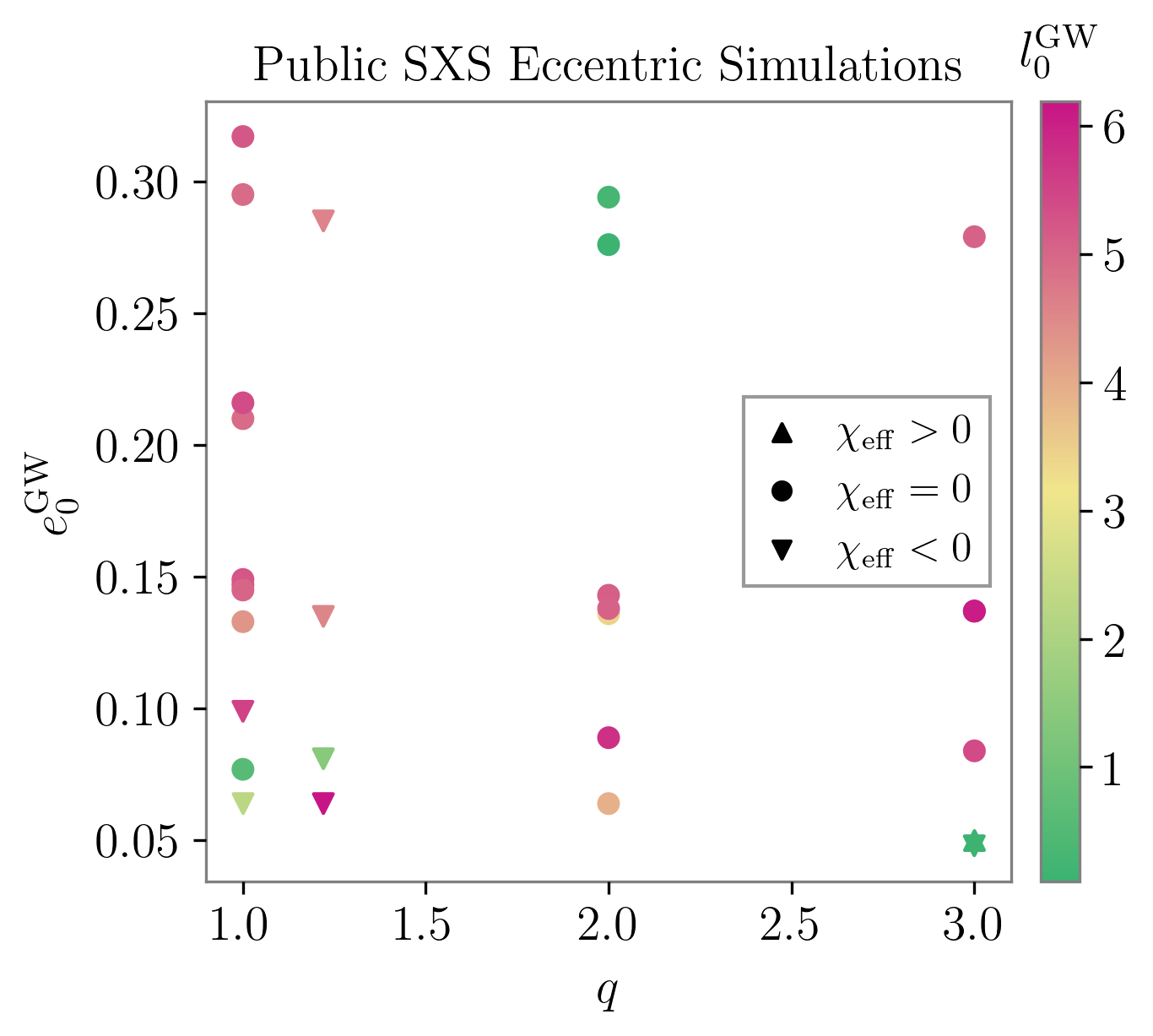}
    \caption{Parameter space distribution of the 28 eccentric SXS NR waveforms used in this work, shown as a scatter plot with mass ratio $q=m1/m_2$ on the $x$-axis, initial eccentricity measured with the \texttt{gw\_eccentricity} package at the initial orbit-averaged frequency $e^{\mathrm{GW}}_0$ on the $y$-axis, and the dimensionless effective spin $\chieff=(\chi_1m_1+\chi_2m_2)/M$ indicated by the color scale.
    }
    \label{fig:NRsims}
\end{figure}
We assess the accuracy of the \phTE model by computing mismatches against 28 eccentric BBH NR waveforms produced with the SpEC code~\cite{Hinder_2019, Boyle:2019kee, SpECWebsite} from the SXS Collaboration. In Figure~\ref{fig:NRsims}, we present the distribution of these simulations across the parameter space.  
As shown, the NR available waveforms correspond to relatively low initial eccentricities, where the initial eccentricity is computed as the initial orbit-averaged frequency using the \texttt{gw\_eccentricity} package~\cite{Shaikh:2023ypz}, and are primarily concentrated in the non-spinning regime.

\begin{figure*}
    \centering
    \includegraphics[width=\textwidth]{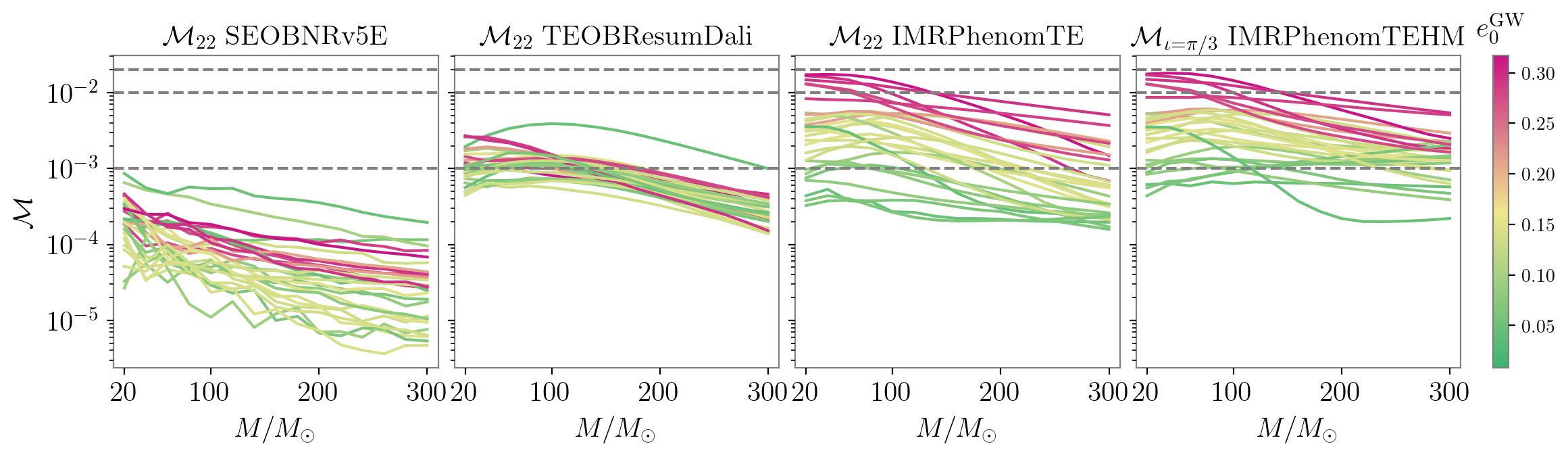}
    \caption{Mismatches for the 28 available SXS eccentric simulations shown in Fig.~\ref{fig:NRsims}, comparing against different eccentric aligned-spin models. The first three columns show the $(2,2)$-mode mismatches, $\mathcal{M}_{22}$, for the \seobe, \teobdali and \phTE models, respectively, while the fourth column presents the mismatches including all available higher modes $\mathcal{M}$ in \phTE for an inclination of $\pi/3$. Line colors indicate the initial eccentricity of the NR simulations computed with the \texttt{gw\_eccentricity} package, $e_{\mathrm{GW}}$, as shown in the color bar legend. All mismatches are calculated over total masses $M\in\ [20,300]\ M_{\odot}$. 
    }
    \label{fig:SXSmismatches}
\end{figure*}

Results for the mismatches against all NR simulations are presented in Fig.~\ref{fig:SXSmismatches}. The mismatches are computed over total masses $M\in[20,300]\ M_{\odot}$ using the faithfulness function defined in Eq.~\eqref{eq:faith}. Specifically, for the first three columns we compute the $\mathcal{M}_{22}$ mismatches by optimizing over $e$, $l$, $Mf_{\mathrm{ref}}$, $\varphi$, and $t_c$ at the lowest mass, $20\ M_{\odot}$. For higher masses, we retain the optimized $e$, $l$, and $Mf_{\mathrm{ref}}$, while optimizing only over $\varphi$ and $t_c$. For these mismatches, we only include the $(2,|2|)$ modes of the NR simulations. Each column corresponds to a different model: \seobe~\cite{Gamboa:2024hli}, \teobdali~\cite{Gamba:2024cvy} and \phTE, respectively.
In the last column, higher order modes in NR up to $l \le 8 $ and all the available modes in the \phTE model are included. We follow the same procedure as for the (2,2)-mode mismatches, but additionally optimize the polarization phase for each mass. Optimizing over higher modes for the EOB models is computationally expensive; thus, Refs.~\cite{Ramos-Buades:2021adz,Gamboa:2024hli} adopt the same parameters optimized for the $(2,2)$-mode. To ensure a fair comparison, we limit our mismatches to the (2,2)-mode for the EOB models, as including higher modes would not be consistent across all models due to the high computational cost of optimizing them for the EOB models, which we can afford for \phTE. 

\begin{figure}
    \centering
    \includegraphics[width=\columnwidth]{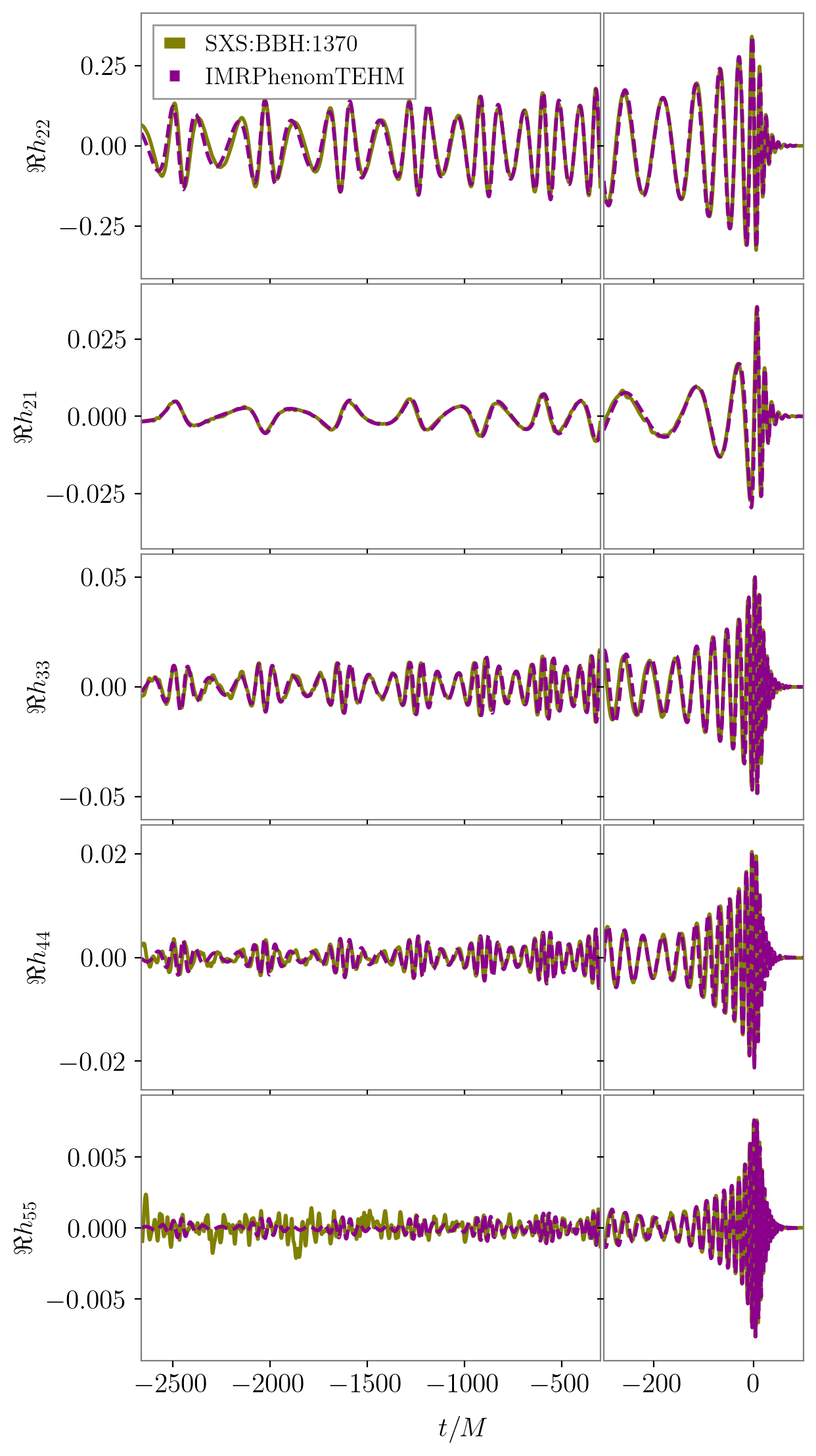}
    \caption{Real part of all available waveform modes in the \phTE model for the nonspinning NR simulation SXS:BBH:1370 with mass ratio $q=2$, effective spin $\chieff=0$, initial eccentricity $e^{\text{GW}}_0=0.294$, and initial mean anomaly $l^{\text{GW}}_0=0.29$ rad (green), and \phTE waveform evaluated at the optimal parameters (magenta). The optimal values for \phTE are $e_{\mathrm{ref}}^{\mathrm{EOB}}=0.288$ and $l_{\mathrm{ref}}=3.68$ rad at $Mf_{\mathrm{ref}}=Mf_{\mathrm{min}}=0.0057$. Left panels show the inspiral region, while the right panels zooms on the merger-ringdown phase.}
    \label{fig:1370}
\end{figure}

The results from Fig.~\ref{fig:SXSmismatches} show that the \phTE model produces unfaithfulness below $2 \%$ when compared against eccentric NR simulations. The cases with unfaithfulness above $1\%$ correspond to NR simulations with initial eccentricity $e_0\sim 0.3$, where we expect that a degradation in the accuracy of the model due to the eccentricity expanded expressions used to construct the waveform, see Sec. \ref{sec:construction}. Additionally, when comparing the (2,|2|)-modes against NR we also compute the mismatches for the \seobe and \teobdali models, which show a lower unfaithfulness and the results are consistent with the ones presented in Refs.~\cite{Gamboa:2024hli, Nagar:2021xnh}.
Finally, \phTE accurately captures the physics of higher modes, as evidenced by the fact that $\mathcal{M}_{22}$ mismatches are comparable to the full $\mathcal{M}$ mismatches at a fixed inclination of $\iota=\pi/3$. We do not include HM mismatches for the \seobe and \teobdali models due to their computational cost. For a reference, results for HM mismatches for these models can be found in Ref.~\cite{Gamboa:2024hli}, where they adopt the same parameters optimized for the $(2,2)$-mode. Unlike for those models, the degradation due to HMs is not significant for the \phTE model.
Due to the limited coverage of the parameter space by the available NR eccentric simulations, and to further investigate the model's performance across a broader parameter space, we extend our analysis in the next section by computing mismatches against the \seobe model.

The accuracy of the higher order modes is further illustrated in Fig.~\ref{fig:1370}, where we compare the real part of the available \phTE modes for the SXS:BBH:1370 simulation against the best-fitting \phTE waveform. The intrinsic parameters of this simulation correspond to $q=2$, $\chieff=0$, $e^{\text{GW}}_0=0.294$, and $l^{\text{GW}}_0=0.28$, while the optimal values for \phTE are $e_{\mathrm{ref}}=0.251$ and $l_{\mathrm{ref}}=3.68$ at $Mf_{\mathrm{ref}}=0.0067$.
The two waveforms show remarkable agreement, even for this case, which corresponds to one of the highest mismatches of the dataset, $\mathcal{M}=1.43\%$.

\subsection{Optional model parameters and choice of default settings}\label{subsec:modellingstrategies}
\begin{figure}[ht]
    \centering
    \includegraphics[width=\columnwidth]{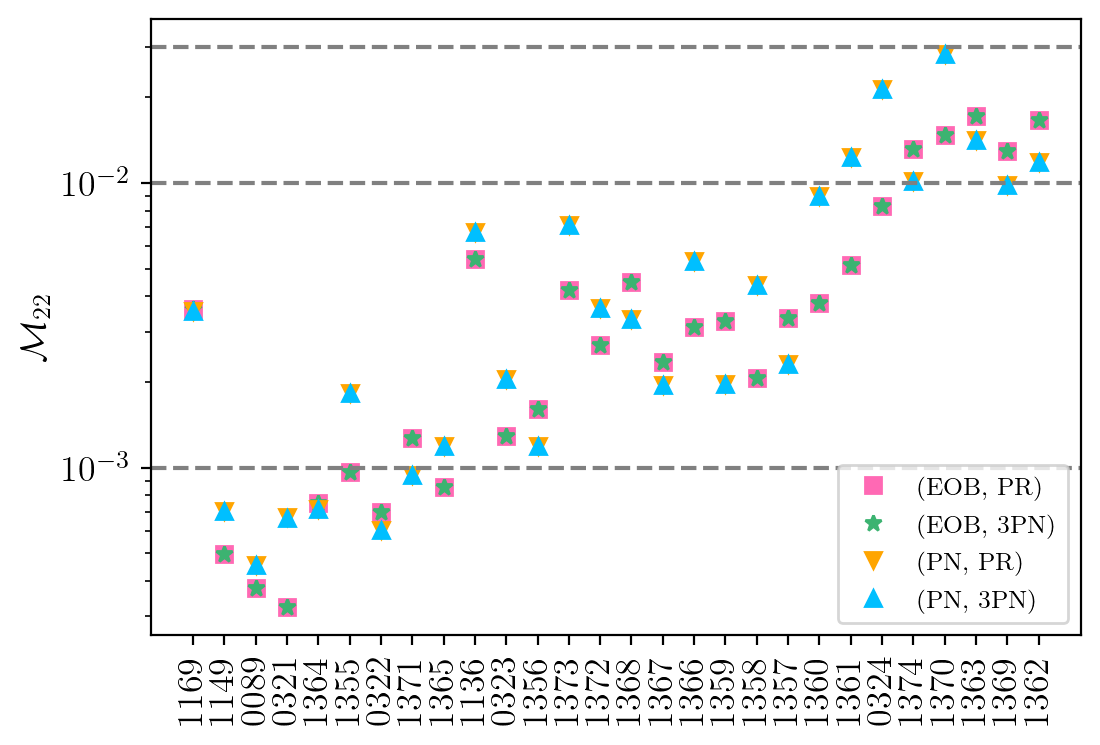}
    \caption{Mismatches for the $(2,|2|)$ modes $\mathcal{M}_{22}$ against the 28 eccentric NR simulations shown in Fig.~\ref{fig:NRsims}, with a total mass of $M=20M_{\odot}$. The $x$-axis corresponds to the simulation labels, arranged by increasing initial eccentricity. We compare different model configurations by varying the gauge choice (EOB vs. PN), the PN order of spinning eccentric corrections to the waveform modes' amplitudes (2PN vs. 3PN), and the application of the phase redefinition in Eqs.~\eqref{eq:xi} and \eqref{eq:psi} (PR). The default options for the spinning PN order for the modes (2PN) and the phase redefinition (off) are not indicated in the legend. }
    \label{fig:modeloptions}
\end{figure}
In the previous section, we presented the mismatches between the default model and the publicly available eccentric NR simulations across a range of masses. In this section, we focus on assessing the impact of the modeling choices introduced in Sec.~\ref{sec:construction} on the model's performance by recomputing the $(2,2)$-mismatches $\mathcal{M}_{22}$.
The model provides three main options.
\begin{enumerate}
    \item \emph{Gauge of the secular evolution equations.} We implement 3PN spinning secular evolution equations for $e$, $l$ and $x$ in the modified-harmonic gauge (denoted as PN) and in EOB coordinates. The default option in the \phTE model are EOB coordinates, as it shows better agreement with NR simulations as well as a closer eccentricity definition to the one described in Ref.~\cite{Ramos-Buades:2022lgf}.
    \item \emph{Order of the spinning contributions to the modes}. Ref.~\cite{Henry:2023tka} provides the full 3PN spinning contribution to the $(2,2)$, $(2,1)$ and $(3,3)$ modes, which are incorporated into the model. However, the default version includes only up to the 2PN spinning contributions to be consistent with the PN information included in the quasicircular \phTHM model, and thus have a more accurate quasicircular limit.
    \item \emph{Phase redefinition.} To remove phase gauge dependencies, it is common to apply the phase shift defined in Eq.~\eqref{eq:psi}. The default version does not apply this phase shift to maintain consistency with the non-eccentric \phTHM model, assuming that this redefinition is already accounted for in the NR calibration of the model.
\end{enumerate}
To assess the impact of the model's different options, we compute the $(2,2)$ mismatch $\mathcal{M}_{22}$ for all NR simulations, and present the results at the lowest mass, $M=20M_{\odot}$, in Fig.~\ref{fig:modeloptions}. The $x$-axis represents the simulation labels, ordered by increasing initial eccentricity. We evaluate all combinations of the gauge choice (EOB\textbf{$\equiv$default}, or PN) with the PN order of spin corrections in the modes (2PN\textbf{$\equiv$default}, or 3PN), and the phase redefinition (PR off\textbf{$\equiv$default}, or PR on).  
First, the results show that the mismatches increase with higher eccentricities, consistent with the trend observed in Fig.~\ref{fig:SXSmismatches}. 
Among the different modeling options, the dominant factor affecting the mismatch is the choice of gauge, while the phase redefinition and higher-order PN spinning corrections have a negligible impact in the eccentric case. Thus, we keep the 2PN spin terms and not applying the phase redefinition as the default options of the model, so that the \phTE model has a more accurate quasicircular limit.
Regarding gauge choices, for low initial eccentricities, both PN and EOB gauges yield comparable accuracy, with EOB generally performing slightly better, though not significantly. However, for higher eccentricities, EOB shows a notable improvement, with mismatch reductions of up to 50\%. Specifically, with the EOB gauge, mismatches remain below 2\%, as also seen in Fig.~\ref{fig:SXSmismatches}, whereas they increase to 3\% in certain cases when using PN. Given this improvement, along with the better alignment of $e_{\mathrm{EOB}}$ with the eccentricity definition used in the \texttt{gw\_eccentricity} python package \cite{Ramos-Buades:2022lgf,Shaikh:2023ypz},  and in the EOB eccentric waveform models, we select EOB as the default gauge for the model.

\subsection{Robustness across parameter space}\label{subsec:robustness}
The publicly available eccentric NR simulations provide a limited coverage of the full eccentric aligned-spin parameter space. As shown in Fig.~\ref{fig:NRsims}, most simulations are non-spinning and limited to small mass ratios. In contrast, Ref.~\cite{Gamboa:2024hli} presents a more comprehensive set of simulations, allowing for model validation over a broader parameter space. Additionally, various robustness tests were conducted during the LVK internal review of the \seobe model. Thus, we assess the accuracy and robustness of the \phTE model by comparing it to \seobe across parameter space.
We generate 5000 waveforms using \seobe on a random uniform grid spanning in mass ratio $q\in[1,20]$, total mass $M\in[10, 300]\ M_{\odot}$,  reference orbital eccentricity $e\in[0,0.4]$, reference mean anomaly $l\in[0,2\pi]$, z-component of the dimensionless spin vectors $\chi_i\in[-1,1]$, inclination $\iota\in[0,\pi]$, azimuthal phase $\varphi\in[0,2\pi]$ at a dimensionless reference frequency $M\Omega_{\mathrm{ref}}=0.03$. We first compute $(2,2)$ mismatch $\mathcal{M}_{22}$ by optimizing over the eccentricity $e$ and mean anomaly $l$ at the $f_{\mathrm{min}}$ which matches the duration of the NR simulation, and then calculate the mismatch $\mathcal{M}$ including all available multipoles in both models optimizing additionally over $\varphi$.

\begin{figure*}
    \centering
    \includegraphics[width=\linewidth]{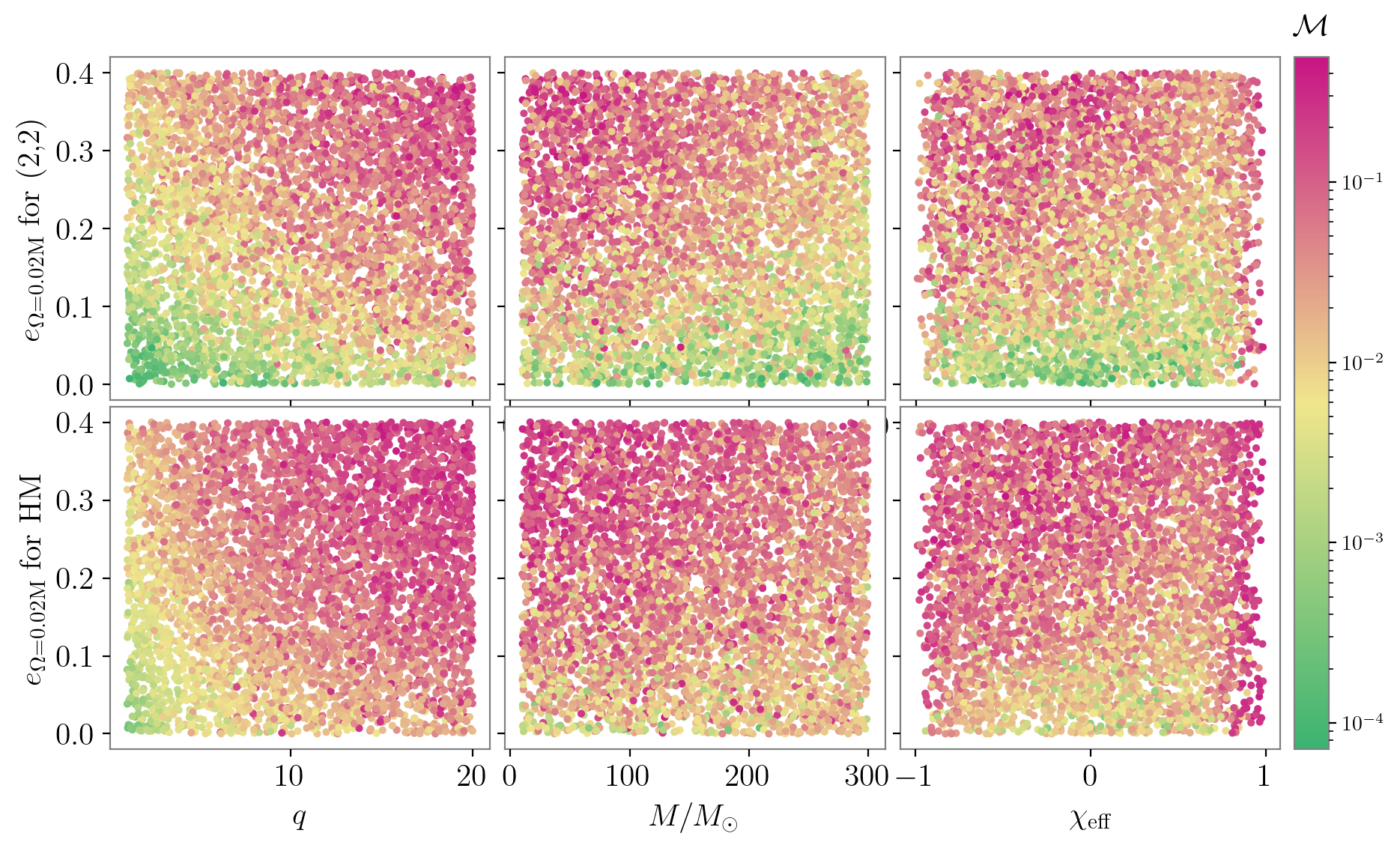}
    \caption{Mismatch distribution between the \seobe and the \phTE models computed over 5000 eccentric waveforms for the dominant $(2,|2|)$-modes (top row, $\mathcal{M}_{22}$), and including all available multipoles for each model (bottom row, $\mathcal{M}$). The configurations are uniformly distributed in mass ratio $q\in [1,20]$, total mass $M\in [10,300]\ M_{\odot},$ $z$-component dimensionless spin vectors $ \chi_i\in[-0.995,0.995],$ azimuthal phase $ \varphi\in[0,2\pi]$, inclination angle $\iota\in[0,\pi]$, and mean anomaly $l$ and eccentricity $e$  defined at $\omega=0.02M$ $e_{\omega=0.02M}$ for the \seobe model. For the \phTE model, waveforms are obtained after optimizing over $e$, $l$, and $f_{\min}$ for $\mathcal{M}_{22}$, with $\varphi$ added for $\mathcal{M}$.
    }
    \label{fig:SEOBmismatches}
\end{figure*}
The top row of Fig.~\ref{fig:SEOBmismatches} shows the $\mathcal{M}_{22}$ mismatches, while the bottom row presents the all-modes mismatch $\mathcal{M}$. The $y$-axis, shared across all plots, represent the eccentricity at the initial dimensionless orbital frequency $\Omega$. Each column corresponds to a different parameter: mass ratio, total mass, and effective spin, respectively.
As expected, we observe low mismatches (below 1\%) in the regions of the parameter space where the underlying aligned-spin QC models have been tested, and where PN expansions are expected to be more reliable. In contrast, mismatches increase in unexplored regions of the parameter space where neither of the two models has been validated due to the lack of NR simulations. A visual inspection of the worst mismatch cases does not reveal any unexpected behavior in either model; rather, we find that the waveforms correspond to different binary systems, and consequently, different evolutions. Therefore, we caution against drawing conclusions on the accuracy of any model from those cases of high mismatches.
The main takeaway from Fig.~\ref{fig:SEOBmismatches} is that \phTE exhibits a reasonable and coherent behavior across the full parameter space, with comprehensive transitions from regions to low to higher mismatches. It demonstrates excellent agreement with \seobe for small mass ratios, even at relatively high initial eccentricities (close to 0.4), and progressively degrades as the mass ratio and eccentricity increase, both due to the PN expansions and the underlying QC models.
For very low total mass systems, we observe higher mismatches at high eccentricities due to differences in frequency evolution, which become more pronounced in long waveforms. Conversely, for high-mass binaries, where only merger-ringdown phase is detectable, the agreement between the underlying non-eccentric models is remarkable. Finally, we note that the mismatch increases for extreme positive effective spin, as expected from the behavior of the non-eccentric baseline models~\cite{Pompili_2023,Estelles:2020twz}.
 
In conclusion we have investigated the robustness of the \phTE model across parameter space by comparing with the \seobe model and found agreement at comparable masses and large discrepancies in corners of parameter space. The visual inspection of the high mismatch cases reveal no unphysical behavior in the \phTE multipoles, but rather waveform systematics between the models. This robustness study is complemented with all the parameter estimation runs presented in Sec. \ref{sec:PE}, during which the model is evaluated a larger number of times $\mathcal{O}(10^6-10^7)$. We leave for future work a more comprehensive study of waveform systematics in the nonprecessing-spin eccentric parameter space.

\subsection{Timing results}\label{subsec:benchmarks}
One of the most outstanding features of the \phTE model compared to other available IMR time-domain eccentric models~\cite{Islam:2021mha, Setyawati:2021gom, Gamboa:2024hli, Gamba:2024cvy} is its computational efficiency. 
Eccentric models generally come with a higher computational cost due to the intricate time evolution of the eccentric quantities. 
For example, EOB eccentric models require numerical integration of the secular evolution equations, which involve evaluating lengthy PN expansions at each time step, apart from the usual integration of the Hamiltonian and equations of motion. While EOB models have proved to provide high accuracy in modeling eccentric systems, they are generally slower.
In contrast, the \phTE model is based on phenomenological approaches that rely on closed-form expressions, making it faster at the cost of some accuracy. While the \phTE model may not capture the full range of eccentric effects with the same precision as EOB models, it offers a much more computationally efficient alternative.

The \phTE model achieves a high computational efficiency by implementing a highly optimized approach to solving the orbit-averaged eccentric dynamics and evaluating the waveform modes.
The method for solving the coupled differential equations from Eqs.~\eqref{eq:EDOs} is detailed in Sec.~\ref{subsubsec:orbfreq}. In summary, the key to accelerating the computation is to extract an optimal integration grid based on the QC frequency from the quasicircular \phTHM model, which determines the rate at which the frequency evolves. This is achieved by evaluating the \phTHM frequency on a grid constant step in the $\theta$ variable, see Eq.~\eqref{eq:TaylorT3}, during the inspiral region of the \phTHM model, and then apply a constant grid of $\mathrm{d}t=0.5M$ until the peak frequency of the \phT model. Subsequently, this solution is interpolated and used by the solver to evaluate the right hand side of the evolution equations.

For waveform mode evaluation, we also leverage the usage of non-uniform time grids. The ODE solver extracts the minimum number of points needed to accurately describe the secularly evolving quantities. The PN expressions for the modes, as given in Refs.~\cite{Ebersold:2019kdc} and \cite{Henry:2023tka}, take the form
\begin{equation}
    H_{lm}=|H_{lm}|e^{-im\lambda} =\left(A_{n}(x,e)e^{-inl}\right)e^{-im\lambda},
\end{equation}
where the oscillatory component arises from the mean anomaly $l$ and the orbit-average orbital phase. Henceforth, we focus on the calculation of the amplitudes, whose evaluation can be separated into two parts: (1) an ``orbit-averaged'' component $A_{n}$ that only needs to be computed at the grid points of the secular solver, and (2) an oscillatory component from the exponentials of $l$, which is shared across all modes and only needs to be computed once. The former is computed on a coarser grid, interpolated and evaluated at the corresponding input constant time step needed for the Fourier transform of the signal, while the latter only needs a single computation. We leave further optimizations in other parts of the code as future work.
This optimization in mode evaluation significantly reduces computational cost, particularly at low masses, where even the underlying QC model is slow due to the necessity of evaluating all quantities at each time step. Notably, no changes have been made so far to the original \phTHM model, where similar techniques could be applied to further accelerate computations. We also leave this improvement for future work.
\begin{figure*}
    \centering
    \includegraphics[width=\linewidth]{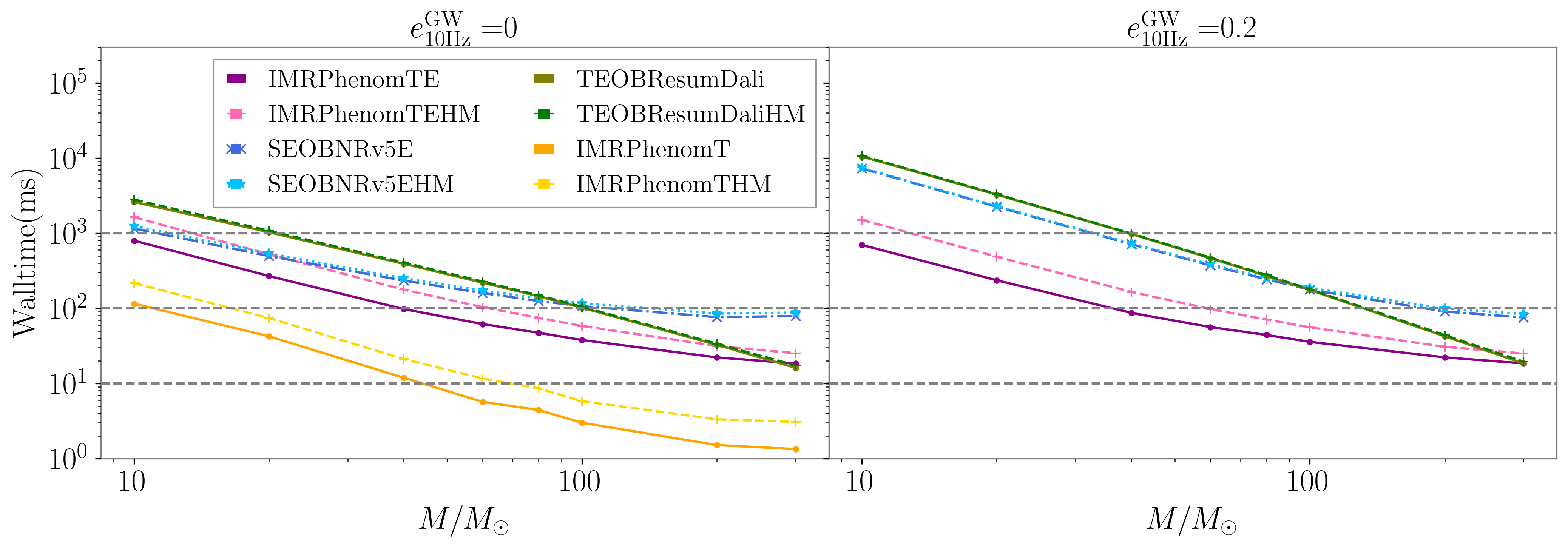}
    \caption{Benchmark results comparing the computational efficiency of different waveform models. The plot shows the wall time (in milliseconds) required to generate waveforms for a binary with mass ratio $q=3$, spins $\chi_1 = 0.4$, $\chi_2=0.3$, and mean anomaly $l=1.2$ at $10$ Hz for different initial eccentricities ($e=0,0.2$), each represented in a separate column. Solid lines correspond to the evaluation of only the dominant harmonic, while dashed lines include all higher modes. We compare \phTE, \seobe, \teobdali for all cases, including \phTHM, as implemented in the \texttt{phenomxpy} package in the $e=0$ case. The benchmark is performed over a range of total masses, $M=\{10,20,40,60,80,100,200,300\}M_{\odot}$.}
    \label{fig:bench}
\end{figure*}
We present benchmark results in Fig.~\ref{fig:bench}, evaluating a waveform with $q=3$, $\chi_1 = 0.4$, $\chi_2=0.3$, and $l=1.2$ at $10$Hz for different initial eccentricities, one per column ($e=0,0.2$). The plot shows the wall time, in milliseconds, required to evaluate each waveform model, considering only the dominant harmonic (solid lines) and including all higher modes (dashed lines), as indicated in the legend. The benchmark is performed for different total masses using \phTE, \seobe, \teobdali for both cases, including \phTHM as implemented in \texttt{phenomxpy} for the non-eccentric case. 
For total masses below $100M_{\odot}$, \phTE is significantly faster than the other two eccentric models. 
For the $e=0.2$ case at the lowest mass $M=10M_{\odot}$, we find \phTE waveform generation times of around $800$ ms for the dominant harmonic and $1000$ ms when including all modes, compared to $8000$ ms for \seobe and $10^{4}$ ms for \teobdali. These results demonstrate that \phTE is approximately an order of magnitude faster than the state-of-the-art EOB models, a trend that approximately holds up to total masses of $M=100M_{\odot}$. 
At higher masses, only \teobdali shows slightly faster performance evolution than the other two models, exhibiting a steeper decrease in wall time. The wall time difference between \seobe and \phTE remains stable, while \teobdali eventually reaches the same wall time as \phTe - approximately 11 ms at $M=300M_{\odot}$- performing slightly faster than \phTE. However, at this scale, these differences become negligible.
We find a similar trend in the non-eccentric case, with smaller absolute differences in wall time when comparing different models. The considerable slowdown compared to QC models arises from the fact that, even though the eccentric expansions vanish, the evolution of the orbit-averaged quantities still requires numerical integration.
Finally, comparing \phTE with its QC counterpart, \phTHM, we observe that the slowdown at low masses originates from \phTHM itself. Any optimization applied to \phTHM would thus have a substantial impact on \phTE, making this a promising avenue for future improvements.

The advantage of a computationally efficient eccentric model is the ability to perform parameter estimation studies with a wide range of configurations, as we demonstrate in Sec.~\ref{sec:PE}.

\section{Parameter Estimation studies}\label{sec:PE}
A key application of waveform models is the Bayesian inference of source parameters from GWs. 
In this section, we assess the performance of the eccentric, aligned-spin \phTE model through PE studies. We conduct synthetic injections of three NR signals introduced in Sec.~\ref{subsec:SXS} into zero detector noise, and analyze two real GW events detected by the LVK Collaboration - GW150914~\cite{GW150914, LIGOScientific:2016vlm} and GW190521~\cite{GW190521}. 
We compare our findings with results from the literature, specifically those obtained using the \seobe model~\cite{Gamboa:2024hli}. 

We use the Python package \texttt{\texttt{bilby\_pipe}}~\cite{bilby_pipe_paper,bilby}, with the nested sampler \texttt{dynesty}~\cite{dynesty}. 
We adopt priors on the inverse mass ratio ($1/q$) and chirp mass ($\mathcal{M}$) to ensure a uniform distribution in the component masses. The priors on the reference eccentricity $e_{\text{ref}}$ and mean anomaly $l_{\text{ref}}$ are uniformly distributed, with $l_{\text{ref}}\in[0,2\pi]$ and $e_{\text{ref}}\in[0,e_{\max}]$, where $e_{\max}$ specified for each case. 
For the spin components $\chi_i$, we use priors corresponding to the projections of a uniform and isotropic spin distribution along a direction perpendicular to the binary's orbital plane~\cite{PhysRevD.91.042003}.
For the luminosity distance $d_L$, we generally use the commonly adopted prior proportional to $d_L^2$~\cite{gwtc1, gwtc2, gwtc21, gwtc3}, except for GW190521, where we specify a prior uniform in comoving volume prior as in Ref.~\cite{GW190521}.
The remaining priors, including the extrinsic parameters and the binary's orbital phase $\varphi$, are the same as in Ref.~\cite{gwtc1}. The specific prior boundaries for each parameter are detailed in the corresponding sections for each case.

Comparing eccentric parameters across different waveform models- or even within the same model but in different coordinates, as is our case- requires additional post-processing due to the gauge dependence of eccentricity in General Relativity.
Thus, we present our results using a commonly established definition of eccentric parameters across models. 
We adopt the GW eccentricity $e^{\mathrm{GW}}$ and mean anomaly $l^{\mathrm{GW}}$ as introduced in Ref.~\cite{Ramos-Buades:2022lgf,Shaikh:2023ypz}, which are efficiently computed using the \texttt{gw-eccentricity} Python package~\cite{Shaikh:2023ypz}.
We thus evaluate the waveform at each sample of the posterior distribution and apply the \texttt{gw-eccentricity} package to measure $e^{\mathrm{GW}}$ and $l^{\mathrm{GW}}$ at the chosen reference time.

A summary of the PE runs performed in this study is presented in Table~\ref{tab:pesummary}, with details about the waveform model used, the computational resources required, and the specific minimum and reference frequencies for each run. 
For all runs, we set the number of accepted steps to 60 (\texttt{naccept}=60) and the number of live points to 1000 (\texttt{nlive}=1000) for the \texttt{dynesty} sampler. Additionally, we enable distance marginalization to reduce computational cost, and we set the sampling rate to 4096 Hz.
The computational efficiency of our model allows us to investigate different configurations/physical descriptions of our model and its impact on waveform systematics.
In particular, we assess the effect of using the two available gauge choices for our model -EOB (default) and PN- as well as the role of higher order modes. 

Incorporating higher order modes requires lowering the starting frequency to ensure all relevant modes remain within the analysis frequency band. HMs with $m>2$ have higher frequency content for the same time interval, so one needs to adjust the starting frequency based on the highest $m$-mode ($m_{\max}$) included in the analysis,
\begin{equation}
f_{\mathrm{min}}=\frac{2f_{\mathrm{start}}}{(m_{\max}-2)+2}.
    \label{eq:fstart}
\end{equation}
As shown in Tab.~\ref{tab:pesummary}, the minimum frequency is adjusted according to Eq.~\eqref{eq:fstart} based on the maximum number of $l$-modes included and the starting frequency of the analysis.
As introduced in Sec. \ref{sec:construction}, the \phTE model is capable to specify a starting frequency distinct from the reference frequency. Hence, changes in the starting frequency, due to the higher order modes or conditioning of time-domain waveforms to perform Fourier transforms, do not modify the physical configuration.

\begin{table*}[ht]
\centering
\begin{tabular}{@{}ccccc@{}}
\toprule
\multirow{2}{*}{\shortstack{\textbf{Event} \\ (Detectors, $\Delta T$ (s))\\ $[f_{\min},f_{\max}]$}} & \multirow{2}{*}{\textbf{Model (sampler)}} & \multicolumn{1}{c}{\textbf{Waveform settings}} & \textbf{Computing resources} & \textbf{Runtime} \\ 
\cmidrule(r){3-3} \cmidrule(r){4-4} 
  & &  \textbf{$f_\mathrm{ref}$($f_\mathrm{min}$) (Hz)} & \textbf{cores} $\times$ \textbf{nodes} &  \\ 
  \addlinespace[1.4pt]
\midrule
\multirow{4}{*}{\shortstack{\textbf{\texttt{SXS:BBH:1355}} \\ (LHV, 4) \\ $[20,2048]$}} 
  & \phTe (EOB)  & 20 & 50 $\times$ 1 & 14h 38min \\ 
  & \phTe (PN) &  20 & 50 $\times$ 1 & 13h 57min \\ 
  & \phTE (EOB, $l\leq 4$)  & 20(10) & 112 $\times$ 1 & 21h 49min \\ 
  & \phTE (PN, $l\leq 4$) &  20(10) & 112 $\times$ 1 & 21h 07min \\ 
 \midrule
\multirow{4}{*}{\shortstack{\textbf{\texttt{SXS:BBH:1359}} \\ (LHV, 4) \\ $[20,2048]$}} 
  & \phTe (EOB)  &  20& 50 $\times$ 1 & 15h 17min \\ 
   & \phTe (PN)  &  20 & 50 $\times$ 1 & 15h 10min \\
    & \phTE (EOB, $l\leq 4$) &  20(10) & 112 $\times$ 1 & 21h 57min \\ 
   & \phTE (PN, $l\leq 4$) & 20(10) & 112 $\times$ 1 & 21h 04min \\ 
 \midrule
\multirow{4}{*}{\shortstack{\textbf{\texttt{SXS:BBH:1363}} \\ (LHV, 4) \\ $[20,2048]$}} 
 & \phTe (EOB)  & 20 &50 $\times$ 1 & 15h 15min \\ 
   & \phTe (PN) &  20 & 50 $\times$ 1 & 16h 47min \\
   & \phTE  &  20(10) & 112 $\times$ 1 & 18h 34min \\ 
   & \phTE (20Hz) &  20 &112 $\times$ 1 & 8h 20 \\ 
 \midrule
\midrule
\multirow{3}{*}{
\shortstack{\textbf{GW150914} \\ (LH, 8) \\ $[20,896]$}} 
& \phTHM  & 20(10) & 112 $\times$ 1 & 9h 02min \\
 & \phTE (EOB, $l\leq 4$)  &  20(10) & 112 $\times$ 1 & 1d 3h \\ 
 & \phTE (PN, $l\leq 4$)  & 20(10) & 112 $\times$ 1 & 1d 2h \\ 
 % \addlinespace[0.2em]
 % \cdashline{2-7}
 % \addlinespace[0.2em]
 % & \seobe  & LH & 8 & 10 & 64 $\times$ 1 & 3d 20h \\ 
\midrule
\multirow{3}{*}{
\shortstack{\textbf{GW190521} \\ (LHV, 8) \\ $[11,521]$}}
& \phTHM ($l\leq 4$)  & 5.5 & 112 $\times$ 1 & 2h 40min \\
& \phTE  ($l\leq 4$) & 5.5 & 112 $\times$ 1 & 11h 06min  \\ 
& \phTE  ($l\leq 4$, $r<10M$)& $5.5^*$& 112 $\times$ 1 & 7h 32min  \\ 
% \addlinespace[0.2em]
% \cdashline{2-7}
% \addlinespace[0.2em]
%  & \seobe  & LHV & 8 &  5.5 & 32 $\times$ 1 & 3d 13h \\ 
\bottomrule
\end{tabular}
\caption{Summary of the PE runs performed in this study. The first column lists the NR SXS simulation used for injections and the GW events, along with the detectors considered (L for Livingston, H for Hanford, and V for Virgo) and the duration of the injected signal ($\Delta T$ in seconds). The second column specifies the waveform model, including the highest $l$-modes and coordinates used. The \textbf{Waveform settings} column provides the reference frequency, with the minimum frequency in brackets when different. The star $^*$ next to the minimum frequency indicates that no additional waveform conditioning was applied, meaning the minimum frequency is exactly as listed. The table also includes the computing resources and runtime, enabling the total computational hours to be calculated by multiplying the resources by the runtime. }
\label{tab:pesummary}
\end{table*}

\subsection{Numerical relativity injections}\label{subsec:NRinjections}
In this section, we conduct zero-noise injections for three of the publicly available eccentric NR waveforms and perform PE studies to assess the model's accuracy in recovering the injected parameters. For comparison with Refs.~\cite{Ramos-Buades:2023yhy,Gamboa:2024hli}, we selected the same three public SXS NR simulations: \texttt{SXS:BBH:1355}, \texttt{SXS:BBH:1359} and \texttt{SXS:BBH:1363}, corresponding to eccentricities at the initial orbit averaged frequency for each waveform of 0.077, 0.145, and 0.317 respectively. We included all NR modes up to $l=8$, set the total mass to $M=70M_{\odot}$, fixed the inclination angle between the system's orbital angular momentum and the line of sight $\iota=0$, and used a coalescence phase $\varphi=0$ at a luminosity distance of $d_{L}=2307$~Mpc. This configuration yields a three-detector matched-filtered SNR, $\mathrm{SNR}^{\mathrm{N}}=20$, when using the zero-detuned high power Advanced LIGO(for LIGO Livingston and Hanford) and Virgo PSDs at design sensitivity~\cite{Aasi_2015,Acernese_2015,dcc:2974}.

The computational efficiency of the \phTE model enables us to systematically run PE studies including higher modes in in less than a day, as shown in Tab.~\ref{tab:pesummary}. With these NR injections, we investigate the impact of HMs on signal recovery, and explore the influence of the coordinate choice, EOB(default) or PN.
Note that the choice of simulations -all featuring equal-mass, non-spinning black holes and zero inclination- would allow us to use only the dominant $(2,|2|)$ harmonics, as only modes with $m=2$ are visible for a face-on binary, and other analyses of these events using EOB models are typically performed with only the $(2,|2|)$ mode due to the high computational cost~\cite{Ramos-Buades:2021adz, Gamboa:2024hli}.
For each simulation, we performed four different PE runs (see Tab.~\ref{tab:pesummary} for details): using either EOB or PN coordinates and including either only the dominant $(2,2)$ harmonic or all modes up to $l=4$. 
Due to the characteristics of the binaries (equal-mass and no inclination), we do not include $l=5$ modes, as we do not expect higher modes to have a significant impact on these specific runs. 

We adopt the same priors as in Refs.~\cite{Ramos-Buades:2023yhy,Gamboa:2024hli}: $1/q\in[0.05,1]$, $\mathcal{M}\in[5,100]M_{\odot}$, $\chi_i\in[0,0.99]$. 
The eccentricity prior has an upper bound of $e_{\max}=0.3$ for \texttt{SXS:BBH:1355} and \texttt{SXS:BBH:1359}, whereas for \texttt{SXS:BBH:1363}, we extend the upper bound to $e_{\max}=0.5$ to prevent the posterior from railing. These priors are set at a reference frequency of 20 Hz.

\begin{table*}
    %\centering
    \begin{tabularx}{\textwidth}{>{\centering\arraybackslash}X >{\centering\arraybackslash}c 
>{\centering\arraybackslash}c >{\centering\arraybackslash}X 
>{\centering\arraybackslash}X >{\centering\arraybackslash}X 
>{\centering\arraybackslash}X}
        \toprule
        \multirow{2}{*}{\textbf{Event}} & \multirow{2}{*}{\textbf{Parameter}} & \textbf{Injected} & \phTe &   \phTe & \phTE & \phTE \\
        & & \textbf{value} & (EOB) & (PN)  &(EOB) &   (PN) \\
        \midrule
        \multirow{10}{*}{\texttt{SXS:BBH:1355}}
        & $M / M_\odot$ & 70.0 & $70.70^{+3.10}_{-2.67}$  & $70.71^{+3.03}_{-2.63}$ &$70.56^{+2.84}_{-2.64}$ &$70.64^{+2.82}_{-2.65}$\\
        
        & $\mathcal{M} / M_\odot$ & 30.47 & $30.34^{+1.19}_{-1.16}$   & $30.35^{+1.17}_{-1.13}$ &$30.43^{+1.17}_{-1.13}$ & $30.46^{+1.15}_{-1.13}$
                                        \\
        
        & $1 / q$ & 1.0 & $0.78^{+0.19}_{-0.23}$ & $0.79^{+0.19}_{-0.23}$ & $0.84^{+0.15}_{-0.23}$ & $0.84^{+0.15}_{-0.22}$
                      \\
        
        & $\chi_{\text{eff}}$ & 0.0 & $0.00^{+0.10}_{-0.10}$   & $0.00^{+0.10}_{-0.11}$ & $0.00^{+0.10}_{-0.11}$ & $0.00^{+0.10}_{-0.11}$
                                  \\
        
        & $e_{20\mathrm{Hz}}\ \left(e^{\mathrm{GW}}_{20\mathrm{Hz}}\right)$ & - (0.07) & $0.05^{+0.05}_{-0.04} \left(0.05^{+0.05}_{-0.04}\right)$  & $0.04^{+0.04}_{-0.03} \left(0.05^{+0.05}_{-0.04}\right)$   &  $0.04^{+0.05}_{-0.04} \left(0.04^{+0.05}_{-0.04}\right)$ & $0.03^{+0.04}_{-0.03} \left(0.04^{+0.05}_{-0.04}\right)$
                                \\
        
        & $l_{20\mathrm{Hz}}\ \left(l^{\mathrm{GW}}_{20\mathrm{Hz}}\right)$ & - (1.96) & $2.04^{+2.01}_{-1.40} \left(1.94^{+2.36}_{-1.37}\right)$   & $2.06^{+1.98}_{-1.37}\left(1.95^{+2.19}_{-1.35}\right)$ & $2.04^{+2.33}_{-1.43}\left(1.95^{+2.62}_{-1.42}\right)$ &  $2.06^{+2.49}_{-1.46}\left(1.96^{+2.78}_{-1.43}\right)$
                               \\
        
        & $\iota [\text{rad}]$ & 0.0 & $0.61^{+0.62}_{-0.43}$  & $0.61^{+0.59}_{-0.44}$  &$0.47^{+0.55}_{-0.35}$ &  $0.48^{+0.53}_{-0.35}$
                      \\
        
        & $d_L[\text{dMpc}]$ & 2307 & $1827^{+442}_{-682}$  &  $1827^{+450}_{-676}$  & $1967^{+358}_{-613}$ & $1964^{+363}_{-592}$
                    \\
        
        & $\varphi$ & 0.0 & $3.11^{+2.87}_{-2.79}$   & $3.18^{+2.79}_{-2.83}$ & $3.12^{+2.83}_{-2.79}$ &  $3.10^{+2.84}_{-2.79}$
                        \\
        
        & $\text{SNR}^N$ & 20.0 & $19.07^{+0.10}_{-0.19}$   & $19.06^{+0.10}_{-0.18}$ & $19.04^{+0.11}_{-0.19}$ & $19.04^{+0.11}_{-0.19}$\\

        & $\log_{10}B_{\mathrm{EOB}/\mathrm{PN}}$ & - & $0.20^{+0.17}_{-0.17}$   & - & $0.06^{+0.17}_{-0.17}$ & -\\

        \bottomrule
        \multirow{10}{*}{\texttt{SXS:BBH:1359}}
        & $M / M_\odot$ & 70.0 & $70.26^{+2.89}_{-2.64}$  & $70.31^{+2.85}_{-2.61}$ & $70.43^{+2.79}_{-2.51}$ & $70.48^{+2.69}_{-2.47}$ \\
        & $\mathcal{M} / M_\odot$ & 30.47 & $30.21^{+1.12}_{-1.18}$   & $30.24^{+1.15}_{-1.17}$ & $30.42^{+1.14}_{-1.11}$ & $30.45^{+1.12}_{-1.09}$\\
        & $1 / q$ & 1.0 & $0.80^{+0.18}_{-0.23}$   & $0.81^{+0.17}_{-0.23}$ & $0.85^{+0.14}_{-0.22}$ & $0.85^{+0.13}_{-0.22}$\\
        & $\chi_{\text{eff}}$ & 0.0 & $0.00^{+0.10}_{-0.10}$    & $-0.01^{+0.10}_{-0.10}$  & $0.00^{+0.10}_{-0.10}$ & $0.00^{+0.10}_{-0.10}$\\
        & $e_{20\mathrm{Hz}}\ \left(e^{\mathrm{GW}}_{20\mathrm{Hz}}\right)$ & - (0.13) & $0.12^{+0.05}_{-0.05} \left(0.12^{+0.05}_{-0.05}\right)$   & $0.09^{+0.04}_{-0.04} \left(0.12^{+0.05}_{-0.06}\right)$ & $0.10^{+0.05}_{-0.05} \left(0.10^{+0.05}_{-0.05}\right)$ & $0.08^{+0.04}_{-0.04} \left(0.10^{+0.05}_{-0.05}\right)$\\
        & $l_{20\mathrm{Hz}}\ \left(l^{\mathrm{GW}}_{20\mathrm{Hz}}\right)$ & - (0.81) & $0.86^{+5.29}_{-0.75} \left(0.89^{+5.31}_{-0.79}\right)$   & $0.88^{+5.27}_{-0.77} \left(0.88^{+5.30}_{-0.78}\right)$ & $0.92^{+5.25}_{-0.82} \left(1.00^{+5.20}_{-0.90}\right)$ & $0.90^{+5.28}_{-0.81}\left(0.99^{+5.20}_{-0.90}\right)$\\
        & $\iota[\text{rad}]$ & 0.0 & $0.61^{+0.59}_{-0.44}$   & $0.61^{+0.59}_{-0.44}$ & $0.48^{+0.54}_{-0.35}$ & $0.49^{+0.52}_{-0.35}$\\
        & $d_L[\text{dMpc}]$ & 2307 & $1824^{+452}_{-667}$   &  $1837^{+446}_{-680}$ & $1984^{+362}_{-616}$ & $1981^{+370}_{-613}$\\
        & $\varphi$ & 0.0 & $3.11^{+2.84}_{-2.80}$   & $3.08^{+2.90}_{-2.79}$  & $3.16^{+2.80}_{-2.84}$ & $3.14^{+2.82}_{-2.84}$\\
        & $\text{SNR}^N$ & 20.0 & $19.00^{+0.11}_{-0.20}$  & $18.99^{+0.10}_{-0.20}$  & $18.93^{+0.10}_{-0.20}$ & $18.91^{+0.11}_{-0.20}$\\
        & $\log_{10}B_{\mathrm{EOB}/\mathrm{PN}}$ & - & $0.17^{+0.18}_{-0.18}$   & - & $0.13^{+0.17}_{-0.17}$ & -\\
        \bottomrule 
        \multirow{10}{*}{\texttt{SXS:BBH:1363}}
        & & & \phTe &   \phTe & \phTE & \phTE \\
        & & & (EOB) &   (PN) &   & (20 Hz) \\
        \cmidrule(r){4-7}
        & $M / M_\odot$ & 70.0 & $71.83^{+4.75}_{-3.46}$ & $71.92^{+4.00}_{-3.27}$ &$75.13^{+4.04}_{-4.02}$ & $72.51^{+3.07}_{-3.23}$ \\
        & $\mathcal{M} / M_\odot$ & 30.47 & $30.68^{+1.97}_{-1.59}$  & $30.86^{+1.72}_{-1.52}$ & $32.39^{+1.66}_{-1.73}$& $31.00^{+1.78}_{-1.78}$\\
        & $1 / q$ & 1.0 & $0.75^{+0.22}_{-0.24}$  & $0.79^{+0.19}_{-0.24}$ & $0.83^{+0.15}_{-0.23}$& $0.80^{+0.18}_{-0.26}$\\
        & $\chi_{\text{eff}}$ & 0.0 & $0.10^{+0.12}_{-0.11}$  & $0.07^{+0.12}_{-0.11}$ & $0.18^{+0.11}_{-0.12}$ & $0.11^{+0.12}_{-0.12}$\\
        & $e_{20\mathrm{Hz}}\ \left(e^{\mathrm{GW}}_{20\mathrm{Hz}}\right)$ & - (0.25) & $0.26^{+0.03}_{-0.05}\ \left(0.26^{+0.03}_{-0.05}\right)$  & $0.21^{+0.02}_{-0.03}\ \left(0.27^{+0.03}_{-0.04}\right)$ & $0.22^{+0.04}_{-0.04}\ \left(0.22^{+0.04}_{-0.04}\right)$& $0.26^{+0.04}_{-0.04}\ \left(0.26^{+0.04}_{-0.04}\right)$\\
        & $l_{20\mathrm{Hz}}\ \left(l^{\mathrm{GW}}_{20\mathrm{Hz}}\right)$ & - (4.27) & $4.58^{+0.67}_{-0.77}\ \left(4.45^{+0.66}_{-0.77}\right)$  & $4.63^{+0.60}_{-0.66}\ \left(4.51^{+0.59}_{-0.66}\right)$ & $4.22^{+0.77}_{-0.80}\ \left(4.09^{+0.76}_{-0.80}\right)$ & $4.68^{+0.96}_{-1.03}\ \left(4.56^{+0.96}_{-1.01}\right)$\\
        & $\iota[\text{rad}]$ & 0.0 & $0.61^{+0.59}_{-0.44}$  & $0.62^{+0.61}_{-0.45}$ & $0.48^{+0.53}_{-0.35}$ & $0.46^{+0.52}_{-0.34}$\\
        & $d_L[\text{dMpc}]$ & 2307 & $1970^{+507}_{-732}$  & $1989^{+503}_{-736}$ &  $2317^{+433}_{-714}$ & $2219^{+261}_{-501}$\\
        & $\varphi$ & 0.0 & $3.15^{+2.83}_{-2.81}$  & $3.15^{+2.82}_{-2.82}$ & $3.19^{+2.81}_{-2.90}$ & $3.03^{+2.96}_{-2.75}$ \\
        & $\text{SNR}^N$ & 20.0 & $18.84^{+0.11}_{-0.20}$  & $18.83^{+0.11}_{-0.20}$ & $18.51^{+0.12}_{-0.20}$ & $18.84^{+0.11}_{-0.20}$\\
        & $\log_{10}B_{\mathrm{EOB}/\mathrm{PN}}$ & - & $0.36^{+0.18}_{-0.18}$   & - & - & -\\
        \bottomrule
    \end{tabularx}
    \caption{Injected, median values, and 90\% credible intervals for the posterior distributions in Fig.~\ref{fig:NRinjections} for the three NR injections (indicated at each row), recovered with \phTE using different settings specified in the columns. 
    The table reports the total mass $M$ and chirp mass $\mathcal{M}$ (both in solar masses), the inverse mass ratio $1/q$, the effective spin parameter $\chi_{\mathrm{eff}}$, the reference eccentricity and mean anomaly ($e_{20\mathrm{Hz}}$ and $l_{20\mathrm{Hz}}$ respectively), the inclination angle $\iota$, the luminosity distance $d_L$, the coalescence phase $\varphi$, and the network matched-filtered SNR, $\text{SNR}^{\mathrm{N}}$, for LIGO Hanford and Livingston, and Virgo detectors. In brackets, we also report the injected and recovered GW eccentricity $e_{\mathrm{GW}}$ and mean anomaly $l_{\mathrm{GW}}$. All values are given at the reference frequency of 20 Hz.}
    \label{tab:NRresults}
\end{table*}

\begin{figure*}
    \centering
    \texttt{SXS:BBH:1355} % Title for the row
    \begin{minipage}{\textwidth}
    \includegraphics[width=0.345\linewidth]{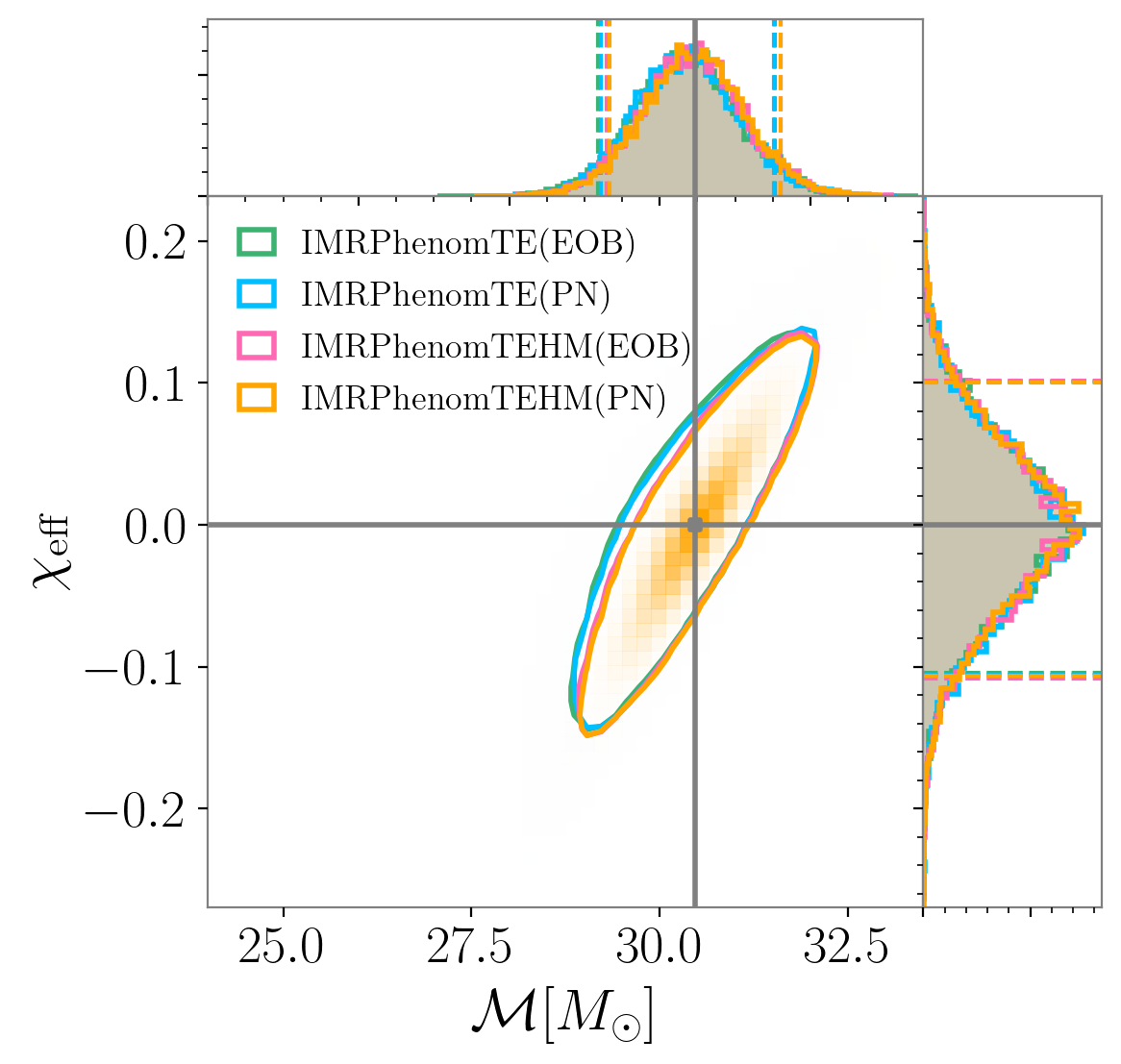}
    \includegraphics[width=0.32\linewidth]{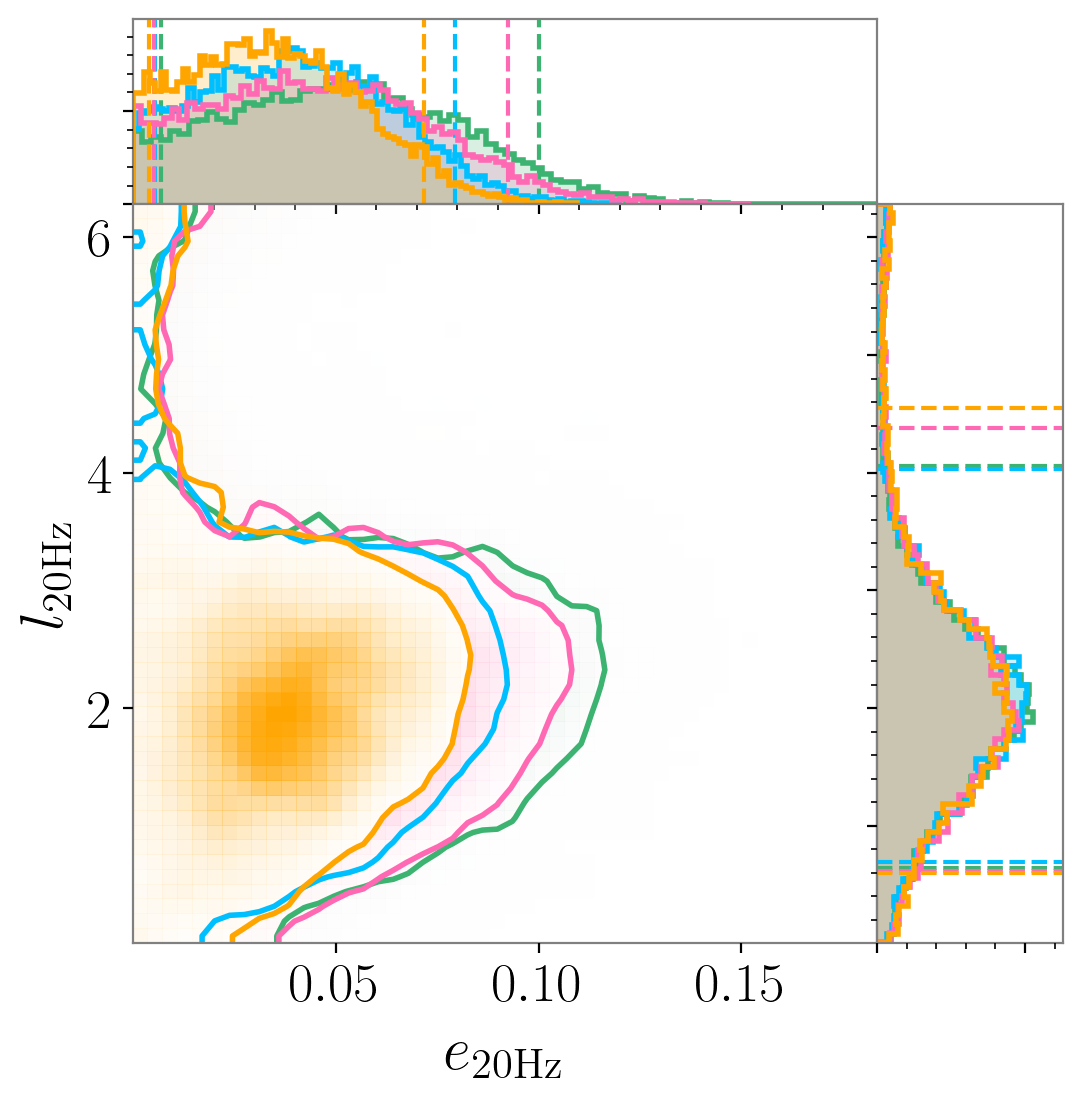}
    \includegraphics[width=0.32\linewidth]{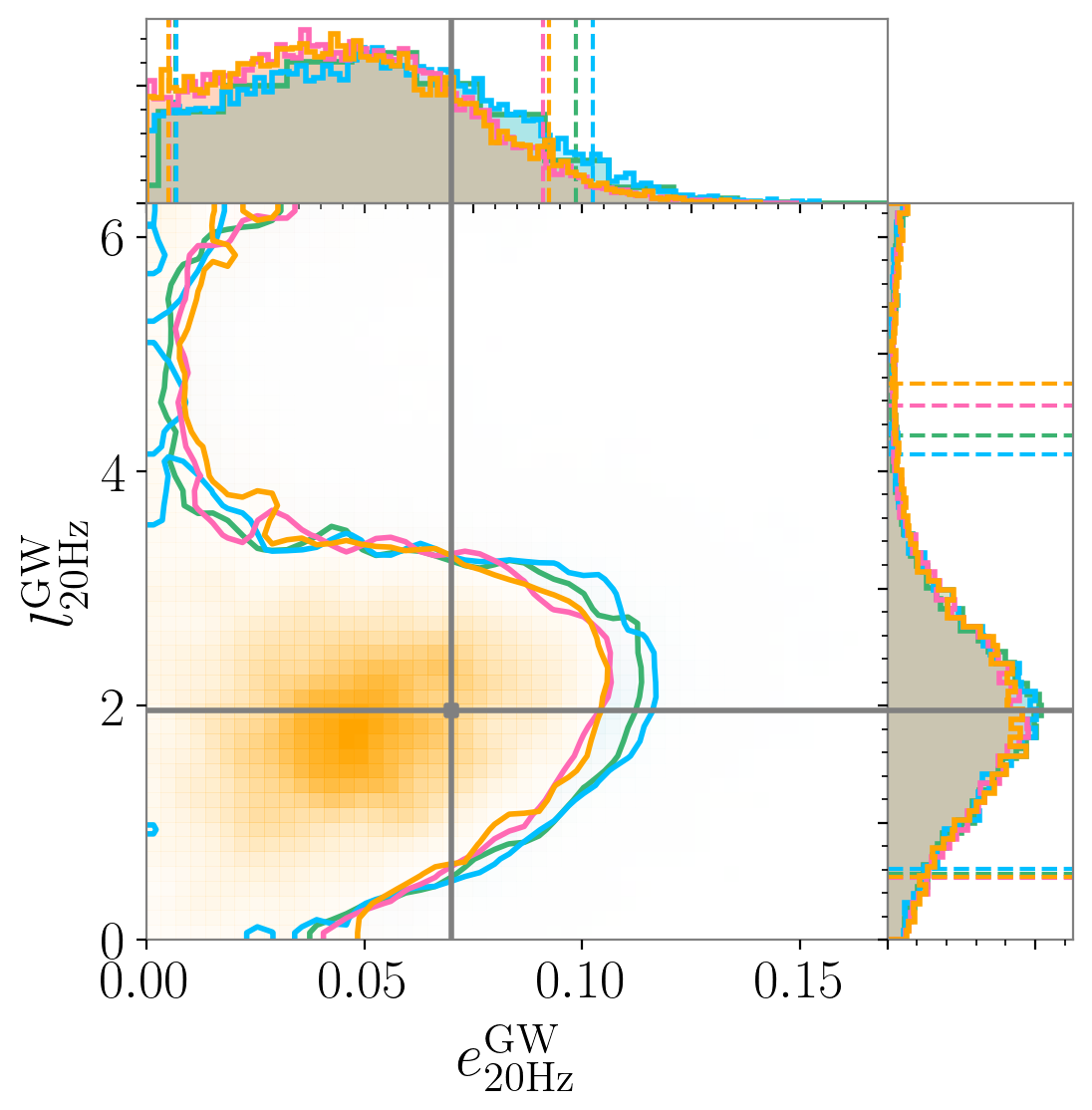}
    \vspace*{0.5em}
    \end{minipage}
    \texttt{SXS:BBH:1359} % Title for the row
    \begin{minipage}{\textwidth}
    \includegraphics[width=0.345\linewidth]{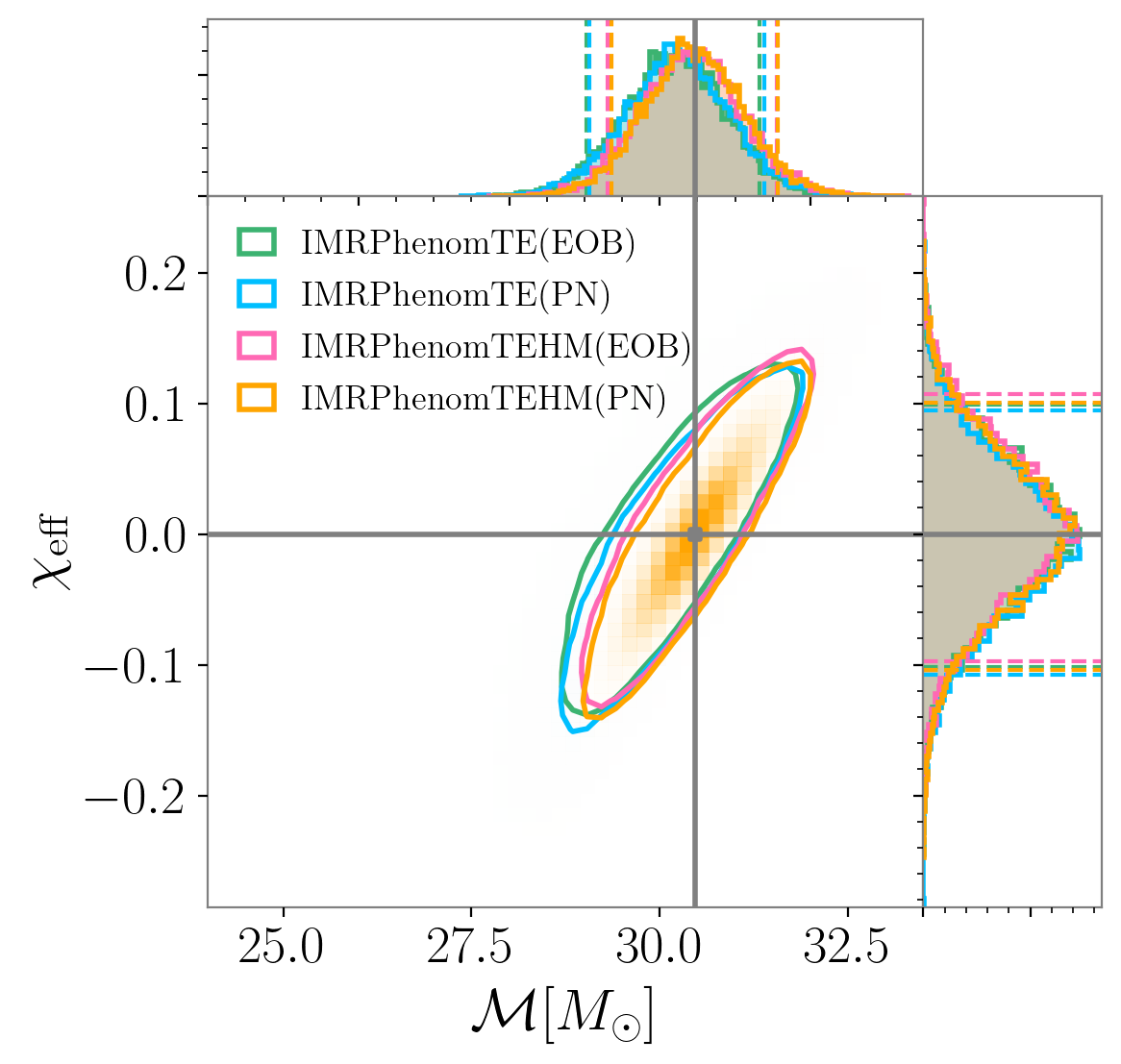}
    \includegraphics[width=0.32\linewidth]{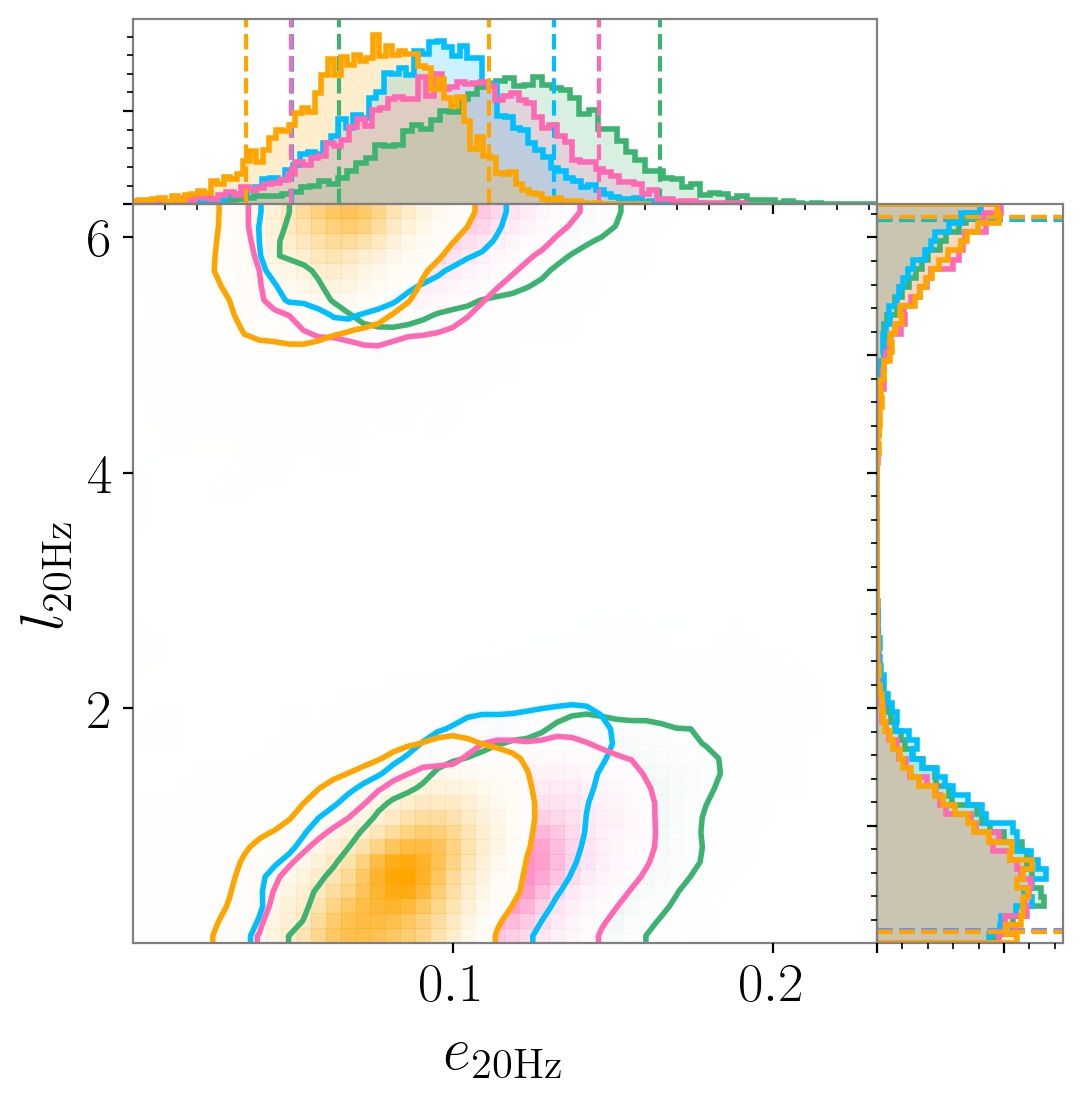}
    \includegraphics[width=0.32\linewidth]{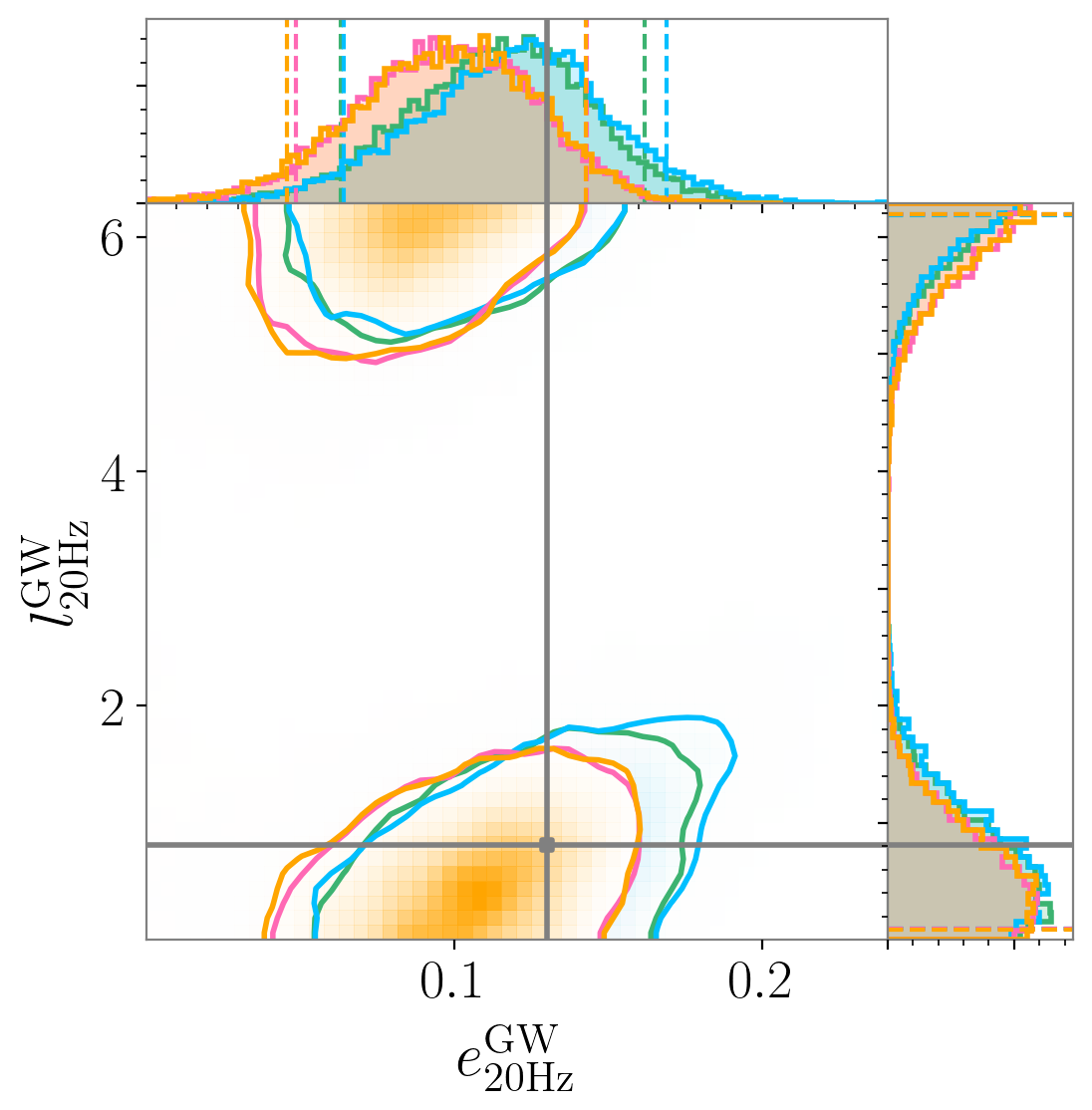}
    \vspace*{0.5em}
    \end{minipage}
    \texttt{SXS:BBH:1363} % Title for the row
    \begin{minipage}{\textwidth}
    \includegraphics[width=0.345\linewidth]{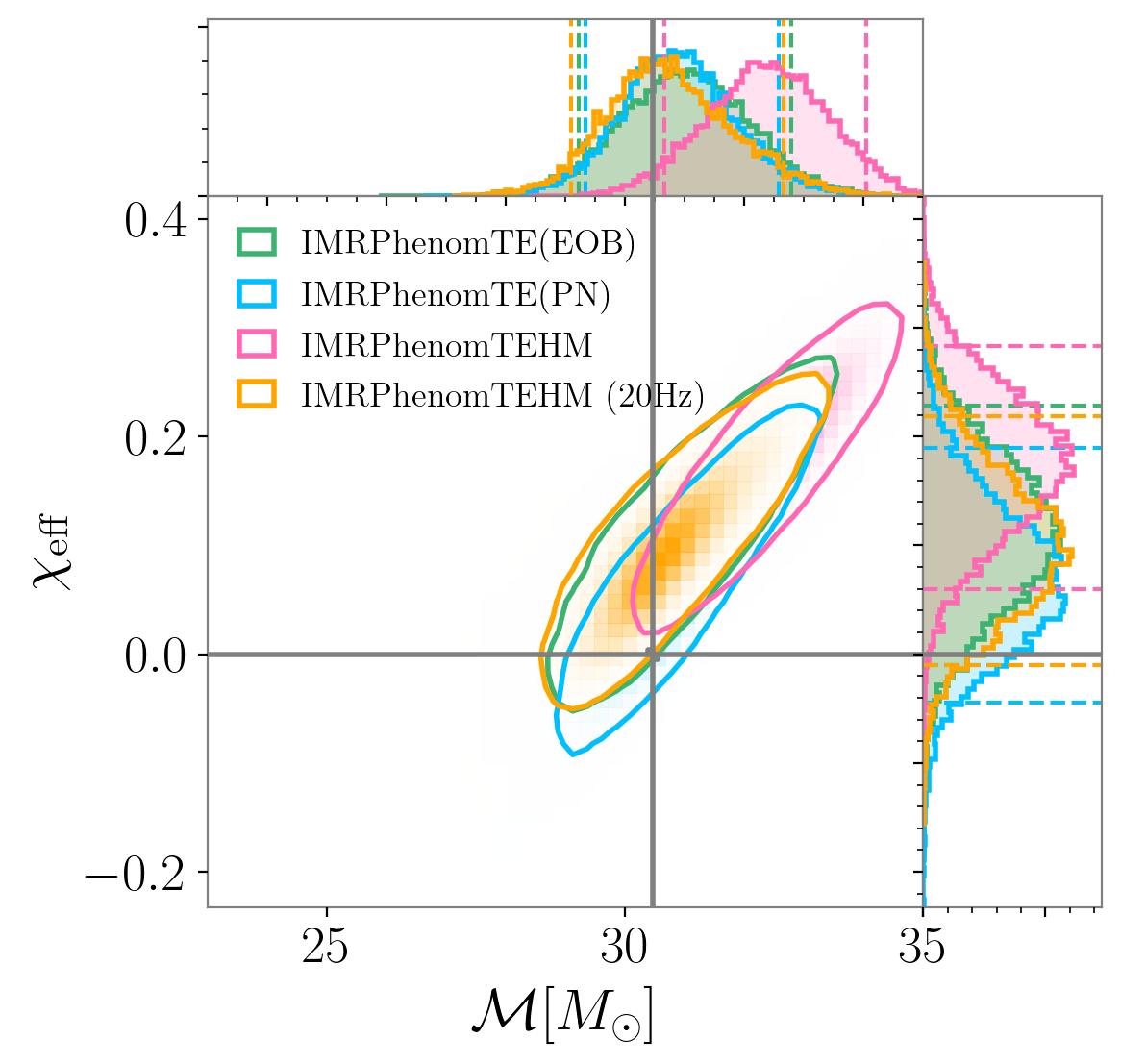}
    \includegraphics[width=0.32\linewidth]{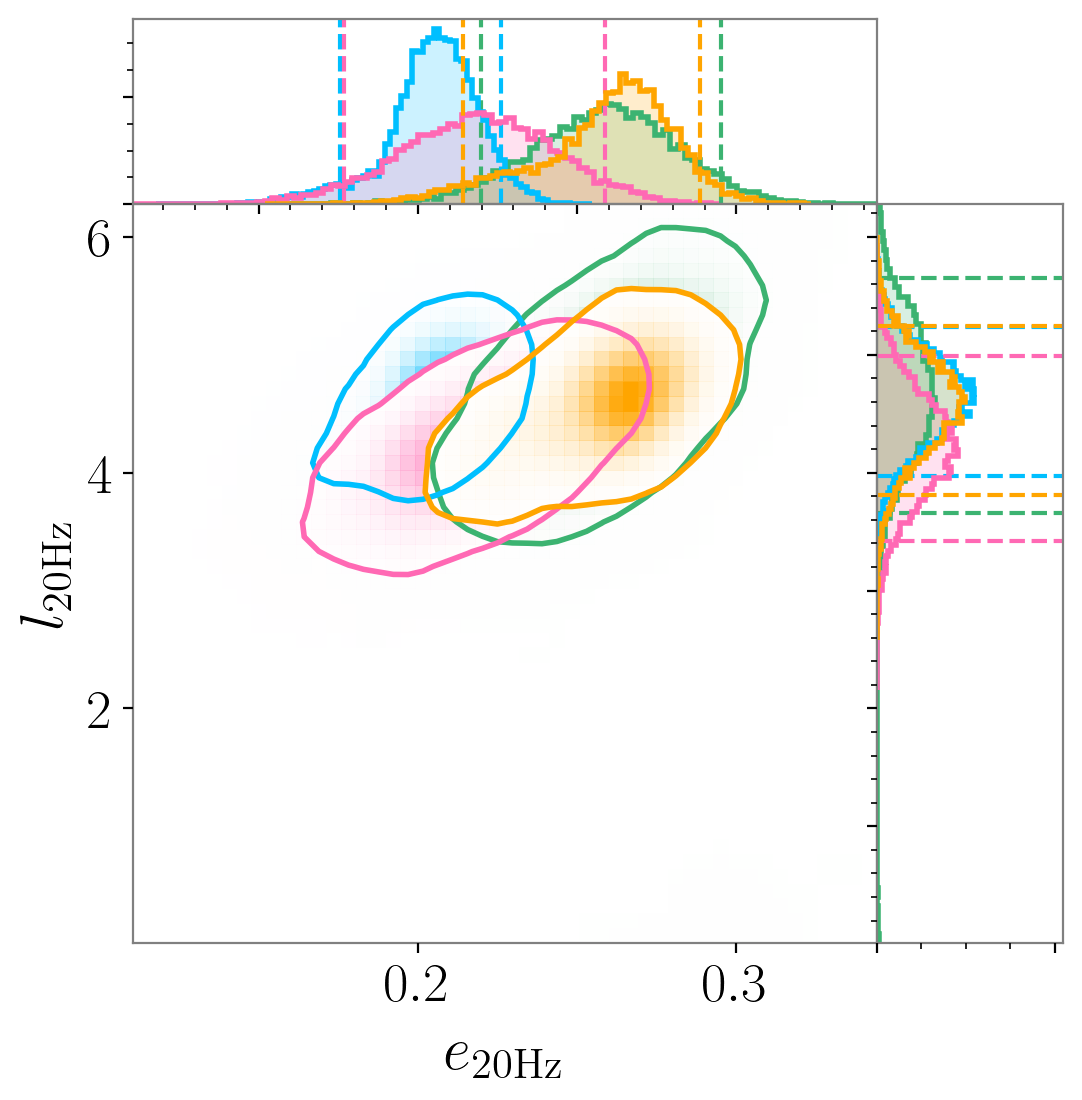}
    \includegraphics[width=0.32\linewidth]{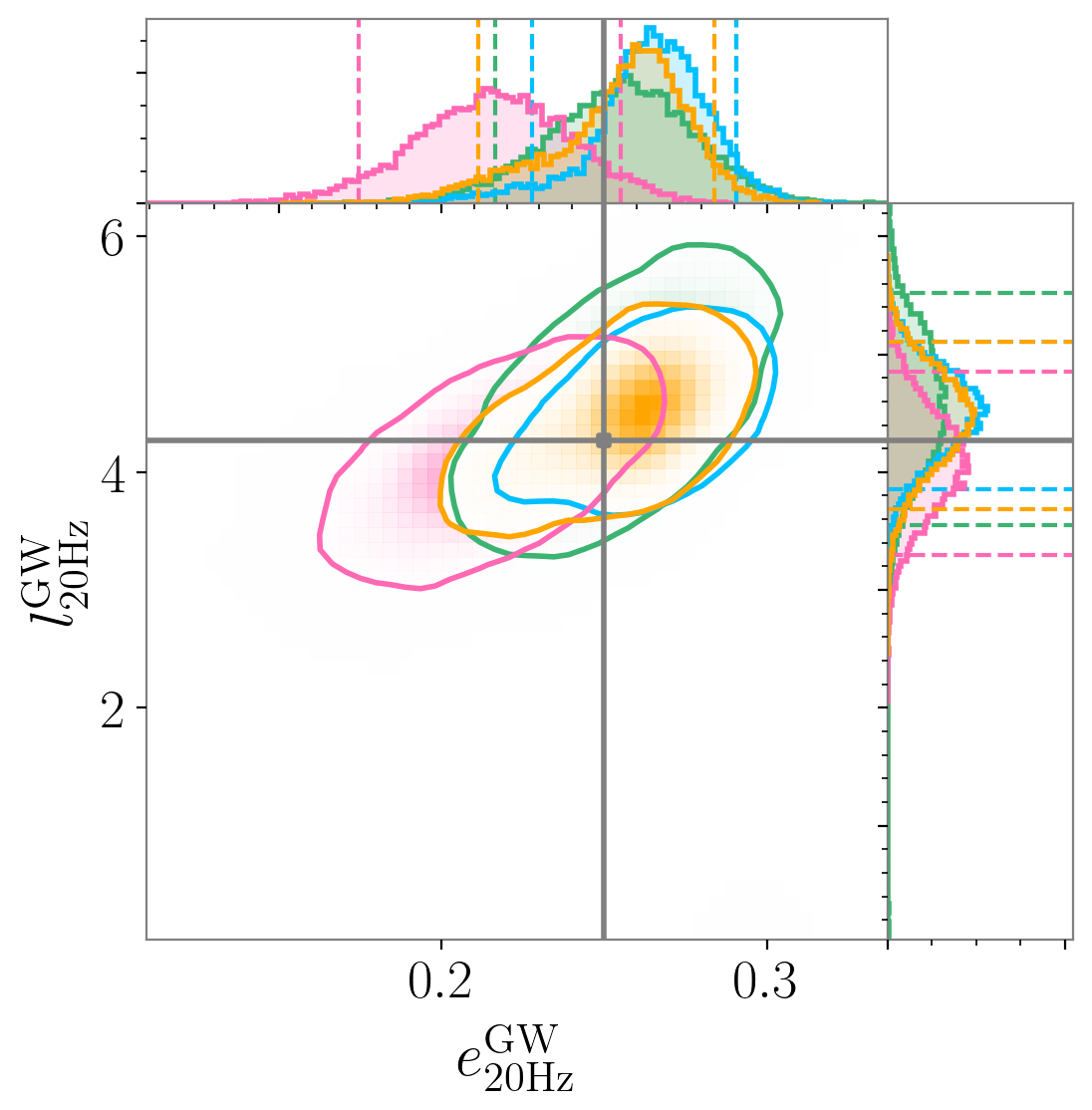}
    \end{minipage}
    \caption{Posterior distributions for three NR injections, with each row corresponding to one simulation. The figure shows marginalized 2D and 1D posterior distributions for key parameters: chirp mass $\mathcal{M}$ and effective spin $\chi_{\mathrm{eff}}$ (\textit{first column}), reference eccentricity $e_{20\mathrm{Hz}}$ and reference mean anomaly $l_{20\mathrm{Hz}}$ (\textit{second column}), and GW eccentricity $e_{\mathrm{GW}}$ and mean anomaly $l_{\mathrm{GW}}$ (\textit{third column}). The injected values are indicated by gray lines. Finally, we also include the log-10 Bayes factor between the EOB (EOB) and the PN (PN) hypothesis, $\log_{10}B_{\mathrm{EOB}/\mathrm{PN}}$All parameters are measured at the reference frequency $f_{\mathrm{ref}}=20$ Hz.}
    \label{fig:NRinjections}
\end{figure*}

Fig.~\ref{fig:NRinjections} displays the posterior distributions for each NR injection across the four analysis cases, showing maginalized 1D and 2D posteriors for the chirp mass $\mathcal{M}$ and the effective spin $\chi_{\mathrm{eff}}$, and the reference eccentricity $e_{20\mathrm{Hz}}$ and mean anomaly $l_{20\mathrm{Hz}}$. Additionally, we include the GW eccentricity and mean anomaly computed using the \texttt{gw\_eccentricity} package, also evaluated at the reference frequency. Table~\ref{tab:NRresults} provides a summary of the injected intrinsic parameters alongside the recovered median values and the 90\% credible intervals for each run.
The results in Fig.~\ref{fig:NRinjections} and Table~\ref{tab:NRresults}  demonstrate the ability of the \phTE model to accurately recover the binary parameters. Moreover, all posterior distributions shown in Fig.~\ref{fig:NRinjections} are Gaussian and unimodal across all NR injections, even for the highest eccentricity case, \texttt{SXS:BBH:1363}. This further supports the accuracy of the model in capturing the eccentric dynamics.
A key observation, consistent with the \seobe injection studies in Refs.~\cite{Ramos-Buades:2023yhy,Gamboa:2024hli}, is that when only using the dominant harmonics, small biases appear in the recovery of the luminosity distance $d_L$ and inclination angle $\iota$. 
Similarly, we find a significantly improved recovery of $d_L$ when HMs are included, as seen in Tab.~\ref{tab:NRresults}. However, their inclusion does not lead to a comparable improvement in the recovery of the inclination angle.

A more detailed analysis is needed to understand the bias observed for \texttt{SXS:BBH:1363} when including HMs, as seen in the bottom row of Fig.~\ref{fig:NRinjections}. We noticed a clear shift in the QC posteriors when including HMs, as shown by the pink curve (HMs results are only shown for the default version in this case). However, this difference stems from the choice of starting frequency for the signal and the template. When including HMs, the starting frequency is lowered to 10 Hz to ensure all modes remained in band. In contrast, when keeping the same minimum frequency as in the analysis with only the $(2,2)$ harmonic, we found equivalent posteriors, as indicated by the orange curve.
This is explained because of the limited length of the NR waveform used to construct the mock signal injection. This NR waveform is generated at a starting frequency of 20Hz, as it cannot be generated at 10Hz. Consequently, the HMs of the signal enter the frequency band during the analysis, which explains part of the biases of the \phTE run starting at 10Hz.
Additionally, this NR simulation starts with a considerable initial eccentricity ($e_{0}^{\text{GW}}=0.317$), a regime where the eccentricity expansions used in the model construction degrade its accuracy (see Fig.~\ref{fig:SXSmismatches}). 
Thus, we observe a small bias in the recovered effective spin parameter, although the injected value lies in the 90\% credible interval, which may also be related to the length issues of the signal.

The right panels of Fig.~\ref{fig:NRinjections} display the posterior distributions for the GW eccentricity $e^{\text{GW}}$ and mean anomaly $l^{\text{GW}}$ at a reference frequency of 20 Hz. 
\phTE successfully recovers the injected parameters within the 90\% credible intervals for both coordinates, while the reference posteriors $e_{20\text{Hz}}$ and $l_{20\text{Hz}}$ show significant differences. Notably, the EOB posteriors align more closely to the GW ones than the PN posteriors, supporting our choice of EOB as the default coordinate system. To further justify this selection, we compute the log-10 Bayes factor between the EOB (EOB) and the PN (PN) hypothesis, $\log_{10}B_{\mathrm{EOB}/\mathrm{PN}}$, shown in Tab.~\ref{tab:NRresults}. While the values are small, they consistently favor EOB coordinates.

These studies demonstrate that our model can recover the injected waveform parameters with a performance comparable to the \seobe model, despite the higher mismatches. This suggests that mismatch calculations alone may not fully capture the model's ability to recover the injected waveforms. Furthermore, the significantly faster performance of \phTE enabled us to explore the impact of various effects on these PE runs. We leave for future work to expand the injection study in this section to further investigate the accuracy and the role of HMs and gauge choices in the recovery of eccentric signals.

\subsection{Analysis of real GW events}\label{subsec:GWevents}
We present an analysis of the GW events GW150914~\cite{GW150914, LIGOScientific:2016vlm}, as the first GW signal from a BBH, and GW190521~\cite{GW190521}, as a possible eccentric candidate, using the \phTE model. 
The strain data for both events were obtained from the Gravitational Wave Open Source Catalog (GWOSC)~\cite{gwosc12, gwosc3}, along with the publicly available PSDs and calibration envelopes from the Gravitational Wave Transient Catalog GWTC-2.1~\cite{gwtc21}. We conducted six different PE runs, as detailed in Tab.~\ref{tab:pesummary}, using \phTE but also \phTHM~\cite{Estelles:2020osj,Estelles:2020twz} via the \texttt{phenomxpy}~\cite{phenomxpy} infrastructure for comparison.
Results for both events are summarized in Tab.~\ref{tab:GW_pes} and Fig.~\ref{fig:GWevents}, and we now proceed to discuss each case in more detail in the following subsections. For comparison, we also include in the table the results obtained using \seobe from Ref.~\cite{Gamboa:2024hli}.

\begin{table*}%[h!]
\centering
\begin{tabular}{@{}llccccccccccc@{}}
\toprule
\textbf{Event} & \textbf{Model} & $M/M_\odot$ & $\mathcal{M}/M_\odot$ & $1/q$ & $\chi_\text{eff}$ & $e^\text{GW}$ & $l^\text{GW}$ & $d_L$ & $\text{SNR}^{\text{N}}$ & $\log_{10}B_{\mathrm{E}/\mathrm{QC}}$ \\ 
\midrule
\multirow{4}{*}{GW150914} 
& \phTHM  & $71.45^{+3.00}_{-3.05}$ & $30.97^{+1.31}_{-1.34}$ & $0.89^{+0.10}_{-0.17}$ & $-0.03^{+0.10}_{-0.11}$ & - & - & $453^{+145}_{-158}$ & $25.11^{+0.08}_{-0.13}$ & - \\
& \phTE (EOB)  & $71.16^{+3.15}_{-3.01}$ & $30.86^{+1.37}_{-1.33}$ & $0.90^{+0.09}_{-0.16}$ & $-0.03^{+0.10}_{-0.11}$ & $0.03^{+0.05}_{-0.03}$ & $3.92^{+2.05}_{-3.60}$ & $446^{+144}_{-159}$ & $25.13^{+0.11}_{-0.15}$ & $-0.50^{+0.16}_{-0.16}$ \\ 
& \phTE (PN) & $71.22^{+3.20}_{-2.97}$ & $30.88^{+1.38}_{-1.31}$ & $0.89^{+0.10}_{-0.17}$ & $-0.03^{+0.10}_{-0.11}$ & $0.03^{+0.05}_{-0.03}$ & $3.94^{+2.04}_{-3.59}$ & $452^{+143}_{-155}$ & $25.13^{+0.10}_{-0.14}$ &  $-0.68^{+0.16}_{-0.16}$\\

& \seobe  & $70.9^{+2.62}_{-2.8}$ & $30.72^{+1.15}_{-1.24}$ & $0.88^{+0.09}_{-0.14}$ & $-0.05^{+0.09}_{-0.05}$ & $0.06^{+0.07}_{-0.05}$ & $3.17^{+2.49}_{-2.54}$ & $480^{+116}_{-125}$ & -&$-0.57^{+0.13}_{-0.13}$ \\ 
\midrule
\multirow{4}{*}{GW190521} 

& \phTHM & $255.4^{+30.2}_{-33.1}$ & $109.3^{+14.0}_{-18.1}$ & $0.74^{+0.23}_{-0.27}$ & $0.01^{+0.31}_{-0.39}$ & - & - & $4127^{+1662}_{-1877}$ & $14.37^{+0.21}_{-0.26}$ & - \\ 

& \phTE  & $259.1^{+26.4}_{-28.3}$ & $111.3^{+12.0}_{-15.5}$ & $0.78^{+0.20}_{-0.27}$ & $0.02^{+0.30}_{-0.34}$ & $0.31^{+0.13}_{-0.28}$ & $3.18^{+2.82}_{-2.88}$ & $4275^{+1490}_{-1732}$ & $14.44^{+0.21}_{-0.30}$ & $0.12^{+0.13}_{-0.13}$ \\

& \phTE ($r<10M)$  & $258.5^{+26.8}_{-28.6}$ & $111.1^{+12.2}_{-14.8}$ & $0.78^{+0.19}_{-0.26}$ & $0.01^{+0.30}_{-0.36}$ & $0.29^{+0.14}_{-0.26}$ & $3.02^{+2.42}_{-2.35}$ & $4284^{+1516}_{-1782}$ & $14.42^{+0.23}_{-0.28}$ & $0.28^{+0.13}_{-0.13}$  \\

& \seobe  & $260.7^{+20.3}_{-19.5}$ & $111.4^{+9.7}_{-11.7}$ & $0.73^{+0.20}_{-0.21}$ & $0.05^{+0.20}_{-0.20}$ & $0.29^{+0.16}_{-0.23}$ & $3.14^{+2.54}_{-2.52}$ & $4786^{+1261}_{-1230}$ & - & $-0.36^{+0.11}_{-0.11}$ \\ 
\bottomrule
\end{tabular}
\caption{Median values and 90\% credible intervals for the posterior distributions shown in Fig.~\ref{fig:GWevents} for the two analyzed GW events (indicated in each row), recovered with \phTE using different settings specified in the \textbf{Model} column and Tab.~\ref{tab:pesummary}, as well as with the QC aligned-spin model \phTHM through the \texttt{phenomxpy} infrastructure. For comparison, we also include results obtained using \seobe from Tab~IV, rows 2 and 7 from Ref.~\cite{Gamboa:2024hli}.
The table reports the same parameters as Tab.~\ref{tab:NRresults}, as well as the log-10 Bayes factor between the eccentric (E) and the quasi-circular (QC) hypothesis $\log_{10}B_{\mathrm{E}/\mathrm{QC}}$. All values are given at the reference frequency of 20 Hz for GW150914 and 5.5 Hz for GW190521.}
\label{tab:GW_pes}
\end{table*}

\subsubsection{GW150914}\label{subsec:GW150914}
GW150914 is the first detected GW signal from a BBH coalescence, consistent with a QC, non-spinning, and comparable-mass binary system~\cite{GW150914, LIGOScientific:2016vlm}. With a SNR of approximately 23.7, it remains one of the highest-SNR events observed during the first three observing runs of the LVK Collaboration~\cite{gwtc1, gwtc2,gwosc12, gwtc3}.
For our analysis, we adopt the priors introduced in Sec.~\ref{sec:PE}, with boundaries $1/q\in[0.05,1]$, $\chirpMass\in[20,50]$, and $e_{20\mathrm{Hz}}\in[0,0.3]$, $l_{20\mathrm{Hz}}\in[0,2\pi]$, all defined at a reference frequency of $f_{\mathrm{ref}}=20$ Hz. The specific run settings are detailed in Tab.~\ref{tab:pesummary}.

\begin{figure*}
    \centering
    GW150914
    \begin{minipage}{\textwidth}
    \includegraphics[width=0.32\linewidth]{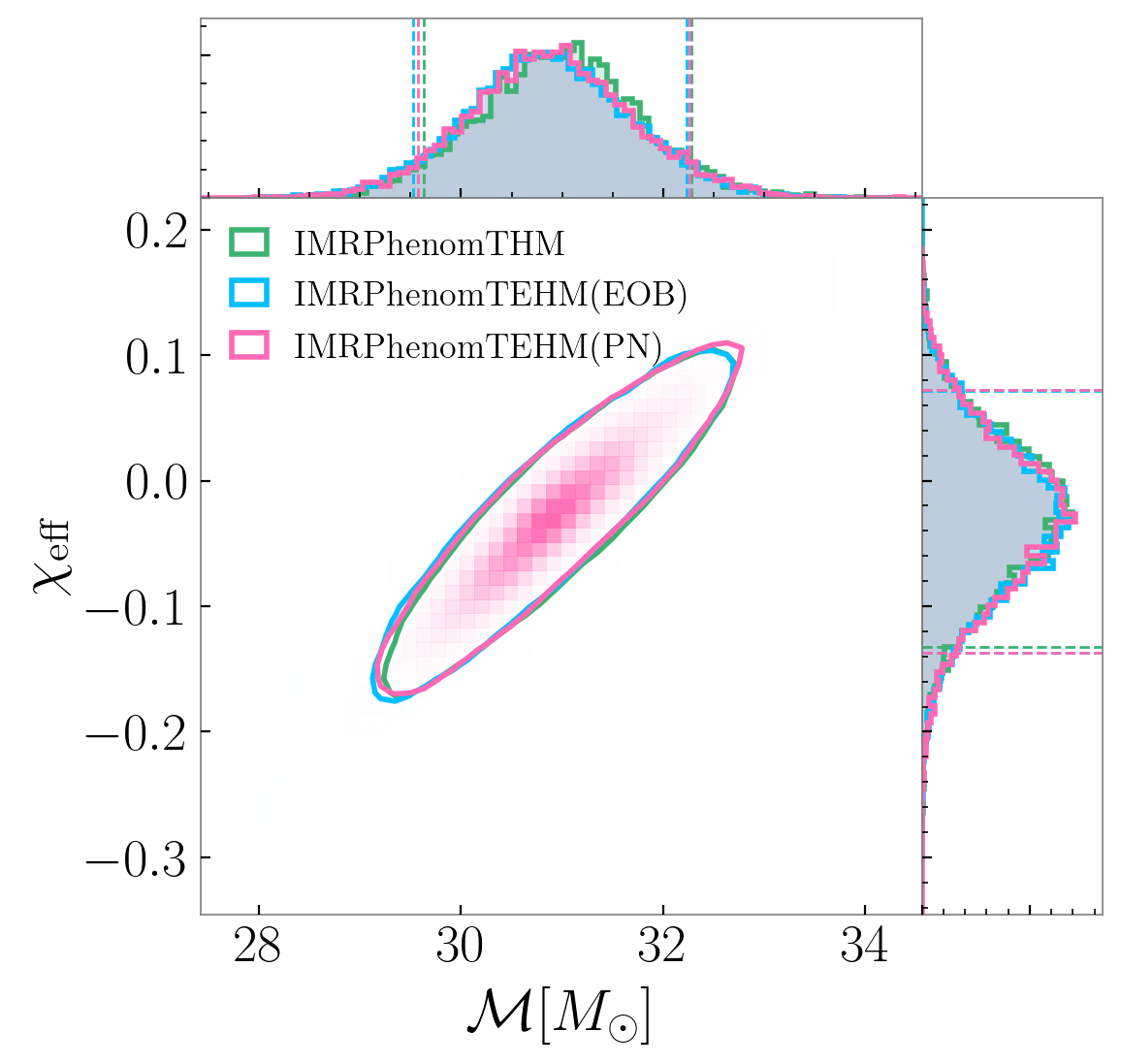}
    \includegraphics[width=0.3\linewidth]{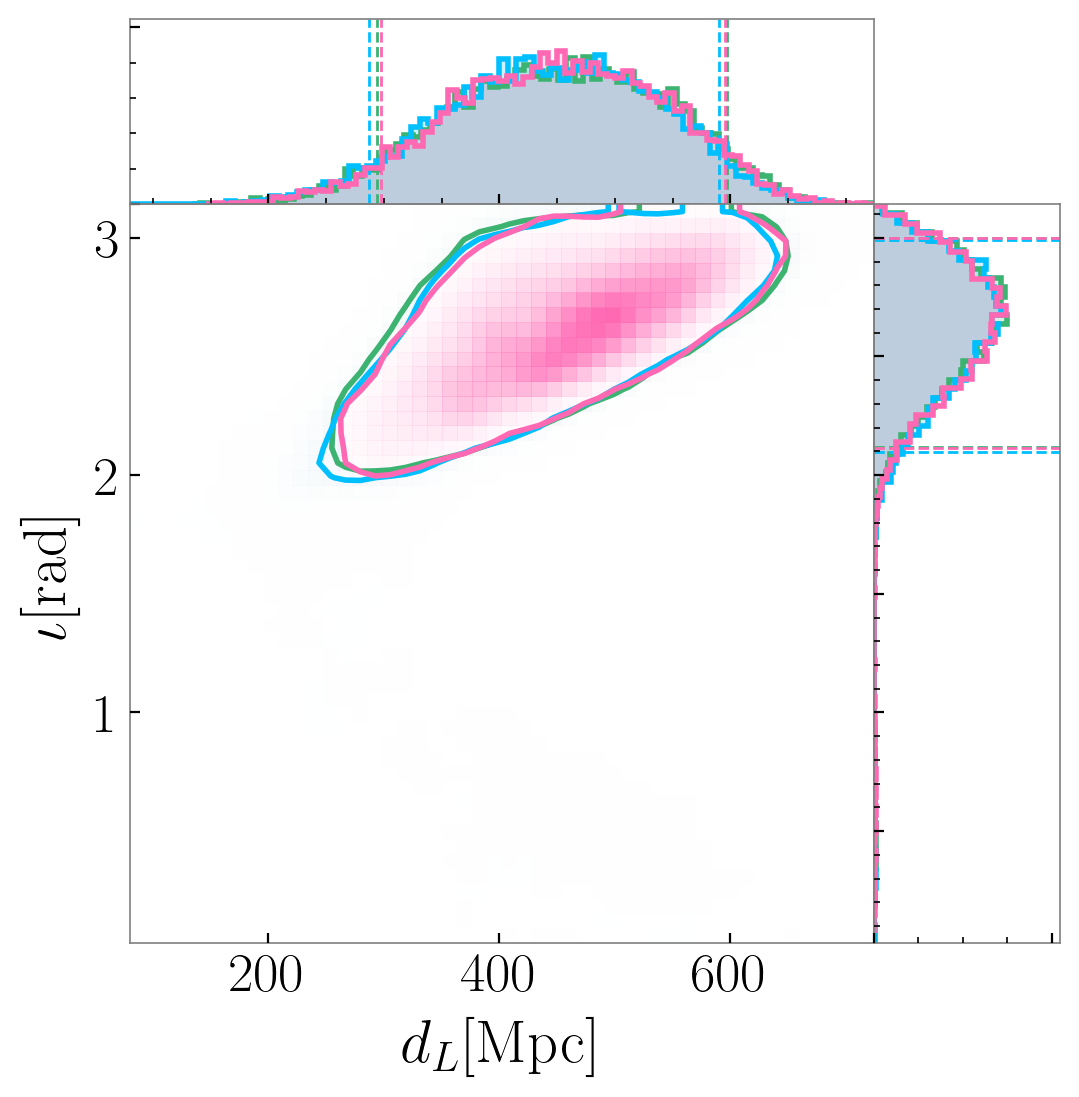}
    \includegraphics[width=0.3\linewidth]{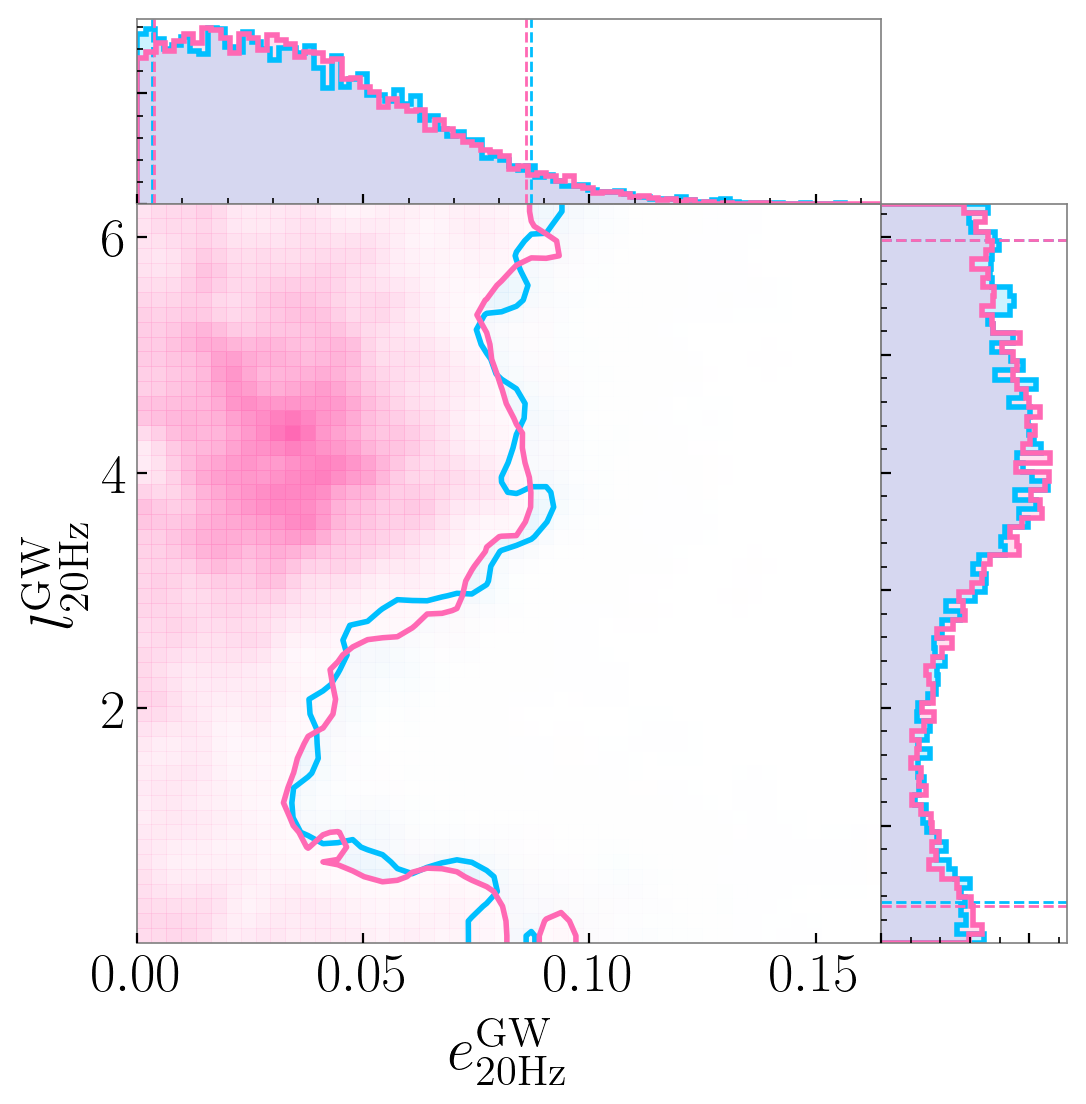}
    \vspace*{0.7em}
    \end{minipage}
    GW190521
    \begin{minipage}{\textwidth}
    \includegraphics[width=0.32\linewidth]{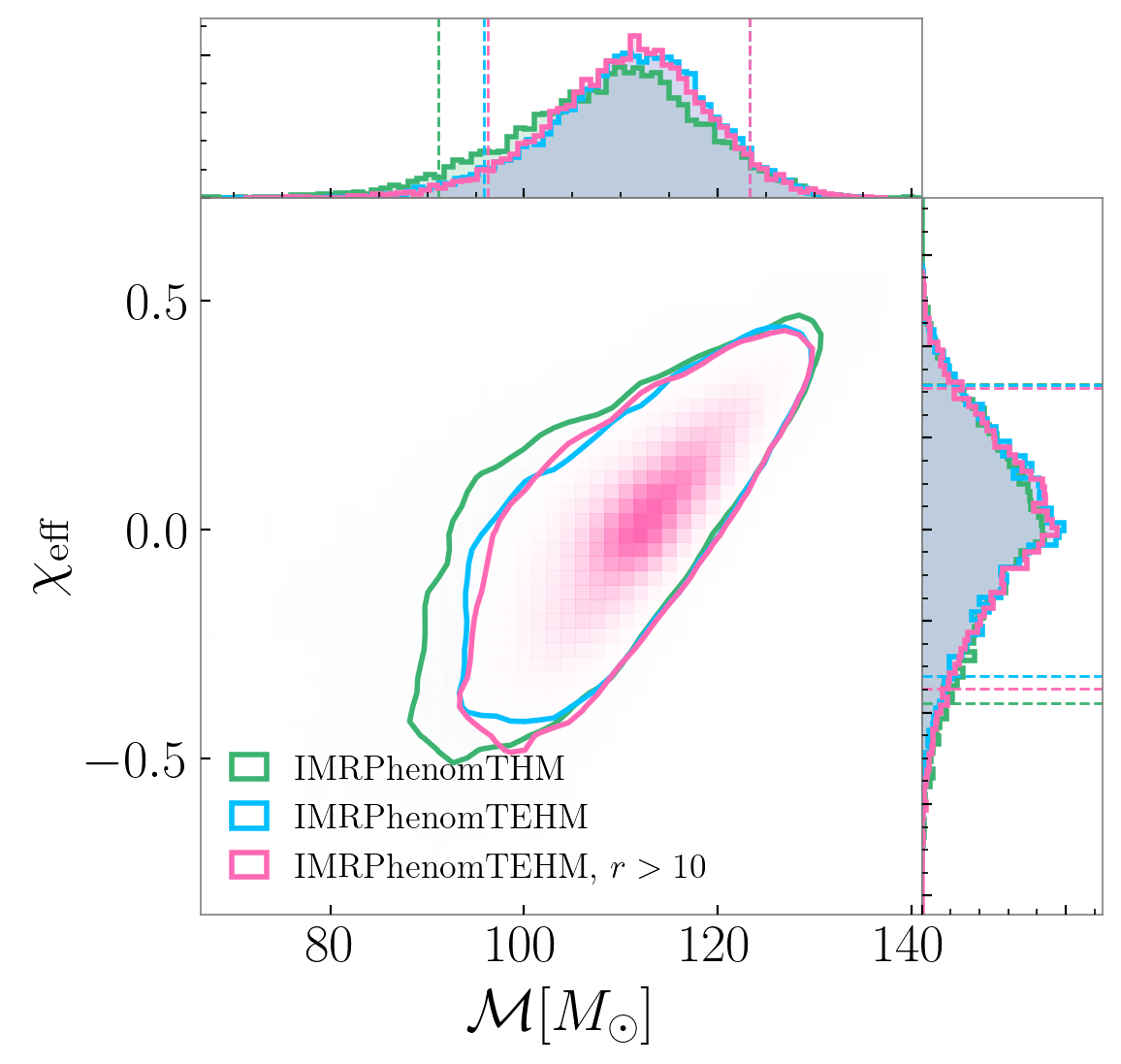}
    \includegraphics[width=0.3\linewidth]{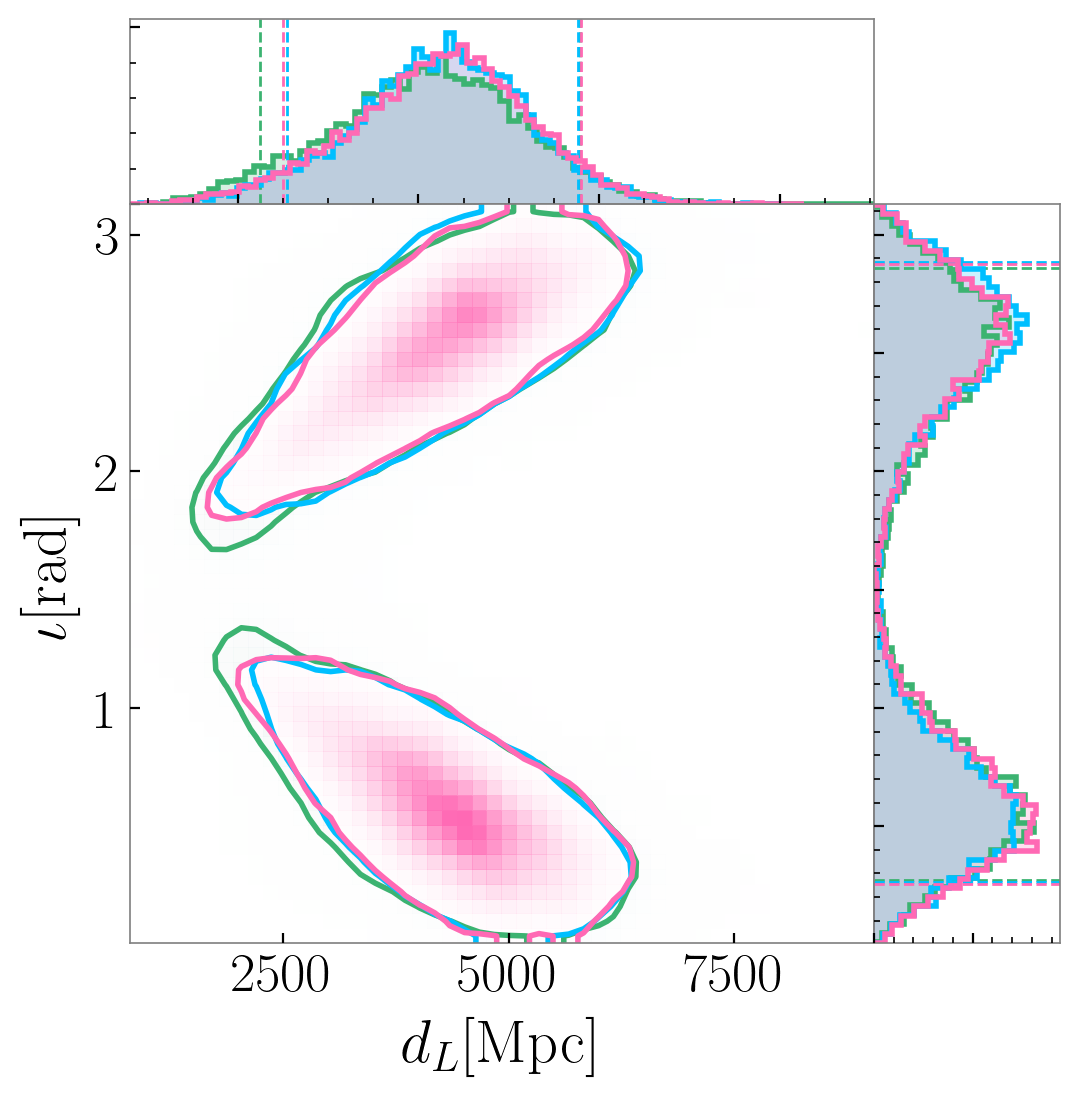}
    \includegraphics[width=0.3\linewidth]{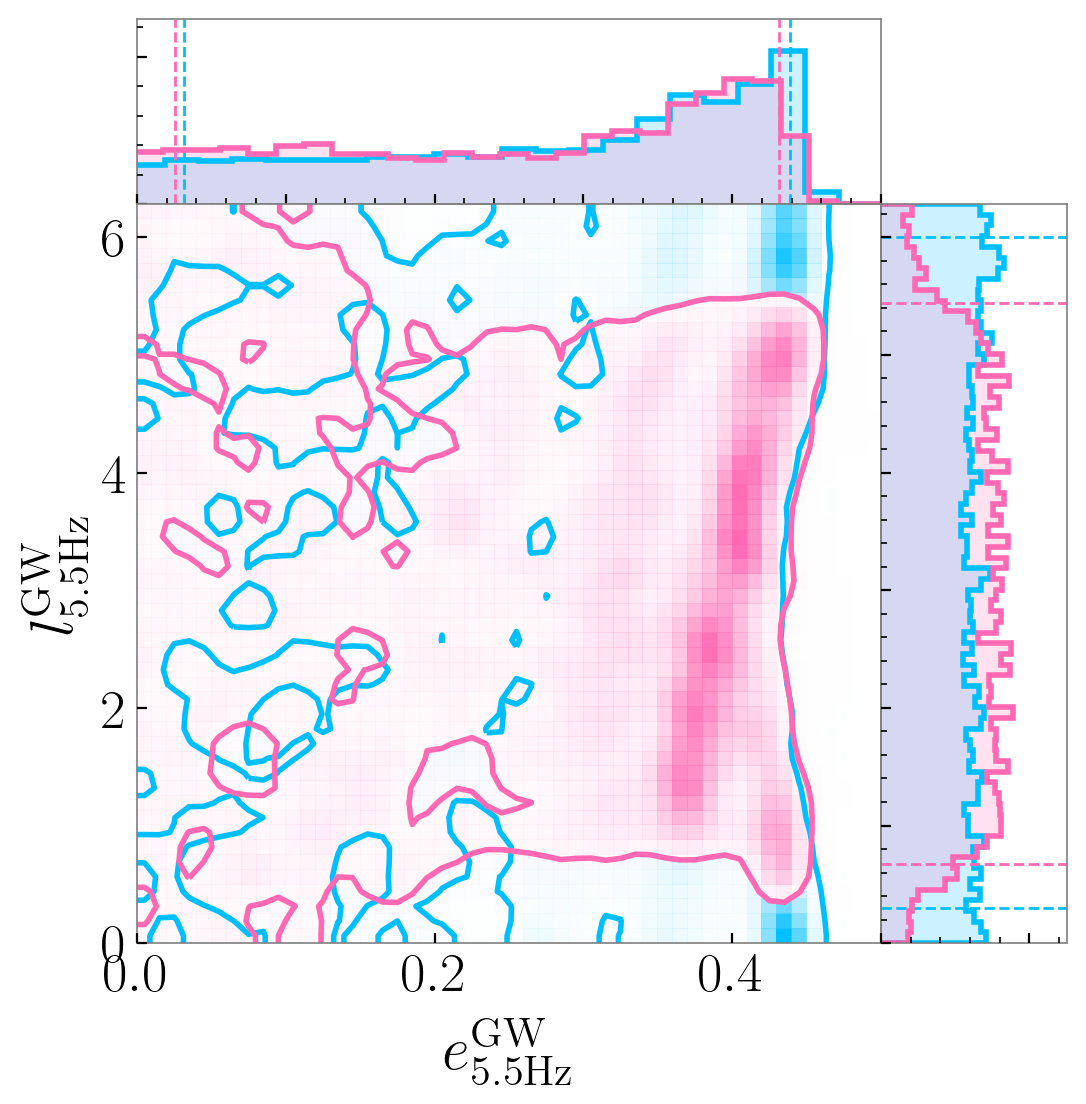}
    \end{minipage}
    \caption{Posterior distributions for the parameter estimation studies of two real GW events, GW150914 (\textit{top row}) and GW190521 (\textit{bottom row}). The figure presents the posterior distributions of chirp mass and effective spin (\textit{first column}), inclination angle and luminosity distance $d_L$ (\textit{second column}), and GW eccentricity and mean anomaly (\textit{third column}). All parameters are measured at the reference frequency, which is $f_{\mathrm{ref}}=20$ Hz for GW150914 and $f_{\mathrm{ref}}=5.5$ Hz for GW190521.}
    \label{fig:GWevents}
\end{figure*}

The top rows of Fig.~\ref{fig:GWevents} and Tab.~\ref{tab:GW_pes} present the results for GW150914 in the same format as those for the NR injections presented in Sec.~\ref{subsec:NRinjections}. The table provides the median values along with the 90\% credible intervals, while the figure display the marginalized 1D and 2D posterior distributions for the effective spin $\chi_{\mathrm{eff}}$, chirp mass $\chirpMass$, and GW eccentricity and mean anomaly $e^{\mathrm{GW}}$, $l^{\mathrm{GW}}$, respectively.

Our results obtained with \phTE show exceptional agreement with results in the literature for this event~\cite{GW150914,LIGOScientific:2016vlm,  gwtc1}: the posteriors support a non-eccentric, non-spinning, comparable-mass BBH, independently of the gauge chosen for the secular evolution equations. Specifically, the recovered chirp mass $\chirpMass$ and effective spin $\chieff$ parameters align closely with the posteriors obtained using the \phTHM QC aligned-spin model, as seen in the upper left plot in Fig.~\ref{fig:GWevents}. Regarding the eccentricity posterior, we obtain a median value of $e^{\text{GW}}_{20\text{Hz}}=0.03^{+0.05}_{-0.03}$, proving support for zero eccentricity, as shown in the second and third columns of Fig.~\ref{fig:GWevents} and in Tab.~\ref{tab:GW_pes}. Additionally, we observe support for any mean anomaly value - although the posterior is not completely flat.

We also compute the log-10 Bayes factor between the eccentric (E) and the quasi-circular (QC) hypothesis, $\log_{10}B_{\mathrm{E}/\mathrm{QC}}$, for each eccentric run, comparing to those obtained with \phTHM. The values are listed in Tab.~\ref{tab:GW_pes}. These results indicate a slight preference for the QC hypothesis, consistent with the findings reported in Refs.~\cite{Ramos-Buades:2023yhy,Gamboa:2024hli} for the SEOBNRv4EHM and \seobe models. Additionally, the LVK analysis of this event~\cite{LIGOScientific:2016vlm} found no evidence of residual eccentricity.

\subsubsection{GW190521}
GW190521 is a particularly intriguing GW event that challenges our understanding of astrophysical BH formation channels, the accuracy of waveform models, and data analysis techniques. The detected signal is an exceptionally short transient, lasting approximately 0.1 seconds and spanning only about four cycles in the 30-80 Hz frequency band. The source was initially identified as a BBH merger with a total mass of approximately 150$M_{\odot}$ in the source frame~\cite{GW190521, Estelles:2021jnz, Romero-Shaw:2020thy}. 
Due to the short signal duration, there exist strong degeneracies between source parameters, hindering a confident identification of the nature of the binary. Consequently, a wide range of alternative interpretations have been explored in the literature, including a hyperbolic collision~\cite{Gamba2023}, a boson star merger~\cite{GW190521_boson_star}, a high-mass black hole-disk system~\cite{Shibata_2021}, the first instance of an intermediate mass-ratio inspiral~\cite{Nitz_2021}, and an eccentric binary merger~\cite{Romero-Shaw_2020}, though other recent studies find no clear evidence of eccentricity~\cite{Ramos-Buades:2023yhy,Gamboa:2024hli}.
As in the case of GW150914, we adopt the priors introduced in Sec.~\ref{sec:PE}, with boundaries $1/q\in[0.17,1]$, $\chirpMass\in[60,200]$, and $e_{5.5\mathrm{Hz}}\in[0,0.5]$, $l_{5.5\mathrm{Hz}}\in[0,2\pi]$, all defined at the reference frequency $f_{\mathrm{ref}}=5.5$ Hz. The specific settings for all the runs performed are detailed in Tab.~\ref{tab:pesummary}.

The bottom rows of Fig. \ref{fig:GWevents} and Tab. \ref{tab:GW_pes} show the results for GW190521.
The recovered chirp mass $\chirpMass$ and effective spin $\chieff$ parameters lie within the posteriors obtained using the \phTHM model, as shown in the bottom left plot in Fig.~\ref{fig:GWevents}. These results are anticipated, since the main difference in the QC parameters arises when including precessing spin effects, as seen for instance in Refs.~\cite{GW190521,  Estelles:2021jnz, Gamboa:2024hli}.
Regarding the eccentricity posterior, we find mostly uninformative posteriors. While the median values in Tab.~\ref{tab:GW_pes} suggest high eccentricities, this simply reflects the properties of the given prior.
Similarly, the mean anomaly posteriors are uninformative and consistent with a uniform prior.
It is worth noting that the in the bottom right panel in Fig.~\ref{fig:GWevents}, $e^{\rm GW}$ posteriors do not reach the prior boundary, which arises because $e^{\rm GW}$ is generally slightly lower than $e^{\rm EOB}$. We leave for future work to explore the impact of directly using prior on $e^{\rm GW}$ to ensure consistency.
These results are expected due to the quasi-circularity condition imposed in the merger-ringdown of \phTE. This implies that any waveform with any eccentricity will exhibit a QC merger-ringdown phase, thus implying support for any case. Therefore, accurately measuring eccentricity in high-mass systems, where only the late part of the evolution falls within the frequency band of current detectors, requires incorporating eccentricity effects in that part of the waveform, which would likely involve calibration to NR. We plan to address these improvements in future work.
We also compute the log-10 Bayes factor between the eccentric (E) and the quasi-circular (QC) hypothesis, $\log_{10}B_{\mathrm{E}/\mathrm{QC}}$, for each eccentric run, comparing to those obtained with \phTHM. The values are listed in Tab.~\ref{tab:GW_pes}. 
Even though these results indicate a slight preference for the eccentric hypothesis, the uncertainties in the default option allow support for the QC hypothesis as well. This aligns with previous findings using eccentric EOB models~\cite{Ramos-Buades:2023yhy,Gamboa:2024hli,Gupte:2024jfe}, where no evidence for eccentricity can be claimed for this event.

Finally, our posteriors on eccentricity differ from those obtained with the SEOBNRv4EHM and \seobe models, as shown in Refs.~\cite{Ramos-Buades:2023yhy,Gamboa:2024hli}. 
This discrepancy arises because \phTE allows the starting frequency to be different than the reference frequency.
In contrast, the SEOBNRv4EHM and \seobe models are constrained to the have the same starting frequency and reference frequency ( $f_{\mathrm{ref}}=f_{\min}$), and for sufficiently high starting frequency they do not generate waveforms, see Refs. \cite{Ramos-Buades:2021adz,Gamboa:2024hli} for details.
This artificial cut-off in waveform generation affects the priors, preventing uniform support across the full eccentricity range—especially restricting high eccentricities at high masses.
To illustrate this, we impose a similar condition as in the \seobe model by requiring an initial separation $r>10M$, using the 3PN expansion from Ref.~\cite{Henry:2023tka}. The resulting posteriors, shown in pink in the bottom row of Fig.~\ref{fig:GWevents}, display almost no support for eccentricities $\gtrapprox  0.3$ with mean anomaly near 0, i.e., periastron passage. This is expected, as highly eccentric systems starting near periastron have shorter waveforms than those starting near apastron. While our model does not exhibit the same posterior cuts as \seobe due to differences in dynamics, this test confirms that truncating waveform generation can artificially restrict priors, a limitation we avoid by extending the waveform duration.
We consider generating waveforms across the full parameter space a more realistic approach. However, extra caution needs to be taken into account when generating waveforms with initial eccentricities >0.5 at high frequencies due to the underlying PN prescription of eccentric effects and the tapering applied at merger to enforce a quasicircular merger ringdown.

A key aspect of the new \phTE model is computational efficiency. The model is only 3 to 5 times slower than \phTHM while offering a significant speedup compared to other eccentric models (see e.g.~ \cite{Ramos-Buades:2021adz,Nagar_2024,Gamboa:2024hli}). 
The significant reduction in wall time makes \phTE a highly competitive model for parameter estimation studies. This efficiency sets \phTE as a strong candidate to become a standard tool for parameter estimation in current and future GW detectors.

\section{Conclusions}\label{sec:conclusions}

In this work, we presented \phTE, a time-domain, multipolar waveform model designed for aligned-spin BBHs in elliptical orbits. \phTE extends the time-domain phenomenological PhenomT waveform family~\cite{Estelles:2020osj, Estelles:2020twz, Estelles:2021gvs} by incorporating 3PN eccentric corrections for the eccentric dynamics into the state-of-the-art quasicircular aligned-spin model \phTHM. 
The model includes the waveform modes $(l,|m|)=\{(2,2), (2,1), (3,3), (4,4), (5,5)\}$, with PN corrections expanded in eccentricity up to $\mathcal{O}(e^6)$. 
\phTE thus provides a reliable description of eccentric binaries with mass ratios $q\in[1,20]$, aligned spins $\chi_i\in[-0.995,0.995]$, and eccentricities up to $e = 0.4$ at 10 Hz.
By default, we solve the secular evolution equations in EOB coordinates, but the model also supports modified-harmonic coordinates. Additionally, it allows for the inclusion of up to 3PN spinning eccentric effects in the GW modes~\cite{Henry:2023tka,Henry2025}.

The construction of the \phTE model is outlined in Sec.~\ref{sec:construction}, with a summary provided in Sec.~\ref{subsec:summary_construction}. The underlying QC model, \phTHM, which serves as the foundation for the eccentric extension, is summarized in Sec.~\ref{subsec:THM}, while Sec.~\ref{subsec:PN} describes the incorporation of the eccentric corrections in both the QC dynamics and waveform modes. 
We assess \phTE's accuracy in the QC limit by computing mismatches against \phTHM in Sec.~\ref{subsec:nonecclimit}, achieving values below $10^{-5}$ across the entire QC parameter space - a crucial characteristic to avoid biases in parameter estimation studies due to an inaccurate implementation of the quasicircular limit~\cite{Bonino:2022hkj, Ramos-Buades:2023yhy}. 
In Sec.~\ref{subsec:SXS} we compute mismatches against 28 publicly available SXS eccentric simulations, and compare to other state-of-the-art eccentric models, \teobdali and \seobe. 
We find mismatches below $2\%$ for most cases, with higher values than those for the \seobe and \teobdali models, yet sufficiently low to accurately recover posterior distributions in Bayesian inference studies, as confirmed in Sec.~\ref{sec:PE}.
To further investigate the robustness and accuracy of the \phTE model, we computed mismatches against \seobe in Sec.~\ref{subsec:robustness}. 
Our results indicate that higher mismatches occur in regions of parameter space with limited coverage by NR waveforms, thus dominated by waveform systematic errors, whereas for mass ratios below 10 and eccentricities up to 0.4 at 10 Hz, mismatches remain below ~5\%. 
Furthermore, in Sec.~\ref{subsec:benchmarks}, we estimate the model’s computational efficiency, finding that \phTE is the fastest time-domain eccentric IMR waveform model to date compared to \seobe and \teobdali.
Section~\ref{sec:PE} demonstrates the model’s applicability in PE studies. In Sec.~\ref{subsec:NRinjections}, we injected three NR eccentric simulations, successfully recovering both the quasicircular and eccentric injected parameters within their 90\% credible intervals. Finally, in Sec.~\ref{subsec:GWevents}, we reanalyzed the GW events GW150914 and GW190521, obtaining results consistent with the literature, particularly those from \seobe~\cite{Gamboa:2024hli}.
 
An important feature of \phTE is the ability to set a reference frequency different from the minimum frequency, significantly enhancing its practical applications. 
This ensures that one can freely choose the starting frequency for each physical configuration, preventing waveform generation errors that arise when initializing the PN/EOB evolution at small separations. Additionally, extending the waveform duration can also improve the conditioning routines on time-domain signals for Fourier transforming the waveforms. Consequently, this prevents the need for artificial cuts in priors for PE studies, as demonstrated for the high-mass short-duration event, GW190521, and providing consistent results with Refs.~\cite{Ramos-Buades:2023yhy,Gamboa:2024hli}.

A key advantage of \phTE is its computational efficiency, making it highly competitive for parameter estimation studies. We find that the model is only a few times slower than \phTHM, while showing a significant improvement in speed compared to the eccentric \seobe or \teobdali models. The reduction in wall time (see e.g. Fig.~\ref{fig:bench}) makes \phTE a practical choice for large-scale Bayesian inference studies, positioning it as a strong candidate for standard parameter estimation pipelines in both current and future GW detector networks. Given these benchmarks and the proven accuracy of the model, we are currently undertaking a systematic re-analysis of GW events of interest reported by the LVK, which we plan to present in a follow-up paper.

Looking ahead, further improvements to \phTE are planned to enhance its accuracy and applicability. While the current accuracy is sufficient for existing detectors, refinements are necessary for next-generation observatories.
One key limitation of \phTE is the treatment of eccentric corrections in the mode amplitudes. Currently, we rely on PN expansions up to $\mathcal{O}(e^6)$, which break down at high eccentricities due to the lack of closed-form tail terms. Future work will explore non-expanded expressions, leveraging recent re-summation techniques for tail contributions \cite{Gamboa:2024imd}.
Additionally, highly eccentric waveforms at high frequencies may introduce nonphysical effects due to enforcing a quasi-circular merger-ringdown transition. Calibration to eccentric NR waveforms in this phase is necessary, particularly for high-mass and highly eccentric systems, when they have not circularized before merger.
Although \phTE can generate waveforms beyond its formal validity range without failing, careful interpretation is required in the cases described above. A detailed study of these regimes is left for future work.
Expanding the model’s applicability to include spin precession alongside eccentricity is another promising direction. A ``twisting-up'' approach~\cite{PhysRevD.84.024046,PhysRevD.84.124011, PhysRevD.86.104063, PhysRevD.91.024043, PhysRevD.85.084003}, guided by PN evolution accounting for both eccentricity and spin-precession effects, could incorporate both effects. The modular structure of phenomenological models makes this a feasible short-term goal.
These developments are crucial for next-generation GW detectors, where detection and accurate characterization of generic binaries, including spin precession and eccentricity, will provide valuable insights into the formation channels of these sources.

\section*{Acknowledgements}
The authors thank Quentin Henry for valuable discussions and for providing additional material that contributed to this work.
We also sincerely thank Gonzalo Morras for his helpful comments on the manuscript.

We thankfully acknowledge the computer resources at MareNostrum and the technical support provided by Barcelona Supercomputing Center (BSC)  through funding from the Red Española de Supercomputación (RES).
The authors are also grateful for computational resources provided by the cluster HAWK provided by Cardiff University and supported by STFC grants ST/I006285/1 and ST/V005618/1.
This research has made use of data or software obtained from the Gravitational Wave Open Science Center (gwosc.org), a service of the LIGO Scientific Collaboration, the Virgo Collaboration, and KAGRA.
This material is based upon work supported by NSF's LIGO Laboratory which is a major facility fully funded by the National Science Foundation.
LIGO is funded by the U.S. National Science Foundation. Virgo is funded by the French Centre National de Recherche Scientifique (CNRS), the Italian Istituto Nazionale della Fisica Nucleare (INFN) and the Dutch Nikhef, with contributions by Polish and Hungarian institutes.

Maria de Lluc Planas is supported by the Spanish Ministry of Universities via an FPU doctoral grant (FPU20/05577).
A. Ramos-Buades is supported by the Veni research programme which is (partly) financed by the Dutch Research Council (NWO) under the grant VI.Veni.222.396; acknowledges support from the Spanish Agencia Estatal de Investigación grant PID2022-138626NB-I00 funded by MICIU/AEI/10.13039/501100011033 and the ERDF/EU; is supported by the Spanish Ministerio de Ciencia, Innovación y Universidades (Beatriz Galindo, BG23/00056) and co-financed by UIB.
CG is supported by the Swiss National Science Foundation (SNSF) Ambizione grant PZ00P2\_223711.
This work was supported by the Universitat de les Illes Balears (UIB); the Spanish Agencia Estatal de Investigación grants PID2022-138626NB-I00, PID2019-106416GB-I00, RED2022-134204-E, RED2022-134411-T, funded by MCIN/AEI/10.13039/501100011033; the MCIN with funding from the European Union NextGenerationEU/PRTR (PRTR-C17.I1); Comunitat Autonòma de les Illes Balears through the Direcció General de Recerca, Innovació I Transformació Digital with funds from the Tourist Stay Tax Law (PDR2020/11 - ITS2017-006), the Conselleria d’Economia, Hisenda i Innovació grant numbers SINCO2022/18146 and SINCO2022/6719, co-financed by the European Union and FEDER Operational Program 2021-2027 of the Balearic Islands; the “ERDF A way of making Europe”.

% ~~~~~~~~~~ References ~~~~~~~~~~ %

% Try this if the last two columns before the bib are not breaking nicely.
%\vspace{0.1in}

%\vfil

\let\c\Originalcdefinition %
\let\d\Originalddefinition %
\let\i\Originalidefinition

\bibliography{bib_TEHM}
% ~~~~~~~~~~ END DOCUMENT ~~~~~~~~~~ %

\end{document}

%% file: UsePackages.tex
\usepackage[T3,T1]{fontenc}
\usepackage{mathrsfs} % For \mathscr fonts
\usepackage{bm} % For \bm fonts
\makeatletter
\@ifclassloaded{beamer}
  { % if this is a beamer document, use sans-serif fonts
    \typeout{UsePackages: Detected beamer}
    \usepackage{tgheros}
    
  }
  { % otherwise:
    \typeout{UsePackages: Did not detect beamer}
    \ifx\asybeamer\undefined % use times for articles
    \typeout{UsePackages: Detected article}
    \usepackage[varg]{txfonts} % For math
    \usepackage{tgtermes} % For text
    \else % use beamer fonts for asy documents with beamer
    \typeout{Fonts: Detected asy for beamer}
    \usepackage{tgheros}
    
    \fi
  }
\usepackage{microtype} % For nicer alignment of text
\def\MT@register@subst@font{
  \MT@exp@one@n\MT@in@clist\font@name\MT@font@list
  \ifMT@inlist@\else\xdef\MT@font@list{\MT@font@list\font@name,}\fi}
\makeatother

\usepackage{graphicx} %
\usepackage[x11names,svgnames,rgb]{xcolor} %
\usepackage{xspace}
\usepackage{braket}
\usepackage{accents}
\usepackage{siunitx}
\usepackage{mathtools}
\DeclareSymbolFontAlphabet{\mathrm}{operators}

\definecolor{CiteColor}{rgb}{0.18039, 0.18824, 0.57255}
\definecolor{UrlColor} {rgb}{0.741, 0.173, 0.000}
\definecolor{DarkUrlColor} {rgb}{0.500, 0.110, 0.000}
\definecolor{LinkColor}{rgb}{0.25098, 0.47843, 0.04706}
\makeatletter %
\newcommand{\ShowFont}{%
  \typeout{The main font is \f@encoding \space \f@family \space %
    \f@series \space \f@shape \space at \f@size pt.}%
  \typeout{The math font sizes are \tf@size pt (main), \sf@size pt %
    (script), and \ssf@size pt (scriptscript).}%
  \typeout{The linewidth is \the\linewidth}} %
\makeatother %
\usepackage{xargs}